\documentclass[prc,aps,floatfix,twocolumn,showpacs,nofootinbib]{revtex4-1}

\usepackage{amsmath}
\usepackage{stmaryrd}
\usepackage{amsfonts}
\usepackage{mathdots}
\usepackage{graphicx}
\usepackage{color}
\usepackage{dsfont} 
\usepackage{enumerate}

\usepackage{mathrsfs} 

\usepackage{hyperref}
\hypersetup{colorlinks=true,
            linkcolor=blue,
            citecolor=blue,
            filecolor=green,
            urlcolor=cyan 
}

\newcommand{\eins}{{\mathchoice {\rm 1\mskip-4mu l} {\rm 1\mskip-4mu l}
   {\rm 1\mskip-4.5mu l} {\rm 1\mskip-5mu l}}}

\usepackage[T1]{fontenc}
\usepackage[utf8]{inputenc}

\allowdisplaybreaks     % allow for breaking of equations

\newcommand{\nuc}[2]{$^{#1}${#2}}

\newcommand{\iunit}{\text{i}}

\newcommand{\op}[1]{\hat{#1}}
\newcommand{\bra}[1]{\langle #1 \vert}
\newcommand{\ket}[1]{\vert #1 \rangle}
\newcommand{\norm}[2]{\langle #1 \vert #2 \rangle}
\newcommand{\elma}[3]{\bra{#1} #2 \ket{#3}}
\newcommand{\redu}[3]{\langle #1 \vert\vert #2 \vert\vert #3 \rangle}

\definecolor{greenbb}{RGB}{34,139,34}
\definecolor{bluebb}{RGB}{46,84,206}
\definecolor{redbb}{RGB}{228,30,43}

\newcommand{\norzero}{\mbox{\normalfont      \bfseries 0}}

\newcommand{\larzero}{\mbox{\normalfont\large\bfseries 0}}

\newcommand{\etal}{\emph{et al.}}
\newcommand{\nn}{\nonumber}

\definecolor{mbscolor}{rgb}{0.60, 0.0, 0.65}

\newcommand{\emb}{{\rule{0cm}{0cm}}}

\graphicspath{{.}{figures/}{newfigures/}} 

\begin{document}

\title{Projection on particle number and angular momentum: Example of triaxial Bogoliubov quasiparticle states}

\author{Benjamin Bally}
\affiliation{Departamento de F\'isica Te\'orica, Universidad Aut\'onoma de Madrid, E-28049 Madrid, Spain}

\author{Michael Bender}
\affiliation{Universit{\'e} de Lyon, Institut de Physique des 2 Infinis de Lyon, IN2P3-CNRS-UCBL, 4 rue Enrico Fermi, F-69622 Villeurbanne, France}

%
%=======================================================================
%
\begin{abstract}
\begin{description}

\item[Background] Many quantal many-body methods that aim at the description of self-bound nuclear or mesoscopic electronic
systems make use of auxiliary wave functions that break one or several of the symmetries of the Hamiltonian in order to include
correlations associated with the geometrical arrangement of the system's constituents.
Such reference states have been used already for a long time within self-consistent methods that are either based on effective valence-space 
Hamiltonians or energy density functionals, and they are presently also gaining popularity in the design of novel \textit{ab-initio} methods. 
A fully quantal treatment of a self-bound many-body system, however, requires the restoration of the broken symmetries through 
the projection of the many-body wave functions of interest onto good quantum numbers.

\item[Purpose] The goal of this work is three-fold. First, we want to give a general presentation of the formalism
of the projection method starting from the underlying principles of group representation theory. Second, we want to 
investigate formal and practical aspects of the numerical implementation of particle-number and angular-momentum 
projection of Bogoliubov quasiparticle vacua, in particular with regard of obtaining accurate results at minimal 
computational cost. Third, we want to analyze the numerical, computational and physical consequences of intrinsic
symmetries of the symmetry-breaking states when projecting them. 

\item[Methods] Using the algebra of group representation theory, we introduce the projection method for the general 
symmetry group of a given Hamiltonian. For realistic examples built with either a pseudo-potential-based energy density functional
or a valence-space shell-model interaction, we then study the convergence and accuracy of the quadrature rules 
for the multi-dimensional integrals that have to be evaluated numerically and analyze the consequences of conserved 
subgroups of the broken symmetry groups.

\item[Results] The main results of this work are also threefold. First, we give a concise, but general, presentation of the 
projection method that applies to the most important potentially broken symmetries whose restoration is relevant for nuclear spectroscopy.
Second, we demonstrate how to achieve high accuracy of the discretizations used to evaluate the multi-dimensional 
integrals appearing in the calculation of particle-number and angular-momentum projected matrix elements while 
limiting the order of the employed quadrature rules.
Third, for the example of a point-group symmetry that is often imposed on calculations 
that describe collective phenomena emerging in triaxially deformed nuclei, we provide 
the group-theoretical derivation of relations between the intermediate matrix elements that are
integrated, which permits for a further significant reduction of the computational cost of the 
method. These simplifications are valid whatever the number parity of the quasiparticle states 
and therefore can be used in the description of even-even, odd-mass, and odd-odd nuclei.
 
\item[Conclusions] 
The quantum-number projection technique is a versatile and efficient method that permits to restore 
the symmetry of any arbitrary many-body wave function. Its numerical implementation is relatively simple and accurate. 
In addition, it is possible to use the conserved symmetries of the reference states to reduce the computational burden
of the method.
More generally, the ever-growing computational resources and the development of nuclear \textit{ab-initio}
methods opens new possibilities of applications of the method. 

\end{description}
\end{abstract}

%\date

\maketitle

%
%=======================================================================
% 
\section{Introduction}

The concept of symmetry is essential to the analysis, discussion, and understanding of many natural 
phenomena~\cite{Henley96a}.  In physics, the notion of symmetry is associated with the existence 
of physical transformations that leave either the laws of physics or the properties of physical systems
invariant~\cite{Caulton15a}. We will focus here on  invariances of the interactions between a system's 
fundamental constituents under global space-time symmetry transformations,
such as translations in time and space, rotations, inversion of space and time, etc, which depend on a number of global 
parameters. Through Noether's first theorem~\cite{Noether18a,Kosmann11a}, such global invariances 
are connected to the conservation of energy, momentum, angular momentum,  parity, particle 
number, etc. 

From a mathematical perspective, the concepts of symmetry can be expressed within the language 
of group representation theory~\cite{Wigner59a,Hamermesh62a,Tinkham92a,McWeeny02a,Lowdin67a}. 
In particular, for finite quantal systems such as the atomic nucleus, the labels of the 
irreducible representations (irreps) of the general symmetry group of the Hamiltonian can be 
used as good quantum numbers that characterize its eigenstates.  As a consequence, 
there are selection rules for electromagnetic transitions in nuclei, and also for nuclear 
transmutations induced by the weak and strong nuclear forces. 

When solving the nuclear many-body problem exactly, the resulting 
many-body wave function will automatically be an eigenstate of all 
symmetry operators that commutate with the Hamiltonian and among each 
other. In one way or another, however, the microscopic modeling 
of nuclei almost always requires an \textit{ansatz} for the nuclear many-body
wave function $\ket{\Psi}$ and also implies the construction of an effective
many-body Hamiltonian $\hat{H}_{\text{eff}}$ adequate for the model 
vector space covered by all possible $\ket{\Psi}$. While $\hat{H}_{\text{eff}}$
is in general constructed to preserve the physical symmetries of the 
``bare'' Hamiltonian, it is not straightforward to ensure that the model
wave functions $\ket{\Psi}$ conserve the same symmetries.

The latter problem is particularly prominent in variational methods, within which
the physical state is approximated by the trial wave function that gives the lowest 
expectation value of the effective Hamiltonian  within the given variational set. 
Conserving the physical symmetries by artificially restricting the 
variational space to symmetry-conserving product states might at first
sight appear advantageous as one keeps quantum numbers and selection rules. 
However, when making the simple \textit{ansatz} of a single product wave function 
for the variational state as it is done in self-consistent mean-field methods, 
more often than not one finds that a symmetry-breaking wave function 
gives a larger binding energy than a symmetry-conserving one. 
This problem is known for long as the ``symmetry dilemma'', a notion 
first coined by L{\"o}wdin~\cite{Lykos63a} in the context of atomic physics.

This finding has many similarities with the phenomenon of spontaneous symmetry 
breaking \cite{Strocchi05a} that is well known for infinite systems such as the ones treated in condensed-matter 
physics. While for those symmetry breaking is an observable feature of the physical systems 
under study, for finite self-bound systems such as atomic nuclei finding a symmetry-breaking state
that has the lowest energy for a symmetry-conserving Hamiltonian might appear as an 
unwanted byproduct of the modeling.  

Time has shown, however, that it is often possible to attribute a physical meaning to such symmetry-breaking 
model states. It has been understood in ever increasing detail since the 1950s \cite{Bohr76a,Rowe70a,BM98a,RS80a} 
that the pattern of excited states of many atomic nuclei can be easily and intuitively explained by making the assumption that  the 
many-body wave function can in one way or another be separated into a part representing a specific 
geometrical arrangement of nucleons and a part that represents the orientation of this arrangement 
in space, and where only the latter respects the symmetries of the Hamiltonian. By contrast, the part
of the wave function that represents the relative arrangement of the nucleons then typically
breaks some, but not necessarily all, of these symmetries. We will call those that remain 
\emph{intrinsic symmetries} in what follows. They leave a characteristic fingerprint on the excitation 
spectrum of the system and are customarily used to characterize the distribution of  nucleons as
having a ``spherical'', ``axially deformed'', ``triaxial'', ``octupole deformed'', \ldots shape 
\cite{Rowe70a,BM98a,RS80a,Frauendorf01a}, although for atomic nuclei the shape  of the nucleon distribution 
as such is experimentally not directly observable as a consequence of the nuclear Hamiltonian's global invariances. 
The same concepts are also used to interpret fine details in the patterns of excited states at high spin 
in  terms of the orientation of angular momenta relative to the ``intrinsic shape'' of the nucleon distribution 
\cite{Frauendorf01a}.

Over fifty years of experience with variationally optimized symmetry-breaking product states have 
shown that they overall provide a predictive description of phenomena that are commonly interpreted in
terms of nuclear shapes. Similarly, the use of Bogoliubov-type quasiparticle vacua instead of
Slater determinants as variational wave function allows for the modeling of pair correlations in nuclei 
at the expense of breaking the global gauge symmetry associated with particle-number conservation.
Both of these successes explain the wide popularity of mean-field-based models, be 
they self-consistent or not. Still, limiting the modeling of nuclei to symmetry-breaking mean-field calculations 
has its limits; such calculations rarely grasp all correlations associated with the broken symmetry, they 
often fail in the limit of weak symmetry breaking, and the connection of observables calculated for the 
intrinsic nucleon distribution to what is observed in the laboratory frame is not straightforward and 
requires additional modeling.

These limitations, however, can be overcome when doing the calculation in two steps.
First, one allows a symmetry-breaking state to explore all degrees of freedom that lower 
the energy, and then restores the broken symmetries by projecting this trial
state on good quantum numbers \cite{Peierls57a,Rowe70a,RS80a,BR86a,MacDonald70a}. Both can 
even be done simultaneously in a so-called ``variation-after-projection''' (VAP) 
calculation, where the symmetry-broken state is optimized to minimize the energy obtained 
after its projection \cite{Sheikh00a}. This strategy has to be contrasted with carrying 
out one after the other, as much more frequently done in the literature, in a so-called 
``projection-after-variation'' (PAV) calculation, where the non-projected energy is 
variationally optimized. In general, both do not lead to exactly the same results. While the 
former is clearly preferable on formal grounds, it is numerically much more costly such that so far it has 
only  applied in frameworks that either make simplifying assumptions for the variational 
states \cite{Kanada98a} or that use very small valence spaces \cite{Gao15a}, or for 
the technically simple case of particle-number restoration \cite{Anguiano01a,Egido16a}. 
It is also possible to design an intermediate strategy, used either under the name of 
Restricted VAP (RVAP) \cite{Rodriguez05a} or minimization after projection (MAP) \cite{Bender06a},
where the minimum is searched within a small space of suitably constructed projected states 
that are each obtained from a PAV calculation. Choosing either of these strategies 
to generate the final projected state, however, does not alter the formal properties of the actual 
projection technique involved in the process. 

The projection technique has been applied in many contexts, but most concern a framework
where it is applied to simple product states obtained either from a Hartree-Fock (HF),
HF+Bardeen-Cooper-Schrieffer (HF+BCS) or Hartree-Fock-Bogoliubov (HFB) calculation.
Many early results were obtained within truncated valence spaces 
\cite{Ripka68a,MacDonald70a,Mang75a,Schmid04a}. Such calculations continue
to offer a computationally efficient approximation to full shell-model diagonalizations for systems 
with inhibitively large valence spaces \cite{Otsuka01a,Shimizu12a,Hara95a,Sun16a}. 

Over the past two decades, it has also 
become popular to systematically apply these techniques in the context of energy density 
functional (EDF) methods that are based on reference product states that are constructed from
all occupied single-particle states 
\cite{Bender04a,Niksic06a,Kimura07a,Bender08a,Rodriguez10a,Guzman12a,Yao11a,Satula10a,Bally14a,Borrajo15a,Rodriguez15a,Egido16a,Egido16b,Robledo18a,Shimada15a,Ushitani19a,Bender19EDF3}.

Because of its many successes, the strategy to use symmetry-unrestricted reference states is
also progressively used within \emph{ab-initio} many-body methods that are based on correlated
trial states \cite{Soma13a,Hergert14a,Signoracci15a,Neff08a}. In this context, the
restoration of the broken symmetries, however, is not yet as widely-used as for mean-field 
methods, but developments in this direction are underway 
\cite{Neff08a,Duguet14a,Hergert16a,Duguet17a,Qiu17a,Qiu19a,Yao20a}.

Quantum numbers of prime interest for nuclear spectroscopy are total 
angular momentum and its third component, parity, proton and neutron 
number, as well as isospin. The latter is not a conserved symmetry of the Hamiltonian, 
but its breaking is usually so weak that it still can be used as a meaningful
label of nuclear states, and also the proper description of its actual breaking can be 
facilitated when using an extension of the symmetry-breaking-plus-symmetry-restoration
scheme that has been sketched above \cite{Satula10a}. For some applications, mainly
to reaction processes, it can also be relevant to restore translational and/or Galilean 
invariance \cite{Rodriguez04a,Rodriguez04b}. We will limit the discussion below to projection on 
angular momentum and particle number. Their restoration is arguably the most widely discussed in 
the literature, and the specificities of their respective group structure are representative also
for other cases of interest. The goals of this article are as follows:
\begin{itemize}
\item
Clarifying the formal origin and the formal properties of the projection operators
as used in nuclear structure calculations, questions that have been rarely, and
to the best of our knowledge never systematically, been addressed in the
nuclear physics literature so far.

\item
Analyzing how applying a numerical projection operator extracts the targeted components
from typical symmetry-breaking states, and how this information can be used to reduce
the numerical cost of projection.

\item
Discussing how intrinsic symmetries of a symmetry-breaking state influence the 
decomposition of this state into symmetry-conserving components and how this
feature can be used to reduce the numerical cost of symmetry restoration. 
On the one hand, these questions concern intrinsic symmetries 
that are inherent to product states and that distinguish configurations describing
systems with even and odd particle number on a very fundamental level. On the 
other hand, intrinsic symmetries related to the distribution of nucleons and their 
angular momentum in the nucleus can also be exploited to further reduce the numerical
cost of projection. Of prime interest for the latter are  subgroups of the double point symmetry group 
$D_{2h}^{TD}$ as defined in Refs.~\cite{Doba00a,Doba00b}.

\end{itemize}

The paper is organized as follows. 
Section~\ref{sec:proj} introduces the projection method on the grounds of
group theory and explains how it allows to build correlated symmetry-restored 
states starting from a arbitrary symmetry-breaking state.
Section~\ref{sec:PNP} then presents formal properties of particle-number
restoration and its numerical implementation as encountered when applied to 
Bogoliubov quasiparticle vacua that describe systems with either even or 
odd particle number. 
Section~\ref{sec:AMP} presents formal properties of angular-momentum
restoration and its numerical implementation as encountered when
applied to Bogoliubov quasiparticle vacua that describe nuclei with even and odd
total particle number. Section~\ref{sec:simple} describes how point-group symmetries of the
intrinsic Bogoliubov quasiparticle vacua can be used to simplify the numerical evaluation 
of the angular-momentum restoration.

While the formulation of the projection technique is straightforward for
methods employing Hamilton operators,
it has been pointed out that it can become ill-defined
as soon as one makes approximations that violate the Pauli principle 
when calculating the total energy \cite{Donau98a,Anguiano01a} or when making 
\textit{ad-hoc} assumptions when setting  up a multi-reference energy density 
functional that does not correspond to the expectation value of a genuine Hamiltonian 
\cite{Dobaczewski07a,Robledo07a,Lacroix09a,Bender09a,Duguet09a,Robledo10a}. 
Regularization schemes have been proposed to overcome such problems \cite{Lacroix09a,Satula14a}, 
but at present it remains unclear if they lead to energies that are 
well-defined under any circumstances. To avoid any ambiguities, we will
introduce the projection method using Hamilton operators and present
only results obtained with such.

%
%=======================================================================
%
\section{Projection method}
\label{sec:proj}
%
%-----------------------------------------------------------------------
%
\subsection{Basic definitions}
\label{sect:proj:def}

In this section, we will briefly recall those elements of group
theory that are needed to define the projection operators, and that 
provide an insight into the interpretation of the projection method.
For a thorough introduction into group theory as needed in many-body 
quantum mechanics and further background information such as the proof 
of the theorems and other relations used in what follows, we refer 
to Refs.~\cite{Lowdin67a,Wigner59a,Hamermesh62a,McWeeny02a,Tinkham92a}.

In nuclear structure physics, we have to deal with discrete 
symmetries, such as parity, and also continuous symmetries such as global 
gauge and rotational invariances. The former are associated with finite 
groups, whereas the latter are represented by Lie groups. 
With the exception of $SU(2)$ (the group related to angular-momentum and 
isospin), all groups of interest are Abelian. 
In addition, all Lie groups considered here will be compact. 

Let us consider a group $G$ that is either a finite group of order $n_G$ or a 
compact Lie group of volume\footnote{We define the volume of 
the group $G$ as the integral over its domain of definition: 
$v_G =\int_{G} dv_G(g)$, $dv_G(g)$ being the invariant measure for $G$.}
$v_G$. Let us then consider the unitary representation that associates to each 
element $g \in G$ the unitary operator $\hat{U}(g)$ acting on the 
Hilbert space $\mathcal{H}$. 

The different irreducible representations of $G$ will be 
labeled by greek letters, e.g.\ $\lambda, \mu$ or $\nu$.
For the types of groups considered here, we know by theorem that all 
irreps of $G$ are finite-dimensional \cite{Hamermesh62a} and we note 
$d_\lambda$ the dimension of the irrep $\lambda$.

To illustrate our discussion, let us first consider the simple case where 
the Hilbert space $\mathcal{H}$ contains one copy of each irrep of $G$, i.e.\ it
can be decomposed as 
\begin{equation}
\label{eq:hdec}
 \mathcal{H} = \bigoplus_{\lambda} S^{\lambda} \, ,
\end{equation}
with $S^{\lambda}$ being the invariant subspace of dimension $d_\lambda$ associated with the 
irrep $\lambda$.
Choosing an orthonormal basis\footnote{\label{fn:labels}Throughout 
this section, we use a generic notation where 
the indices of states within an \textit{irrep} run from one to the irrep's dimension
$d_\lambda$. This convention, which is independent of the specificities
of a given group, has to be distinguished from the frequently found 
group-dependent labeling where indices are associated with the eigenvalue 
of some operator that differentiates between the members of a given \textit{irrep} 
of a specific group. The latter notation, where the labels carry a direct
physical interpretation, will be used later on when discussing the restoration 
of specific symmetries in Sections~\ref{sec:PNP} and~\ref{sec:AMP}.}
$\big\{ \ket{\Psi^{\lambda i}}$, 
$i \in  \llbracket 1, d_\lambda  \rrbracket \big\}$ for $S^{\lambda}$, 
the transformation of the basis 
functions under the action of $\hat{U}(g)$ reads
\begin{equation} 
 \label{eq:tranbasis}
 \forall \, g \in G , \,  
 \hat{U}(g) \ket{\Psi^{\lambda i}} 
 =  \sum_{j=1}^{d_\lambda} D_{ji}^{\lambda}(g) \ket{\Psi^{\lambda j}} \, , 
\end{equation}
i.e.\ the transformed basis function is in general a superposition of 
several basis functions from the same irrep with the coefficients of the superposition being the matrix elements 
\begin{equation}
 \label{eq:tranelele}
   D_{ji}^{\lambda}(g) \equiv \elma{\Psi^{\lambda j}}{\hat{U}(g)}{\Psi^{\lambda i}} 
\end{equation}
of the unitary matrix $\boldsymbol{D}^\lambda (g)$, which itself is the 
matrix representation of the group element $g$ for irrep $\lambda$. 
Between the irreps of a given finite group hold the orthogonality 
relations
\begin{equation}
\label{eq:GOT:finite}
 \frac{d_\mu}{n_G} \sum_{G} {D_{k l}^{\mu}}\phantom{}^*(g) {D_{ij}^{\lambda}}(g) = \delta_{\mu\lambda} \delta_{ki} \delta_{l j} \, ,
\end{equation}
whereas for a compact Lie group one has
\begin{equation}
\label{eq:GOT:Lie}
 \frac{d_\mu}{v_G} \int_{G} dv_G(g) {D_{k l}^{\mu}}\phantom{}^*(g) {D_{ij}^{\lambda}}(g)  = \delta_{\mu\lambda} \delta_{ki} \delta_{l j} 
\end{equation}
instead. This so-called Great Orthogonality Theorem \cite{Hamermesh62a}
is the fundament on which the projection technique is built, as it permits us to define the 
linear operators
\begin{equation}
\hat{P}^{\mu}_{k l} 
\equiv  \frac{d_\mu}{n_G} \sum_{G} {D^{\mu}_{k l}}\phantom{}^*(g) \hat{U}(g)
\end{equation}
for finite groups, and
\begin{equation}
   \hat{P}^{\mu}_{k l} 
\equiv \frac{d_\mu}{v_G} \int_{G} dv_G(g) {D^{\mu}_{k l}}\phantom{}^*(g) \hat{U}(g) 
\end{equation}
for compact Lie groups, respectively. These operators act on the basis 
functions $\ket{\Psi^{\lambda i}}$ as
\begin{equation}
 \label{eq:projbasis}
  \hat{P}_{kl}^{\mu} \ket{\Psi^{\lambda i}} 
= \delta_{\mu\lambda} \delta_{l i} \ket{\Psi^{\lambda k}} \, ,
\end{equation}
i.e.\ the operator $\hat{P}_{kl}^{\mu}$ either cancels out the basis 
functions corresponding to irreps $\lambda \neq \mu$ or transforms
the basis functions within the irrep $\mu$ one into another.
In Dirac's bra-ket notation, this relation can be rewritten within our basis states as
\begin{equation}
 \label{eq:projbraket}
 \hat{P}^{\mu}_{kl} 
= \ket{\Psi^{\mu k}} \bra{\Psi^{\mu l}} \, .
\end{equation}
It can be deduced from these expressions that by acting with the operators 
$\hat{P}_{kl}^{\mu}$ on an arbitrary superposition of basis functions
\begin{equation}
 \label{eq:extractproj}
  \hat{P}_{kl}^{\mu} \sum_{\lambda} \sum_{i=1}^{d_\lambda} c^{\lambda i} \ket{\Psi^{\lambda i}} = c^{\mu l} \ket{\Psi^{\mu k}} \, ,
\end{equation}
with the coefficents $c^{\lambda i}$ being complex numbers,  
we extract the weight of the $(\mu,l)$ component 
in the original state times the $(\mu,k)$ basis function.
With that, the operators $\hat{P}^{\mu}_{kl}$ can be used to project out 
the components belonging to any irrep $\mu$ contained in such 
a superposed state.
For that reason, these operators  are frequently called projection operators 
in the physics literature. However, Eqs.~\eqref{eq:projbasis} 
and~\eqref{eq:projbraket} imply that
\begin{equation}
 \label{eq:projcarre}
 \hat{P}^{\lambda}_{ij} \hat{P}^{\mu}_{kl} 
 = \delta_{\lambda\mu} \delta_{jk} \hat{P}^{\lambda}_{il} \, , 
\end{equation}
meaning that the operators 
$\hat{P}_{kl}^{\mu}$ are in general not projection operators in the
mathematical sense of being a linear map $p$ with the property $p^2 = p$.
Only the operators $\hat{P}_{kk}^{\mu}$ are true projectors in that sense.
To underline this difference, the operators $\hat{P}_{k l}^{\mu}$ with 
$k \neq l$ are sometimes called \emph{shift operators} \cite{McWeeny02a} 
or \emph{transfer operators} \cite{Tinkham92a} in the literature.
Moreover, using the unitarity of the representation, we see that that under 
hermitian conjugation, the operators $\hat{P}_{kl}^{\mu}$ transform as
\begin{equation}
  \label{eq:projhermi}
  \hat{P}_{kl}^{\mu}\phantom{}^\dagger = \hat{P}_{lk}^{\mu} \, . 
\end{equation}
From this follows that the operators $\hat{P}_{kk}^{\mu}$ are hermitian, 
as expected for true projection operators.
For the special case of Abelian groups the irreducible representations 
are all one-dimensional \cite{Hamermesh62a}. For these, the index $k$ 
labeling the states within a given irrep is redundant and can hence be 
omitted. In such a case, the operators 
$\hat{P}^{\mu} \equiv \hat{P}^{\mu}_{kk}$ are always true projection 
operators.

Based on the aforementioned properties, we can write  
the resolution of the identity within our basis as \cite{Wigner59a}
\begin{equation}
 \label{eq:projident}
  \sum_\mu \sum_{k=1}^{d_\mu} \hat{P}^{\mu}_{kk} 
= \sum_\mu \sum_{k=1}^{d_\mu} \ket{\Psi^{\mu k}} \bra{\Psi^{\mu k}}
= \hat{\eins} \, .
\end{equation}
Up to now, we have concentrated on a generic set of basis functions 
of the irreps, labeled, for example, as $\ket{\Psi^{\mu k}}$. 
However, the Hilbert space $\mathcal{H}$ of interest will in general contain 
multiple subspaces whose elements transform according to the same irrep $\mu$ 
of $G$. Hence, it is necessary to distinguish the basis functions by 
additional quantum numbers and/or labels that are not related to symmetry 
group $G$. For these, we will use the generic label $\epsilon$ such that 
the Hilbert space can then be decomposed as 
\begin{equation}
\label{eq:bighdec}
 \mathcal{H} = \bigoplus_{\mu} \bigoplus_{\epsilon = 1}^{n_\mu} S^{\mu}_{\epsilon} \, ,
\end{equation}
with $S^{\mu}_{\epsilon}$ being one of the $n_\mu$ invariant subspaces of $\mathcal{H}$ associated with
the irrep $\mu$.

Operators $\hat{P}^{\mu}_{kl}$ do not act on the degrees of freedom
related to $\epsilon$, meaning that they cannot distinguish between 
states with same $\mu$ and $k$ or $l$ but different $\epsilon$.  
As a consequence, it is in general necessary to sum over all 
$\epsilon$ when expressing the projection operators in Dirac's bra-ket 
notation in the full Hilbert space
\begin{equation}
 \label{eq:projbraket:alpha}
 \hat{P}^{\mu}_{kl}  = \sum_{\epsilon = 1}^{n_\mu} \ket{\Psi^{\mu k}_{\epsilon}} \bra{\Psi^{\mu l}_{\epsilon}} \, .
\end{equation}
The same is also necessary to obtain the equivalent of 
Eq.~\eqref{eq:projident} in the full Hilbert space
\begin{equation}
 \label{eq:projident2}
  \sum_\mu \sum_{k=1}^{d_\mu} \hat{P}^{\mu}_{kk} 
= \sum_\mu \sum_{\epsilon = 1 }^{n_\mu} \sum_{k=1}^{d_\mu} \ket{\Psi^{\mu k}_{\epsilon}} \bra{\Psi^{\mu k}_{\epsilon}} 
= \hat{\eins} \, .
\end{equation}
In many situations of physical interest, the full symmetry group $G$ 
of the system can be broken down into a direct product of its 
subgroups, e.g. $G = G_1 \times G_2 \times \cdots \times G_m$.
In that case, basis functions of $G$ can be constructed as tensor products 
of basis functions of the constituent groups of the direct product, 
and similarly for projection operators; see Refs.~\cite{Lowdin67a,Wigner59a,Hamermesh62a,McWeeny02a,Tinkham92a} for details.
In addition, the label of the irreps of $G$ 
can be expressed as a set of $m$ labels denoting the irreps of each of the subgroups, i.e.\ $\lambda \equiv \lambda_1, \lambda_2, \ldots, \lambda_m$.

%
%-----------------------------------------------------------------------
%
\subsection{Tensor operators}
\label{sect:tensorop}

Analogously to basis functions, it is possible to identify certain operators 
that have special transformation properties under symmetry operations. 
A set of $d_\lambda$ operators 
$\big\{ \hat{T}^{\lambda}_{i} , i \in \llbracket 1, d_\lambda  \rrbracket \big\}$ 
that transform as
\begin{equation}
\label{eq:tensortrans}
\forall \, g \in G , \, \hat{U}(g) \hat{T}^{\lambda}_{i} \hat{U}^\dagger(g)
= \sum_{j=1}^{d_\lambda} D^{\lambda}_{ji}(g) \, \hat{T}^{\lambda}_{j}
\end{equation}
will be referred to as a set of \textit{tensor operators} of rank $\lambda$.
The ones that transform according to the trivial representation\footnote{i.e.\ the one-dimensional 
representation where all elements of $G$ are associated to the identity operation.}
$\lambda_t$ are called \textit{scalar operators}. For such scalar 
operator $\hat{T}^{\lambda_t}_{1}$, Eq.~\eqref{eq:tensortrans} can be 
straightforwardly recasted into the commutation relation
\begin{equation}
\forall \, g \in G , \, \big[ \hat{U}(g), \hat{T}^{\lambda_t}_{1} \big] = 0 \, .
\end{equation}
The vast majority of observables of interest in nuclear physics 
can be expressed in terms of tensor operators of the general symmetry
group of the Hamiltonian, whether being directly a tensor operator
themself or reducible as a sum of such operators.

%
%-----------------------------------------------------------------------
%
\subsection{Symmetry group of the Hamiltonian}
\label{sect:symmH}

Considering a Hamiltonian $\hat{H}$, $G$ is said to be a 
symmetry group of $\hat{H}$ if the latter commutes with all unitary 
operators $\hat{U}(g)$ associated with the elements $g$ of $G$
\begin{equation}
\label{eq:hcomu}
\forall \, g \in G , \, \big[ \hat{U}(g) , \hat{H} \big] = 0 \, ,
\end{equation} 
i.e.\ the Hamiltonian is a scalar operator with regards to $G$.
The largest group under which $\hat{H}$ transforms as a scalar 
will be called the general symmetry group of the Hamiltonian.
Relation \eqref{eq:hcomu} has several important consequences for 
the eigenstates and eigenvalues of the Hamiltonian $\hat{H}$ 
\cite{Wigner59a}
\begin{enumerate}[(i)]
\item 
With the exception of accidental degeneracies, each eigenspace of $\hat{H}$ 
corresponds to a single irrep of $G$.\footnote{In the exceptional case of an 
accidental degeneracy, the corresponding eigenspace of the Hamiltonian 
$\hat{H}$ can still be decomposed into a direct  sum of irreps. 
The presence of 
systematic degeneracies of different irreps of $G$ signals that one is 
not considering the general symmetry group of $\hat{H}$.} 
Consequently, the degeneracy of the corresponding eigenvalue of $\hat{H}$ 
is determined by the dimension $d_\lambda$ of the irrep.
\item
The irreps' label $\lambda$ can then be used as good quantum number 
to label the eigenstates and eigenvalues of $\hat{H}$. 
In general, however, an additional label will be needed to distinguish
between different eigenspaces of $\hat{H}$ with same $\lambda$.
\item 
There are selection rules for the matrix elements of tensor operators
$\hat{T}^{\lambda}_{i}$ between the eigenstates of $\hat{H}$
that are captured by the Wigner-Eckart theorem associated with the group $G$. 
\end{enumerate}
%

%-----------------------------------------------------------------------
%
\subsection{Symmetry-breaking model wave functions}

As outlined in the introduction, nuclear EDF and other methods make use of 
symmetry-breaking wave functions in order to address a multitude of 
nuclear phenomena in a  computationally-friendly way. But as a result, 
the auxiliary states that these models are built on
break some, if not all, of the symmetries $G$ of the nuclear 
Hamiltonian $\hat{H}$ and therefore lack some of the essential 
quantum mechanical characteristics of the eigenstates of $\hat{H}$.
In particular, we loose the selection rules related to the symmetry $G$ 
for transition moments and other observables, which may severely 
spoil the accuracy and reliability of their evaluation.

Nevertheless, from the commutation relation~\eqref{eq:hcomu} it follows
that for an arbitrary state $\ket{\Theta}$
\begin{equation}
\label{eq:T+U+HUT}
\forall \, g \in G , \, 
\bra{\Theta} \hat{U}^{\dagger}(g) \hat{H} \hat{U}(g) \ket{\Theta}
= \bra{\Theta} \hat{H} \ket{\Theta} \, ,
\end{equation}
meaning that all elements of the set  
\mbox{$G\ket{\Theta} \equiv \big\{\hat{U}(g) \ket{\Theta} , \, 
g \in G \big\}$}, i.e.\ the set of all 
``rotated'' states built from $\ket{\Theta}$, 
are degenerate. This set is formally known as an orbit of $G$ \cite{RW10a}.
Equation~\eqref{eq:T+U+HUT} has the important practical
consequence that the orientation of a symmetry-breaking state
$\ket{\Theta}$ can be freely chosen to whatever is the most advantageous 
for its numerical representation. In addition, Eq.~\eqref{eq:T+U+HUT} suggests 
also that the states in $G\ket{\Theta}$
may strongly interact with each other and therefore that we may gain 
correlation energy by mixing them. This is precisely what 
the projection method accomplishes.

%
%-----------------------------------------------------------------------
%

\subsection{Projection method}
\label{subsect:projection:method}

As an introductory note, we want to point out that the projection method
presented in this section can be applied whether the group $G$ at hand 
is the general symmetry group of the Hamiltonian or if it is only a subpart
of its direct product decomposition (assuming such decomposition exists).

Starting from an arbitrary normalized, and \textit{a priori} 
symmetry-breaking, state $\ket{\Theta}$, there is an elegant way to build 
symmetry-respecting states by diagonalizing $\hat{H}$
within the subspace of $\mathcal{H}$ spanned by all linear combinations 
of the degenerate elements of $G\ket{\Theta}$. This space 
is provided by
\begin{equation}
\label{eq:spanG:d}
\text{span}(G\ket{\Theta}) 
\equiv \Bigg\{ \sum_G f(g) \ket{\Theta(g)}  , \, f(g) \in \mathbb{C} \Bigg\}
\end{equation}
for finite groups, where $\mathbb{C}$ are the complex numbers, and 
\begin{equation}
\label{eq:spanG:c}
\text{span}(G\ket{\Theta}) 
\equiv \Bigg\{ \int_G dv_G(g) \, f(g) \ket{\Theta(g)}  , \, 
f \in L^2(G) \Bigg\} 
\end{equation}
for compact Lie groups, where $L^2(G)$ is the space of square-integrable 
functions over $G$.

First of all, we notice that, by construction, $\text{span}(G\ket{\Theta})$ 
carries a natural representation of $G$ built from the 
restriction\footnote{More generally, for a given operator $\hat{O}$, 
we will for the rest of the section only consider its
restriction to the subspace $\text{span}(G\ket{\Theta})$ that can be written 
$\hat{O}_S \equiv \hat{P}_S \hat{O} \hat{P}_S$ with $\hat{P}_S$ being
a projection operator onto $\text{span}(G\ket{\Theta})$. However, as there is no ambiguity
on the vector space considered and to keep notations simple, we will omit the index $S$ and
use the label $\hat{O}$ both for the operator and its restricted version.} 
of the operators $\hat{U}(g)$ to this
subspace. In addition, we know by theorem \cite{Hamermesh62a} that 
any unitary representation of finite and compact Lie groups is, up 
to equivalency, either irreducible or can be completely decomposed 
as a direct sum of irreps. This implies that in the general case we 
can decompose $\text{span}(G\ket{\Theta})$ into a direct sum 
of the invariant subspaces $S^{\lambda}_{\epsilon}$ of dimension $d_\lambda$ associated 
with the irreps $\lambda$ of $G$
\begin{equation}
\label{eq:subdec}
\text{span}(G\ket{\Theta}) 
= \bigoplus_{\lambda} \bigoplus_{\epsilon=1}^{n_\lambda} 
S^{\lambda}_{\epsilon} \, .
\end{equation}
As already mentioned, the label $\epsilon$ is used to 
distinguish between the different subspaces $S^{\lambda}_{\epsilon}$ 
that carry an irrep with same $\lambda$. The number $n_\lambda$ of 
such different subspaces with same $\lambda$ depends on the state 
$\ket{\Theta}$, in particular of the remaining symmetries it carries, 
but is such that $0\le n_\lambda \le d_\lambda$.
The upper bound for $ n_\lambda$
is a consequence of the decomposition of the regular representation 
(see Peter-Weyl theorem \cite{Peter27a} for compact Lie groups). 
It is also to be remarked that not all possible irreps $\lambda$ 
of $G$ have to appear in the direct sum \eqref{eq:subdec}, 
and for those vanishing irreps we set $n_\lambda=0$.

As $G$ is assumed to be a good symmetry of $\hat{H}$, 
it is possible to find in $\text{span}(G\ket{\Theta})$ a basis of orthonormal 
eigenfunctions of\emb\ $\hat{H}$
\begin{equation}
\label{eq:eigenh}
 \hat{H} \ket{\Psi^{\lambda i}_{\epsilon}} = e^{\lambda}_{\epsilon} \ket{\Psi^{\lambda i}_{\epsilon}}
\end{equation}
such that the decomposition~\eqref{eq:subdec} still holds. In that case, 
the label $\epsilon$ is used to distinguish between the different eigenspaces 
of $\hat{H}$ with same $\lambda$ in $\text{span}(G\ket{\Theta})$. 

The challenge is now to construct these basis functions in an efficient manner
starting uniquely from state $\ket{\Theta}$. One obvious approach would be to simply  
diagonalize the Hamiltonian matrix composed by all rotated states, whose matrix elements are
\begin{equation}
\label{eq:ThetaGHThetaGp}
 \forall \, (g,g') \in G^2 , \elma{\Theta(g)}{\hat{H}}{\Theta(g')} \, .
\end{equation}
Unless when working in small valence spaces that provide an automatic sharp cutoff for the 
spectrum of irreps of continuous groups that the states $\ket{\Theta}$ can be decomposed into, 
for most applications to nuclear structure physics this direct method is in practice very inefficient, 
or even impossible, to implement. It requires to use a discretization of $g$ for which all basis 
functions $ \ket{\Psi^{\lambda i}_{\epsilon}} $ that $\ket{\Theta}$ can be developed into are 
orthonormal with a high numerical precision. The numerical cost of calculating the kernels
\eqref{eq:ThetaGHThetaGp} with such discretization and of diagonalizing the resulting matrix 
will often be prohibitive.

A better strategy is to exploit the properties of projection operators as defined in Sec.~\ref{sect:proj:def}, to first pre-diagonalize the Hamiltonian within in each subspace
$S^\lambda \equiv \bigoplus_{\epsilon=1}^{n_\lambda} S^{\lambda}_{\epsilon}$
of $\text{span}(G\ket{\Theta})$ such that it only remains to diagonalize $\hat{H}$
within each $S^\lambda$. From a numerical point of view, such approach can 
be implemented in a manner that is systematically improvable to a given accuracy.

First, we notice the fact that $\ket{\Theta}$ trivially belongs to $\text{span}(G\ket{\Theta})$ 
\begin{equation}
\ket{\Theta} = \hat{\eins} \, \ket{\Theta} 
= \hat{U}(1_G) \ket{\Theta} \in \text{span}(G\ket{\Theta}) \, ,
\end{equation}
where $1_G$ is the unit element of $G$, to decompose it into the
$\ket{\Psi^{\lambda i}_{\epsilon}}$ defined through Eq.~\eqref{eq:eigenh}
\begin{equation} 
\label{eq:decompoG}
\ket{\Theta} 
= \sum_{\lambda} \sum_{\epsilon=1}^{n_\lambda} 
  \sum_{i=1}^{d_\lambda} c^{\lambda i}_{\epsilon} \,
  \ket{\Psi^{\lambda i}_{\epsilon}} \, ,
\end{equation}
with the coefficients $c^{\lambda i}_{\epsilon}$ being complex numbers 
with the sum rule
\begin{equation}
\label{eq:sum:rule:1} 
\sum_\lambda \sum_{\epsilon=1}^{n_\lambda} 
\sum_{i=1}^{d_\lambda} | c^{\lambda i}_{\epsilon} |^2 
= 1
\end{equation}
to respect the normalization of $\ket{\Theta}$.
These relations imply in particular that we can always write $\ket{\Theta}$ as 
a superposition of basis functions having good symmetry properties. 

Then, acting with the projection operator $\hat{P}^{\lambda}_{ij}$
on the state $\ket{\Theta}$ we place ourself in the subspace $S^\lambda$
of interest 
\begin{equation}
\hat{P}^{\lambda}_{ij} \ket{\Theta} 
= \sum_{\epsilon=1}^{n_\lambda}  c^{\lambda j}_{\epsilon} \,
  \ket{\Psi^{\lambda i}_{\epsilon}} 
\, . 
\end{equation}
It is to be noted that, depending on the decomposition 
\eqref{eq:decompoG}, for certain values of $\lambda$ and $j$ the states 
$\hat{P}^{\lambda}_{ij} \, \ket{\Theta}$  can be the null vector. 
The non-vanishing states $\hat{P}^{\lambda}_{ij} \, \ket{\Theta}$ represent a first step 
in our process as (i) they have good symmetry transformation under the action of
$\hat{U}(g)$, i.e.\ 
the set of states 
$\big\{ \hat{P}^{\lambda}_{ij} \, \ket{\Theta} , \, i \in \llbracket 1, d_\lambda  \rrbracket \big\}$ 
transform according to Eq.\ \eqref{eq:tranbasis}, 
and (ii) they partially diagonalize the Hamiltonian 
\begin{align}
\label{eq:PHP}
\elma{\Theta}{\, \hat{P}^{\mu}_{ij}\phantom{}^\dagger \, \hat{H} \, 
\hat{P}^{\lambda}_{kl} \, }{\Theta} 
& = \delta_{\mu \lambda} \, \delta_{ik} \, \elma{\Theta}{\, \hat{H} \, \hat{P}^{\lambda}_{jl} \, }{\Theta} \, , \\
\elma{\Theta}{\, \hat{P}^{\mu}_{ij}\phantom{}^\dagger \, \hat{P}^{\lambda}_{kl} \, }{\Theta} 
   &= \delta_{\mu \lambda} \, \delta_{ik} \, \elma{\Theta}{\, \hat{P}^{\lambda}_{jl} \, }{\Theta} \, ,
\end{align}
where we have used the properties~\eqref{eq:projcarre} 
of the projection operators and also
that the projection operators commute with
the Hamiltonian
\begin{equation}
  \forall \, \mu,i,j, \, \big[ \hat{H} , \hat{P}^{\mu}_{ij} \big] = 0
\end{equation}
as a consequence of relation~\eqref{eq:hcomu}.
However, neither the Hamiltonian matrix $\boldsymbol{H}^{\lambda}$ 
nor the norm matrix $\boldsymbol{N}^{\lambda}$, whose
elements are
\begin{align}
 H^{\lambda}_{ij} &\equiv \elma{\Theta}{ \, \hat{H} \, \hat{P}^{\lambda}_{ij} \, }{\Theta} \, , \\
 N^{\lambda}_{ij} &\equiv \elma{\Theta}{ \, \hat{P}^{\lambda}_{ij} \, }{\Theta} \, , 
\end{align}
are automatically diagonal. Only in the trivial, 
but important, case 
where $n_\lambda = d_\lambda = 1$ this is necessarily true.  This is for 
example the case for Abelian groups because they have only one-dimensional 
irreps and therefore $0 \leq n_\lambda \leq d_\lambda= 1$.

In the case of non-Abelian groups, in addition to acting with the projection 
operator, we also have to concurrently  
diagonalize the norm and the 
Hamiltonian matrix among the states $\hat{P}^{\lambda}_{ij} \, \ket{\Theta}$. 
We thus represent the eigenstates of $\hat{H}$ as a superposition of states of 
the form
\begin{equation}
\label{eq:supereps}
\ket{\Psi^{\lambda i}_{\epsilon}} 
= \sum_{j=1}^{d_\lambda} f^{\lambda j}_{\epsilon} \, 
  \hat{P}^{\lambda}_{ij} \, \ket{\Theta} \, ,
\end{equation}
where the weight factors $f^{\lambda j}_{\epsilon}$ are complex numbers.
Injecting Eq.~\eqref{eq:supereps} into Eq.~\eqref{eq:eigenh}, we obtain 
the generalized eigenvalue problem (GEP)
\begin{equation}
\label{eq:gepproj}
\boldsymbol{H}^{\lambda} \, \boldsymbol{f}^{\lambda}_{\epsilon}  
= e^{\lambda}_{\epsilon} \, \boldsymbol{N}^{\lambda} \, \boldsymbol{f}^{\lambda}_{\epsilon} \, ,
\end{equation}
with $\boldsymbol{f}^{\lambda}_{\epsilon}$ being a column vector containing the weight factors. The energies 
$e^{\lambda}_{\epsilon}$ are generalized eigenvalues, i.e.\ the roots
of the characteristic equation
\begin{equation}
\text{det}\big( \boldsymbol{H}^{\lambda} - e \boldsymbol{N}^{\lambda} \big) = 0 \, .
\end{equation}
The matrix $\boldsymbol{H}^{\lambda}$ is hermitian, whereas the matrix $\boldsymbol{N}^{\lambda}$, 
being a Gramiam matrix,\footnote{The Gramian matrix $A_{ij} = \langle v_i | v_j \rangle$ is 
the matrix built from the scalar products of all pairs of vectors $| v_i \rangle$ within a given 
set, which in our case is the set 
$\{ \hat{P}^{\lambda}_{ij} \ket{\Theta} , \, j \in  \llbracket 1, d_\lambda  \rrbracket \}$. 
A Gramian matrix is always positive semidefinite, with the
strictly definite case being realized if and only if all the vectors in the set are linearly independent.}
is positive semidefinite. As a consequence, the GEP defined by 
Eq.~\eqref{eq:gepproj} is a hermitian positive semidefinite GEP and 
therefore has a number $n_\lambda$ of finite real eigenvalues 
$e^{\lambda}_{\epsilon}$ equal to the number of non-zero eigenvalues 
of $\boldsymbol{N}^{\lambda}$. 
In particular, for matrices $\boldsymbol{N}^{\lambda}$ that are
strictly definite 
one obtains $n_\lambda=d_\lambda$ finite real eigenvalues 
$e^{\lambda}_{\epsilon}$ when solving Eq.~\eqref{eq:gepproj}.
Otherwise it is necessary to diagonalize $\boldsymbol{N}^{\lambda}$ 
first and to remove all its $d_\lambda - n_\lambda$ zero eigenvalues
in an intermediate step before diagonalizing the Hamiltonian in the 
such reduced subspace \cite{RS80a}.

Equation~\eqref{eq:gepproj} is independent on the label $i$ of the
state $\ket{\Psi^{\lambda i}_{\epsilon}}$, i.e.\ the same equation holds for 
all $d_\lambda$ values it can take. This implies that 
the energies $e^{\lambda}_{\epsilon}$ are $d_\lambda$-fold degenerate, as 
expected for the eigenvalues of $\hat{H}$ from the discussion in 
Sect.~\ref{sect:symmH}. With that, Eq.~\eqref{eq:gepproj} has to be 
solved only for one state $\ket{\Psi^{\lambda i}_{\epsilon}}$ out of 
each eigenspace. 
All other symmetry partners of the basis can then be obtained
through the use of the shift operators 
\begin{equation}
\label{eq:shift}
\forall \, k \in \llbracket 1, d_\lambda  \rrbracket , \,  
\ket{\Psi^{\lambda k}_{\epsilon}} 
= \hat{P}^{\lambda}_{ki} \ket{\Psi^{\lambda i}_{\epsilon}} \, . 
\end{equation}
Having solved the GEP of Eq.~\eqref{eq:gepproj}, we have the weights
$f^{\lambda j}_{\epsilon}$ entering the states 
$\ket{\Psi^{\lambda i}_{\epsilon}}$, Eq.~\eqref{eq:supereps}, and
the corresponding energy $e^{\lambda}_{\epsilon}$, Eq.~\eqref{eq:gepproj},
at our disposal.
Repeating the process for each $\lambda$ that can be found in the
symmetry-breaking state $\ket{\Theta}$, we obtain a set of orthonormal 
basis functions $\{ \ket{\Psi^{\lambda i}_{\epsilon}} , 
\lambda , \, \epsilon \in \llbracket 1, n_\lambda  \rrbracket , \, i \in 
\llbracket 1, d_\lambda  \rrbracket \}$ of $\text{span}(G\ket{\Theta})$ 
that transform according to the restored symmetry and diagonalize the
Hamiltonian in this space
\begin{subequations}
\begin{align}
\hat{U}(g) \, \ket{\Psi^{\lambda i}_{\epsilon}} 
& =  \sum_{j=1}^{d_\lambda} D^\lambda_{ji}(g) \, 
      \ket{\Psi^{\lambda j}_{\epsilon}} \, , \\
\norm{\Psi^{\mu i}_{\xi}}{\Psi^{\lambda j}_{\epsilon}} 
& =  \delta_{\mu\lambda} \, \delta_{ij} \, \delta_{\xi\epsilon}  \, , \\
\elma{\Psi^{\mu i}_{\xi}}{\, \hat{H} \, }{\Psi^{\lambda j}_{\epsilon}} 
& =  \delta_{\mu\lambda} \, \delta_{ij} \, \delta_{\xi\epsilon} \, 
    e^{\lambda}_{\epsilon} \, .
\end{align}
\end{subequations}

%
%------------------------------------------------------------------------
%
\subsection{Discussions}
\label{subsect:projdiscussion}

Thus formulated, see also \cite{BR86a,RW10a,Lowdin67a}, 
the projection method is not simply the extraction of states with 
good quantum numbers from $\ket{\Theta}$, but the efficient 
construction of
states diagonalizing $\hat{H}$ in the subspace $\text{span}(G\ket{\Theta})$,
which automatically have good symmetry properties.

Alternatively, the projection method can also be formulated from a 
variational point of view \cite{BR86a,Sheikh19a}, where the projection operators
emerge naturally from the knowledge of the decomposition of $L^2(G)$ in 
terms of irreps. From that perspective, the projection method can also be 
interpreted as a special case of the
generator coordinate method (GCM) based on the 
set made of the degenerate rotated states 
$G\ket{\Theta} \equiv \big\{\hat{U}(g) \ket{\Theta} \, , \,g  \in G \big\}$, 
where the group element $g$ provides the
generator coordinate and where the weights are partially determined 
by the structure of the group $G$. Depending on the properties of the
corresponding group, the form of the GCM trial wave function is then
given by either Eq.~\eqref{eq:spanG:d} or Eq.~\eqref{eq:spanG:c}.

The diagonalization of $\hat{H}$ in $\text{span}(G\ket{\Theta})$
as such should not be interpreted as an approximation to the diagonalization 
of $\hat{H}$ in the full model space the Hamiltonian has been constructed for. 
The energy spectrum and all other properties of the symmetry-respecting basis 
functions $\ket{\Psi^{\lambda i}_{\epsilon}}$ depend very sensitively on the
choices made for the symmetry-breaking state $\ket{\Theta}$. The main 
purpose of projection is to construct an orthogonal set of symmetry-respecting
states that provide the lowest possible energy for each irrep of interest
contained in $\ket{\Theta}$. To find an approximation to the physical states
of a system, it will be necessary to embed projection into a many-body method
that scans a suitably chosen and sufficiently large model space for the 
optimal symmetry-breaking states $\{ \ket{\Theta} \}$ that give the lowest 
possible energy for each irrep of interest after projection. In general, 
the optimum state $\ket{\Theta}$ will be different for each irrep.
Such a search can be achieved with variational methods for single product 
states \cite{Egido82a,Maqbool11a} and also for superpositions of product states, either 
formulated as a Generator Coordinate Method \cite{Bender03a,Egido16a} 
or as the configuration
mixing in a non-orthogonal basis \cite{Otsuka01a,Shimizu12a,Jiao19a}.
In both cases, the evaluation of matrix elements between projected states 
remains tractable, although non-trivial,\footnote{For example, 
the calculation of overlaps between arbitrary Bogoliubov quasiparticle
states that are required to compute the norm matrix between projected states in multi-reference
EDF methods remained an open problem for decades, with solutions 
known only for special cases.
Only recently, with the use of Pfaffians \cite{Robledo09a,Avez12a,Mizusaki18a,Carlsson20a} or other 
techniques \cite{Bally18a}, the problem was solved in a general and unambiguous manner.}
while the calculation 
of the projected wave function is not.
It is to be noted that, inspired by the effectiveness of the projection method 
within EDF-based methods, recent theoretical schemes have been proposed 
to incorporate symmetry breaking and restoration into \textit{ab initio} 
methods \cite{Duguet14a,Duguet17a,Hermes17a,Qiu17a,Qiu19a,Yao19a}.

In general, any remaining \textit{intrinsic} symmetry of $\ket{\Theta}$  
adopted during its generation 
might leave some characteristic fingerprint on the outcome of the projection 
scheme. Indeed, when concerning the same degrees of freedom as $\hat{U}(g)$, 
the symmetries of $\ket{\Theta}$ will cause linear dependences among the 
rotated states
in $G\ket{\Theta}$ and thus will reduce the dimensionality of 
$\text{span}(G\ket{\Theta})$. 
The most basic example is when we start with a state $\ket{\Theta}$
that is already a basis function of 
a particular irrep $\lambda$, by applying $\hat{U}(g)$ on $\ket{\Theta}$, 
we just generate a space $\text{span}(G\ket{\Theta}) = S^{\lambda}_1$ that 
contains all $d_\lambda$ linearly independent degenerate basis states of 
the single irrep $\lambda$. No correlation energy is generated along the way,
which assures us of the internal consistency of the method.
Another simple example is when we assume that the state 
$\ket{\Theta}$ has, for a given $g' \in G$, the symmetry relation
\mbox{$\hat{U}(g') \ket{\Theta} = \eta \ket{\Theta} $}, 
\mbox{$\eta \in \mathbb{C}$}, it is easy to understand that the number 
of linearly independent rotated elements $\hat{U}(g) \ket{\Theta}$ that 
we can built is reduced and thus also is the dimension of 
$\text{span}(G\ket{\Theta})$.\footnote{This can be seen through
the relation it implies among the components of $\ket{\Theta}$, namely:
$\eta \, c^{\lambda i}_{\epsilon} = \sum_{j=1}^{d_\lambda} D^{\lambda}_{ij}(g') \, c^{\lambda j}_{\epsilon}$.}
Illustrative examples will be discussed below when addressing specific symmetries.

Using the resolution of the identity [Eq.\ \eqref{eq:projident2}], it is possible to obtain for an arbitrary operator $\hat{O}$
a sum rule for its projected matrix elements
\begin{equation}
 \label{eq:sum:rule:partial:O}
 \elma{\Theta}{\hat{O}}{\Theta} = \sum_{\lambda} \sum_{j=1}^{d_\lambda} \elma{\Theta}{\hat{O} \hat{P}^{\lambda}_{jj}}{\Theta} \, .
\end{equation}
In particular, when applied to the norm ($\hat{O}\equiv\hat{\eins}$) or the Hamiltonian ($\hat{O}\equiv\hat{H}$), it leads to useful 
equations
\begin{subequations} 
\begin{align}
\label{eq:sum:rule:partial:1}
\norm{\Theta}{\Theta}
& = \sum_{\lambda} \sum_{j=1}^{d_\lambda} \elma{\Theta}{\hat{P}^{\lambda}_{jj}}{\Theta} \, , \\
\label{eq:sum:rule:partial:H}
\elma{\Theta}{\hat{H}}{\Theta}
& = \sum_{\lambda} \sum_{j=1}^{d_\lambda} \elma{\Theta}{\hat{H}\hat{P}^{\lambda}_{jj}}{\Theta} \, ,
\end{align}
\end{subequations} 
that can be used to evaluate the numerical accuracy of the symmetry projection. 
But as the numerical solution of the GEP \eqref{eq:gepproj} often
introduces substantial numerical noise into some of the basis states,
for the benchmarking of numerical implementations of the projection method 
it may be also of advantage to look also at the sum rules through the direct 
calculation of the coefficients
\begin{equation}
 c^{\lambda i}_{\epsilon} = \norm{\Psi^{\lambda i}_{\epsilon}}{\Theta} 
                          = \sum_{j=1}^{d_\lambda} f^{\lambda j *}_{\epsilon} \elma{\Theta}{\hat{P}^{\lambda}_{ji}}{\Theta} \, ,
\end{equation}
that are then injected in the expression~\eqref{eq:sum:rule:1} for the norm and the equivalent expression for the energy.

In addition,  using the sum rule~\eqref{eq:sum:rule:partial:H}, it is also easy to prove that 
the decomposition of a symmetry-breaking state 
$\ket{\Theta}$ yields some symmetry-respecting states $\ket{\Psi^{\lambda i}_{\epsilon}}$
that have an energy $e^\lambda_\epsilon$ lower than the expectation value  
$\bra{\Theta} \hat{H} \ket{\Theta}$ of the
state one started with. Indeed, 
from the decomposition of $\ket{\Theta}$ into the basis functions
$\ket{\Psi^{\lambda i}_{\epsilon}}$ of $\text{span}(G\ket{\Theta})$,  
equation \eqref{eq:sum:rule:partial:H} can be recast as  
\begin{align}
\label{eq:sum:rule:H}
\elma{\Theta}{\hat{H}}{\Theta} 
& =  \sum_\lambda \sum_{\epsilon=1}^{n_\lambda} \sum_{i=1}^{d_\lambda} 
     | c^{\lambda i}_{\epsilon} |^2 \, e^{\lambda}_{\epsilon} 
 \, .
\end{align}
Then, calling $e_0$ the lowest energy obtained 
among the states $\ket{\Psi^{\lambda i}_{\epsilon}}$, we have the inequality
\begin{equation}
\label{eq:ineqen}
\elma{\Theta}{\hat{H}}{\Theta} 
 \ge \sum_\lambda \sum_{\epsilon=1}^{n_\lambda} \sum_{i=1}^{d_\lambda} | c^{\lambda i}_{\epsilon} |^2 \, e_0 = e_0 \, ,
\end{equation}
where we assume the state $\ket{\Theta}$ be normalized to one such that
we can make use of~\eqref{eq:sum:rule:1}. In general, for a state $\ket{\Theta}$ of unknown 
nature we do not know \textit{a priori} to which irrep(s) $\lambda$ will belong the state(s) 
of lower energy.\footnote{In practice, however, for realistic nuclear Hamiltonians and states that 
are their self-consistent solutions, there is a large body of empirical knowledge about what to
expect.}
In addition, depending on the particularities of the energy spectrum obtained when 
restoring a specific symmetry, the states one is interested 
in to project out are not necessarily among the ones of lowest energy.

Assuming now that $\hat{O}$ is a scalar operator, the decomposition of a projected matrix element reads
\begin{equation}
\elma{\Theta}{\hat{O} \hat{P}^{\lambda}_{jj}}{\Theta}
 = \sum_{\epsilon=1}^{n_\lambda} \sum_{\epsilon'=1}^{n_\lambda}  \left( c^{\lambda j}_{\epsilon} \right)^* c^{\lambda j}_{\epsilon'} 
   \elma{\Psi^{\lambda j}_{\epsilon}}{\hat{O}}{\Psi^{\lambda j}_{\epsilon'}} \, , 
\end{equation}
which in the case of the norm can be simplified to
\begin{equation}
\elma{\Theta}{\hat{P}^{\lambda}_{jj}}{\Theta}
 = \sum_{\epsilon=1}^{n_\lambda} | c^{\lambda j}_{\epsilon} |^2 \, . 
\end{equation}
Obviously, these equations imply that
\begin{equation}
\label{eq:criterion}
\begin{split}
 \elma{\Theta}{\hat{P}^{\lambda}_{jj}}{\Theta} = 0 &\, \Leftrightarrow \,  \forall \, \epsilon, \, c^{\lambda j}_{\epsilon} = 0 \\
                                                   &\, \Rightarrow \, \elma{\Theta}{\hat{O} \hat{P}^{\lambda}_{jj}}{\Theta} = 0 \, ,
\end{split}
\end{equation}
such that the same components contribute to the summation in
\eqref{eq:sum:rule:partial:1} and \eqref{eq:sum:rule:partial:O}.
Note that, when using any of the frequently used prescriptions to calculate energy kernels for a non-Hamiltonian-based EDF 
\cite{Bonche90a,Dobaczewski07a,Robledo07a,Lacroix09a,Bender09a,Duguet09a,Robledo10a},
neither \eqref{eq:criterion} nor \eqref{eq:sum:rule:partial:H} do necessarily have an analogue
for the such calculated energies \cite{Bender09a}. 

%
%=======================================================================
%
\section{Projection on particle number}
\label{sec:PNP}
%
%-----------------------------------------------------------------------
%
\subsection{General considerations}
\label{sec:PNP:general}

For a first practical example of the projection method, we turn our attention 
towards the projection on particle number. While Slater determinants are
eigenstates of the particle-number operator by construction, for Bogoliubov 
quasiparticle states and other types of 
states that are not, a projection on particle number is 
necessary to restore the global gauge invariance, i.e.\ the invariance
under a rotation in Fock space \cite{BR86a},
that is associated with a fixed particle number of the many-body system. 
Bogoliubov quasiparticle states exhibit fluctuations in their number 
of particles, and can only be constrained to have a specific particle 
number on the average. As a benefit of the symmetry breaking,
these states provide a description of pairing correlations at play between
the nucleons while keeping the simplicity and the modest computational
cost of working with product-state wave functions. 
Nevertheless, working with states that are not eigenstates of the 
particle-number operator poses problems with contributions from components 
having the wrong particle number that spoil the calculation of observables.

In addition, it is known for long that the HF+BCS and HFB schemes
breakdown in the limit of weak pairing correlations and that the resulting
sharp transition to a trivial non-paired solution 
is an ``artificial feature of the theory arising mainly from the number 
fluctuation in the wave function'' \cite{Mang66a}. 
Various approximate schemes for variation-after-projection 
on particle number have been designed to lift this problem and to ensure 
the presence of pairing correlations also in the weak-pairing limit.
The most widely-used one is the Lipkin-Nogami (LN) method 
\cite{Nogami64a,Pradhan73a,Bender00a,Valor00a}, which corresponds to a development of 
non-diagonal kernels in terms of matrix elements of the diagonal one.
The LN method is, in fact, an easy to implement non-variational approximation 
to the much more rarely used (variational) second-order Kamlah 
expansion \cite{Kamlah68a,Valor00b}. An alternative
constructive scheme for systematic approximations to VAP calculations
has been proposed in Ref.~\cite{Flocard97a}, but never applied in
realistic calculations. Yet another alternative 
is the Lipkin method \cite{Wang14a}, where, depending on the 
order at which it is implemented, one or a few non-diagonal kernels are 
to be calculated explicitly. This approach can also account for 
the cross terms appearing in simultaneous projection on proton and neutron 
number that are absent in the LN and Kamlah schemes. 
While all of these provide pairing correlations
in the weak-pairing limit, they nevertheless fail to provide their 
realistic description on their own in this delicate case.
Only when combined with exact projection after variation 
\cite{Rodriguez05a}, or, even better, when directly solving the full 
VAP equations \cite{Dietrich64a,Braun98a,Sheikh00a,Sheikh02a,Anguiano01a,Hupin12a} 
this deficiency disappears.

Another problem arises when mixing two or more different 
Bogoliubov quasiparticle vacua, for example 
in GCM calculations \cite{Bonche90a} or when restoring a spatial symmetry
\cite{Hara82a,Yao11a}. In general, the mixed state $\ket{\Psi}$ will not 
have the same average particle number as the states it was constructed 
from. In principle, this can be approximately corrected for by 
using the Lagrange method, i.e.\ by subtracting
$- \lambda \big( \bra{\Psi} \hat{N} \ket{\Psi} - N_0 \big)$
in the variational configuration mixing process, where
$\lambda$ is the Fermi energy and $N_0$ the targeted particle number, from 
the expectation value of the total energy \cite{Hara82a,Bonche90a,Yao11a}.
However, it has to be noted that for a typical size of the Fermi energy
of about $-10$\,MeV such correction takes the order of 1\,MeV already when 
the deviation of mean particle number is as small as 0.1 particles, and 
therefore can easily become larger than the spacing of levels resulting 
from the configuration mixing of interest. In such case, the reliability of 
the configuration-mixing configuration becomes very sensitive to the remaining
deficiencies of the approximate particle-number correction.
In addition, such constraint compromises the orthogonality of the mixed
states \cite{Bonche90a}. As a consequence, it is highly desirable to 
restore the particle number exactly whenever Bogoliubov 
quasiparticle states are mixed.

We will now go into more details about the operation of projection 
on the number of particles. But as the projections on the neutron and 
proton numbers work identically in their respective space, for the sake
of simplicity we will consider for the moment only one generic 
particle species, labeled by $N$, and associated with the group $U(1)_N$. 
How projection on proton and neutron numbers have to be combined 
will then be lined out in Sect.~\ref{subsect:NZproj}.
In addition, to keep notations as simple as possible, we omit all other 
symmetry quantum numbers throughout this section.

%
%-----------------------------------------------------------------------
%
\subsection{Basic principles}

The group associated with global gauge rotations is the unitary group of degree one, labeled $U(1)_N$. 
As a group structure, $U(1)_N$ is a one-parameter Lie group, and therefore is Abelian. 
The parameter will here be labeled by $\varphi_n \in [0,2\pi]$ such that the volume reads
\begin{equation}
 v_{U(1)_N} = \int_0^{2\pi} \! d \varphi_n  = 2\pi \, .
\end{equation}
The gauge rotations are represented by the unitary operators 
\begin{equation}
\hat{U}(\varphi_n) = e^{-\iunit \varphi_n \hat{N}} \, ,
\end{equation}
where $\hat{N}$ is the particle-number operator.

The group $U(1)_N$ being Abelian, it has only one-dimensional irreps
such that the label $k$ used in the general discussion above can be 
dropped. The basis functions $\ket{\Psi^{N_0}}$ 
of the irreps are eigenstates of the particle number operator $\hat{N}$ with 
an eigenvalue ${N_0}$, which will be used to label the irreps. Under gauge
rotations, they transform as
\begin{equation}
 \hat{U}(\varphi_n) \, \ket{\Psi^{N_0}} = D^{N_0} (\varphi_n) \, \ket{\Psi^{N_0}} \, ,
\end{equation}
where
\begin{equation}
 D^{N_0} (\varphi_n) = e^{-\iunit \varphi_n {N_0}} \, .
\end{equation}
While all positive and negative integers
${N_0}$ are possible irreps for $U(1)_N$ from a mathematical point of view, 
only positive or null values for ${N_0}$ have physical meaning for the description
of a many-body system.
Moreover, as we consider states belonging to the Fock space, 
they will automatically be such that ${N_0} \geq 0$.\footnote{\label{fn:negN}
This property propagates to the evaluation of matrix elements 
of any operator. It can be shown, however, that an energy functional that is not constructed as the expectation value 
of a true Hamiltonian can also be decomposed onto negative particle numbers, which is a 
clear indication for the presence of non-physical components in such EDF \cite{Bender09a}.}

For the irrep with particle number ${N_0}$, the corresponding projection operator 
$\hat{P}^{N_0} $ is given by
\begin{equation}
\label{eq:PN}
\hat{P}^{N_0} 
= \frac{1}{2\pi} \int_0^{2\pi} \! d \varphi_n \, e^{-\iunit \varphi_n (\hat{N}- {N_0})} \, .
\end{equation}
It is a true projector in the mathematical sense with the properties
\begin{align}
\label{eq:PP=P}
 \hat{P}^{N_0} \hat{P}^{N_1} & =  \delta_{N_0 N_1} \, \hat{P}^{N_0} \, , \\
\label{eq:P+=P}
 \big( \hat{P}^{N_0} \big)^\dagger  & =  \hat{P}^{N_0} \, ,
\end{align}
as a consequence of the group $U(1)_N$ being Abelian.

Let us now consider a Bogoliubov quasiparticle state $\ket{\Phi}$. All linear superpositions of the gauge-rotated states 
$\big\{\hat{U}(\varphi_n) \ket{\Phi} \, , \, \varphi_n \in [0,2\pi] \big\}$ span a vector space $\text{span}(U(1)_N \ket{\Phi})$ 
that can be decomposed into a direct sum
\begin{equation}
 \text{span}(U(1)_N \ket{\Phi}) = \bigoplus_{N_0 \geq 0} S^{N_0} \, ,
\end{equation}
of one-dimensional subspaces $S^{N_0}$, each associated with a given irrep $N_0$ of $U(1)_N$.

As shown in Sect.~\ref{sec:proj}, the reference state $\ket{\Phi}$ obviously
belongs to $\text{span}(U(1)_N \ket{\Phi})$ and therefore can be written as
\begin{equation}
 \label{eq:decompoN}
 \ket{\Phi} = \sum_{N_0 \ge 0} c^{N_0} \ket{\Psi^{N_0}} \, ,
\end{equation}
where $c^{N_0}$ in general is a complex number and 
$\ket{\Psi^{N_0}}$ is a basis function of $S^{N_0}$. The projected states 
are directly obtained, up to a normalization factor, by simply applying the 
projection operator on $\ket{\Phi}$ 
\begin{equation}
\label{eq:PsiN}
\ket{\Psi^{N_0}} 
\equiv \frac{\hat{P}^{N_0} \ket{\Phi}}
            {\bra{\Phi} \hat{P}^{N_0} \ket{\Phi}^{1/2}}
=      \frac{1}{\sqrt{|c^{N_0}|^2}} \, \hat{P}^{N_0} \ket{\Phi}     
\, ,
\end{equation}
where
\begin{equation}
\label{eq:cN2}
|c^{N_0}|^2 = \bra{\Phi} \hat{P}^{N_0} \ket{\Phi}
\, .
\end{equation}
%

%
%-----------------------------------------------------------------------
%
\subsection{Number parity}
\label{subsect:numberparity}

For the further discussion, it is useful to define the number parity operator 
\begin{equation}
\label{eq:nparNop}
\hat{\Pi}_N = e^{-\iunit \pi \hat{N}} \, .
\end{equation}
The set of operators $\{ \hat{\eins} , \hat{\Pi}_N \}$ forms a cyclic group of order 2, which is also a subgroup
of the operators $\hat{U}(\varphi_n)$. As such, it has two one-dimensional irreps that we label by $\pi_n = \pm 1$.

Eigenstates of the particle-number operator are automatically eigenstates of number parity with eigenvalue 
\begin{equation}
\label{eq:nparN}
 \hat{\Pi}_N  \, \ket{\Psi^{N_0}} = (-1)^{N_0} \, \ket{\Psi^{N_0}} \, .
\end{equation}
As a consequence, those with even particle number have (even) number parity $+1$, 
whereas those with odd particle number have (odd) number parity $-1$.

Broken global gauge symmetry, however, is not necessarily accompanied with broken number parity. 
And indeed, as a consequence of the properties of the Bogoliubov transformations, Bogoliubov quasiparticle states 
have good number parity\footnote{Note that this is not the case anymore for ensemble averages 
of quasiparticle states as they are used in the equal-filling approximation for efficient calculation 
of odd-$A$ nuclei \cite{PerezMartin08a} and also in 
finite-temperature HFB theory \cite{Goodman81a}. For such cases, the projection on number parity and other quantum numbers has 
been outlined in Ref.~\cite{Tanabe05a}.} \cite{RW10a}
\begin{equation}
 \label{eq:numparHFB}
 \hat{\Pi}_N \, \ket{\Phi} = \pi_{n} \ket{\Phi} \, .
\end{equation}
In practice, the number parity of a Bogoliubov quasiparticle state can easily be determined by just counting the number
of single-particle states with occupation strictly equal to one in its canonical basis \cite{BallyPHD}.

By applying the number parity operator $\hat{\Pi}_N$ on the decomposition~\eqref{eq:decompoN} of a quasiparticle state, 
and using Eq.~\eqref{eq:numparHFB}, we obtain that
\begin{equation}
 c^{N_0} = \pi_{n} \, (-)^{N_0} \, c^{N_0} \, .
\end{equation}
As a consequence, a quasiparticle state with number parity $\pi_{n} = +1$
is a superposition of basis states $\ket{\Psi^{N_0}}$ with an even number of
particles, whereas a quasiparticle state with $\pi_{n} = -1$ is a superposition of 
basis states with an odd number of particles.
Note that this statement is completely independent on the average particle-number $\elma{\Phi}{\hat{N}}{\Phi}$
that the state has been constrained to and which can be any (positive)
real number. Bogoliubov quasiparticle states of different number parity have an inherently 
different physical structure \cite{Bender19EDF3}.

The number parity of Bogoliubov quasiparticle states can be exploited in order to reduce the
numerical cost of their particle-number projection. Indeed, as can be easily shown~\cite{BallyPHD}, 
when applied to a state with good number parity,
the integration interval in the projection operator \eqref{eq:PN} can be reduced from $[0,2\pi]$ to $[0,\pi]$.
Therefore, we can define the (simpler) reduced projection operator 
\begin{equation}
 \label{eq:redpnr}
 \hat{\mathcal{P}}^{N_0} = \frac{1}{\pi} \int_0^{\pi} \! d \varphi_n \, e^{-\iunit \varphi_n (\hat{N}- N_0)} \, . 
\end{equation}
Its form is the same for both number parities; however, the reduced 
projection operator \eqref{eq:redpnr} cannot distinguish anymore between 
states of different number parity. Indeed, the operator of 
Eq.~\eqref{eq:redpnr} is now a projection operator for a specific 
\textit{class of states}, where the irrep $N_0$ one projects out has to be 
chosen according to the number parity of the states it acts on, i.e.\ 
$\pi_{n} = (-)^{N_0}$. Conversely, when applying $\hat{\mathcal{P}}^{N_0}$ to a state with
$\pi_{n} \neq (-)^{N_0}$, the resulting matrix elements will in general not be zero, 
although they would be when applying $\hat{P}^{N_0}$ instead.

%
%=======================================================================
%
\subsection{Evaluation of observables}

As far as global gauge rotations are concerned, all relevant operators 
for nuclear structure calculations are tensor operators $\hat{O}$, whose rank
$r$ is given by the difference between the number of creation and 
destruction operators in the second quantized form of $\hat{O}$. They 
transform under gauge rotation as 
\begin{equation}
 \label{eq:tensorn}
 \forall \, \varphi_n \in [0,2\pi] , \,  e^{-\iunit \varphi_n \hat{N}} \, \hat{O} \, e^{+\iunit \varphi_n \hat{N}} 
  = e^{-\iunit \varphi_n r} \, \hat{O} \, ,   
\end{equation}
and they satisfy the commutation relation 
\begin{equation}
 [\hat{N},\hat{O}] = r \, \hat{O} \, .
\end{equation}
Only irreps of $U(1)$ with positive or null values of particle 
numbers play a role in the decomposition of many-body wave functions. 
By contrast, tensor operators can also transform according to irreps 
with negative number of particles.
When being interested in the calculation of spectroscopic properties of a given nucleus,
such as the excitation energies, the charge radius, or the nuclear moments, the operators of interest
are scalar operators, $r = 0$, i.e.\ they conserve the number of particles 
\begin{equation}
 [\hat{N},\hat{O}] = 0 \, .
\end{equation}
Higher-rank tensor operators come into play when looking at reactions
or nuclear decays that change the particle number. For example,
the neutrinoless double-beta decay process in which two neutrons are transformed into 
two protons and two electrons was studied within formalisms that include particle-number projection
such as the well known Gogny MR EDF method \cite{Lopez13a} or  
the newly developed In-Medium Generator Coordinate Method \cite{Yao19a},

For the particle-number projection operators, Eq.~\eqref{eq:tensorn} leads to the relation
\begin{equation}
 \label{eq:scalproj}
 \forall \, N_0 \in \mathbb{N} , \, \hat{P}^{N_0} \, \hat{O} = \hat{O} \, \hat{P}^{N_0 - r} \, , 
\end{equation}
which in turn simplifies the evaluation of operators between projected states, 
as only one projection has to be computed in practice 
\begin{equation}
\begin{split}
\elma{\Phi}{\big( \hat{P}^{N_0} \big)^\dagger \, \hat{O} \, \hat{P}^{N_1}}{\Phi} 
 &= \elma{\Phi}{ \, \hat{O} \, \hat{P}^{N_1} \, }{\Phi} \, \delta_{N_1 N_0 - r} \\  
 &= \elma{\Phi}{ \, \hat{P}^{N_1} \, \hat{O} \, }{\Phi} \, \delta_{N_1 N_0 + r} \, .
\end{split}
\end{equation}
%
%--------------------------------------------------------------------------
%

\subsection{Combined projection on proton and neutron number}
\label{subsect:NZproj}

The group to consider when combining the restoration of proton and neutron number
is the group direct product $U(1)_N \times U(1)_Z$, and the projection operator is 
built as the tensor product $\hat{P}^{N_0} \hat{P}^{Z_0} \equiv \hat{P}^{N_0} \otimes \hat{P}^{Z_0}$ 
of the projection operators  
\begin{alignat}{2}
 \hat{P}^{N_0} &= \frac{1}{2\pi} \int_0^{2\pi} \! d \varphi_n \, e^{-\iunit \varphi_n (\hat{N}- N_0)} &&\, , \\ 
 \hat{P}^{Z_0} &= \frac{1}{2\pi} \int_0^{2\pi} \! d \varphi_z \, e^{-\iunit \varphi_z (\hat{Z}- Z_0)} &&\, , 
\end{alignat}
for $U(1)_N$ and $U(1)_Z$, respectively, and where $\hat{N}$ and $\hat{Z}$ 
are the neutron and proton number operators, respectively.

The groups $U(1)_N$ and $U(1)_Z$ being Abelian, so is $U(1)_N \times U(1)_Z$, 
and therefore possesses only one-dimensional irreps. This implies in 
particular that the application of the projection operator 
$\hat{P}^{N_0} \hat{P}^{Z_0}$ is sufficient to completely diagonalize 
the Hamiltonian in the (one-dimensional) subspace associated with 
a given irrep $(N_0,Z_0)$. With this, one trivially restores also the total 
mass number \mbox{$A_0 = N_0 + Z_0$} and the third component of isospin 
$\hat{\mathscr{T}}_3 = \hat{N}-\hat{Z}$. The restoration of isospin
$\hat{\mathscr{T}}^2$ itself, however, is much more involved.

For the further discussion of practical aspects of particle-number 
projection, we will treat protons and neutrons as distinct species of 
particles, as done in the vast majority of applications of nuclear structure
models. Single-particle states with good isospin serve as elementary 
building blocks to construct separate product states for protons and neutrons.
The nuclear many-body wave function is then built as the tensor product of the 
many-body wave function for each particle species. 
In that case, it is sufficient to project the proton and the neutron parts of the wave 
function separately on proton and neutron number, respectively. 
In addition, it is possible to take advantage of the fact that each part of the wave 
function has a good number parity to reduce the integral of each projection
operator according to Eq.\ \eqref{eq:redpnr}.

None of this is a necessity, though. Single-particle states can be set up 
as mixtures of proton and neutron components, such that the many-body product 
state built from them cannot be factorized anymore. 
This allows for the modeling of various phenomena such as proton-neutron
pairing \cite{Romero19a,Yao19a} and other subtle effects related to isospin, 
in particular when combined with 
subsequent isospin restoration \cite{Satula10a}.\footnote{In the general 
case, the nuclear Hamiltonian contains electromagnetic, and possibly 
also other, terms that are not isospin invariant such 
that the symmetry is physically broken. In that instance, the projection onto good
isospin can still be used to remove the unphysical sources of symmetry breaking in the model
but the method requires a generalization to account 
for the non-commutation of the Hamiltonian
with the projection operators and to include the 
mixing of irreps through the diagonalization of the Hamiltonian in the space
of isospin-projected states \cite{Satula10a}.} 
In contrast with the case case assumed throughout the rest of this paper,
the many-body wave function then only
possesses a good number parity for the total number of nucleons 
$A_0 = N_0 + Z_0$ such that only the projection operator 
\begin{equation}
 \hat{P}^{A_0} = \frac{1}{2\pi} \int_0^{2\pi} \! d \varphi_a \, e^{-\iunit \varphi_a (\hat{A}- {A_0})} \, ,
\end{equation}
where $\hat{A}=\hat{N}+\hat{Z}$, 
can be reduced according to Eq.~\eqref{eq:redpnr},
but not the individual projection operators on $N_0$ and $Z_0$.

%
%-----------------------------------------------------------------------
%
\subsection{Numerical implementation}
\label{subsect:PN:num:imp}

In practice, the projection is carried out computationally.
There are several strategies to evaluate particle-number projected
operator matrix elements that have been used in the literature.
One is to map the gauge-space integral \eqref{eq:PN} onto a contour
integral in the complex plane that then can be evaluated with the 
residue theorem. This approach has been used in several pioneering 
papers \cite{Bayman60a,Dietrich64a} and remains very instructive
for formal analyses of the particle-number restoration method
\cite{Dobaczewski07a,Bender09a,Duguet09a}. A second approach is through the 
use of recurrence relations \cite{Dietrich64a,RS80a,Frank82a,Lacroix10a,Hupin11a}.
Both have the disadvantage that they cannot be easily combined with other 
projections or a configuration mixing calculation.
A more versatile strategy to evaluate matrix elements containing 
the number projection operator is to discretize the integral over
the gauge angle in Eq.~\eqref{eq:PN}.
In this section, we will discuss this discretization and its
specificities such as its convergence with the number of discretization points.

Following the prescription first introduced by Fomenko~\cite{Fomenko70a}, 
we discretize the projector on particle number $N_0$ with the $M_\varphi$-point
trapezoidal quadrature\footnote{Other different, but similar, choices for 
a discretized projection operator are discussed in Appendix~\ref{sec:discpnr}.}
\begin{equation}
\label{eq:flamenko}
\hat{\mathbb{P}}^{N_0}_{M_{\varphi}}
= \frac{1}{{M_{\varphi}}} \sum_{m=1}^{{M_{\varphi}}} 
  e^{-\iunit  \pi \frac{m-1}{{M_{\varphi}}} (\hat{N} - N_0)} \, ,
\end{equation}
explicated here for neutrons. As the first integration point is always 
at $\varphi = 0$,  using a discretization with only integration point in 
total, ${M_\varphi} = 1$, is equivalent to not projecting, 
$\hat{\mathbb{P}}^{N_0}_{M_\varphi=1} \ket{\Phi} = \ket{\Phi}$, 
independent on the particle number $N_0$ projected on.

For numerical reasons it is in general safer to use odd values of 
${M_\varphi}$ than even ones. This avoids to have the angle $\varphi = \pi/2$
in the set of integration points, for which in the projection of a quasiparticle
vacuum one might have to evaluate fractions with both numerator and 
denominator that can become arbitrarily close to zero, see Appendix~B of 
Ref.~\cite{Anguiano01a}.
However, in a calculation that combines particle-number 
projection with other projections or configuration mixings, such numerical 
problem might appear at any gauge angle.\footnote{This difficulty to evaluate 
near-zero values that exactly cancel each other analytically has to be distinguished 
from the so-called ``pole problem'' that appears at the same angle when calculating 
the energy from Hamilton operator neglecting exchange terms \cite{Donau98a} or
from an ill-defined EDF 
\cite{Anguiano01a,Dobaczewski07a,Robledo07a,Lacroix09a,Bender09a,Duguet09a}
for which the near-zero value in the denominator is \textit{not} canceled by the
same factor in the numerator.}

The operator defined through Eq.~\eqref{eq:flamenko} is a 
discretization of the reduced expression of Eq.~\eqref{eq:redpnr}; hence, 
$\hat{\mathbb{P}}^{N_0}_{M_\varphi}$ can be applied only to states 
$\ket{\Phi}$ that 
have a number parity equal \mbox{to $(-)^{N_0}$}. A more general discretized 
operator that can be applied to states of unknown number parity is easily 
defined by replacing the factor $\pi$ in the exponential by a factor $2\pi$. 
From a numerical point of view, however, such operator is not as efficient 
as Eq.~\eqref{eq:flamenko}, as it doubles the number of calculated points 
necessary to reach the same level of convergence, see 
Appendix~\ref{sec:discpnr}.

\begin{table}
\begin{ruledtabular}
\begin{tabular}{cccc}
Label & Z & $\langle \hat{N}\rangle$ & $\langle \Delta \hat{N}^2 \rangle$ \\
\hline
a & 20 & 24  & 4.9 \\
b & 50 & 70  & 7.9 \\
c & 82 & 138 & 16.4
\end{tabular}
\caption{\label{tab:qpstates:n}
Characteristics of the quasiparticle vacua analyzed in
Figs.~\ref{fig:pnr1}-\ref{fig:pnr3}.
}
\label{tab:pnr}
\end{ruledtabular}
\end{table}

\begin{figure}[t!]
\centering  
  \includegraphics[width=7.0cm]{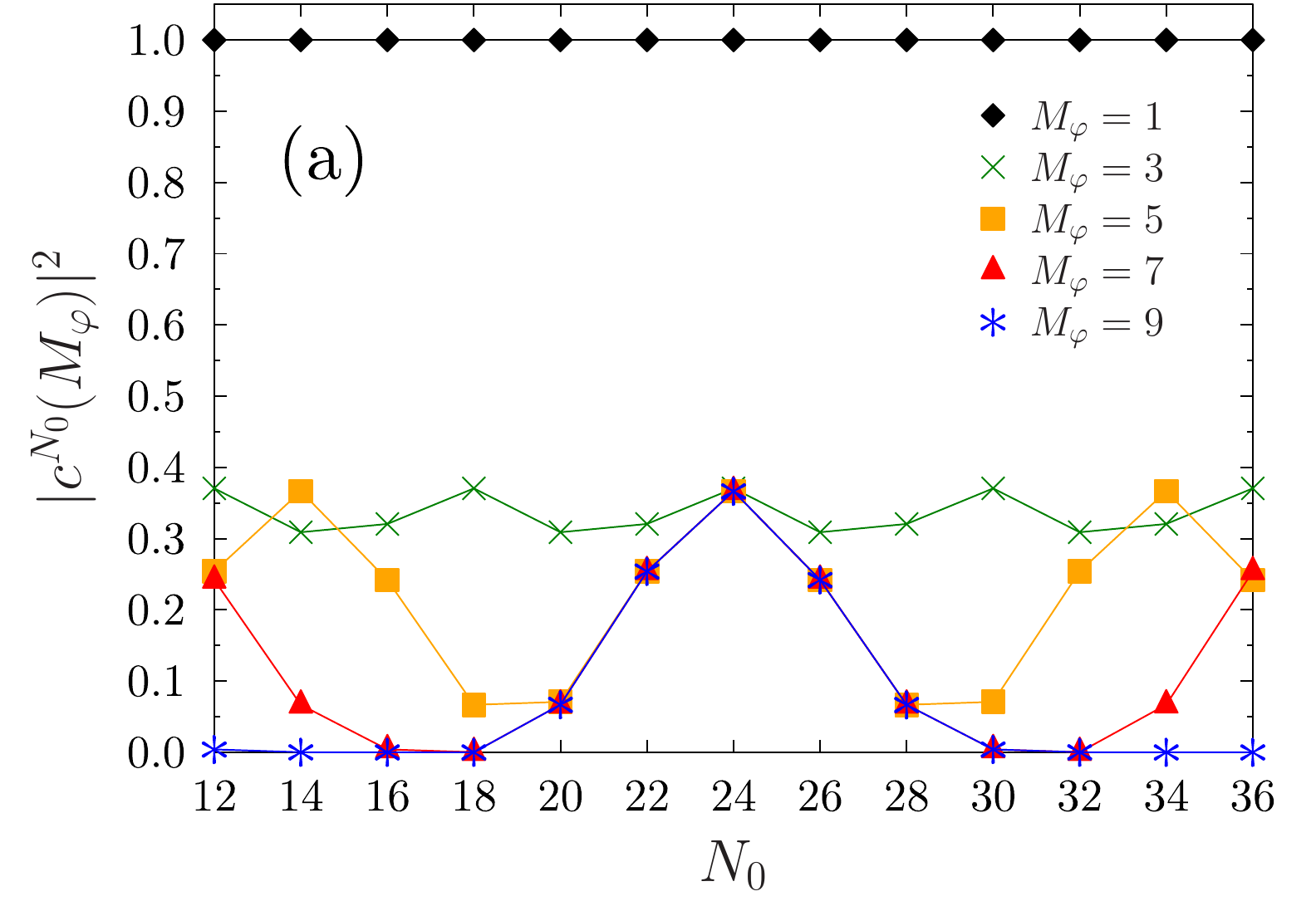}  \\ \vspace*{0.20cm} 
  \includegraphics[width=7.0cm]{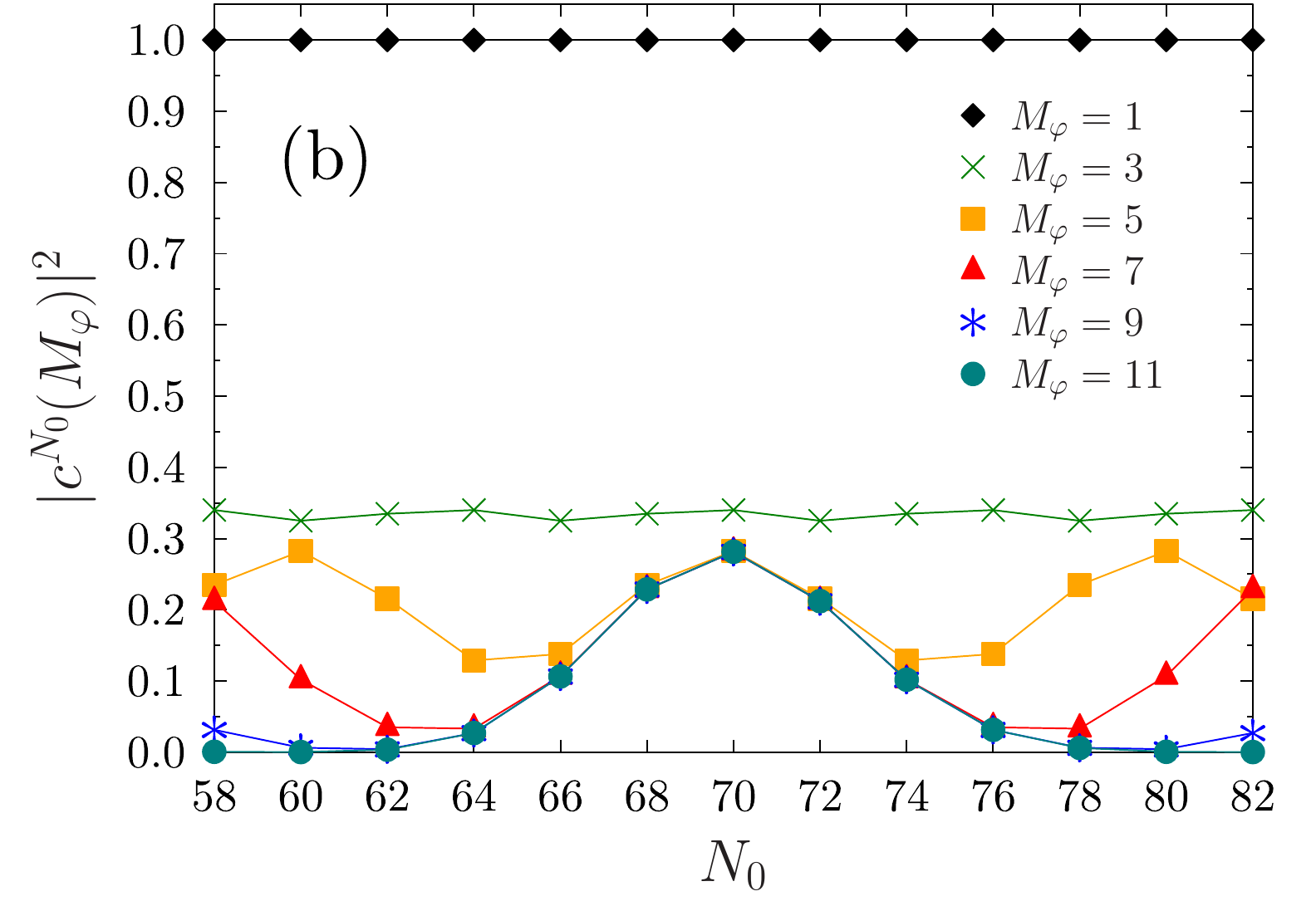} \\ \vspace*{0.20cm}
  \includegraphics[width=7.0cm]{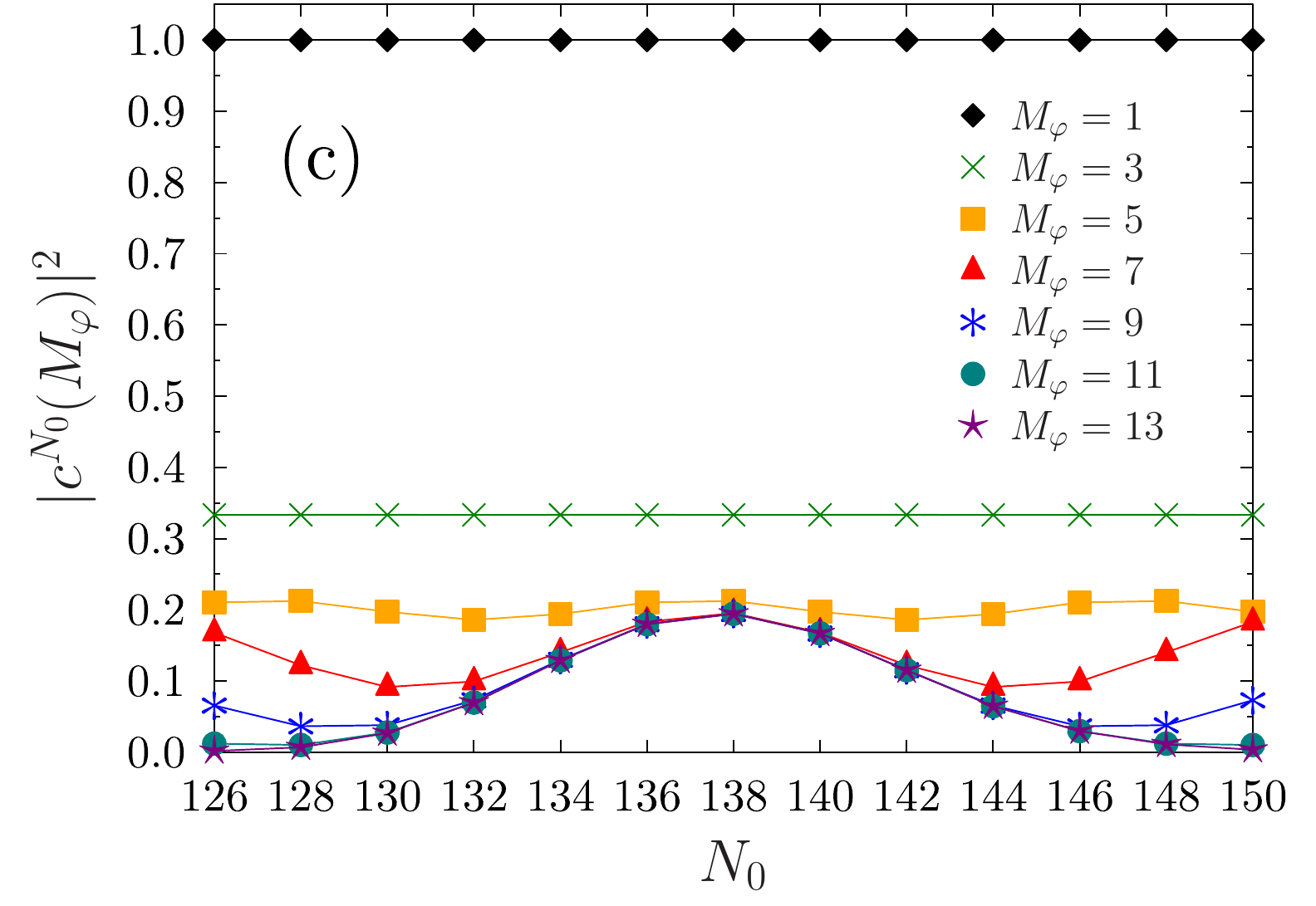}
\caption{
\label{fig:pnr1}
(Color online)
Evolution of the weight $|c^{N_0}(M_\varphi)|^2$ of the numerically projected 
states as a function of the number of neutrons $N_0$ on which one projects for 
different choices for the number of discretization points ${M_\varphi}$ for
the quasiparticle states $\ket{\Phi_a}$ (panel (a)), $\ket{\Phi_b}$ 
(panel (b)) and $\ket{\Phi_c}$  (panel (c)) as specified in the text.
On each panel, the results obtained with the largest value of ${M_\varphi}$
cannot be distinguished from exact results for the values of $N_0$ shown.
Points calculated with same $M_\varphi$ are connected by straight lines 
to guide the eye. 
}
\end{figure}

To illustrate the further discussion of the convergence of results obtained
with the discretized projection operator 
$\hat{\mathbb{P}}^{N_0}_{M_{\varphi}}$~\eqref{eq:flamenko}
in dependence of the number of discretization 
points $M_{\varphi}$, we will examine the decomposition of three (normalized) 
fully paired Bogoliubov-type quasiparticle states $\ket{\Phi}$. The states 
were constructed for \nuc{44}{Ca}, \nuc{120}{Sn} and \nuc{220}{Pb}. 
Also, the states were chosen such that their dispersion of neutron number
\begin{equation}
\label{eq:N:dispersion}
\elma{\Phi}{\Delta \hat{N}^2}{\Phi}
\equiv \elma{\Phi}{\hat{N}^2}{\Phi} - \elma{\Phi}{\hat{N}}{\Phi}^2
\, ,
\end{equation}
which provides a measure for the breaking of global gauge symmetry through the 
presence of pairing correlations, is increasing.\footnote{While the dispersion 
$\elma{\Phi}{\Delta \hat{N}^2}{\Phi}$ increases with particle number for 
the quasiparticle states chosen here, this is not a necessity as the size of 
the dispersion depends very sensitively on the pairing interaction and the 
position of the Fermi energy in a configuration-dependent single-particle spectrum. 
Large values of $\elma{\Phi}{\Delta \hat{N}^2}{\Phi}$ 
usually require large $\elma{\Phi}{\hat{N}}{\Phi}$, but the dispersion
can take very small, even zero, values  in any nucleus for some state that
disfavors the presence of pairing correlations.}\emb\
The actual values for the mean neutron number $\elma{\Phi}{\hat{N}}{\Phi}$ 
and its dispersion are given in Tab.~\ref{tab:pnr}.

For our purpose it is sufficient to consider only projection on neutron 
number, while leaving the proton part of the wave function untouched.

In Fig.~\ref{fig:pnr1}, we plot, for these three states, the weights of 
the numerically projected states
\begin{equation}
| c^{N_0} ({M_\varphi})|^2 
= \elma{\Phi}{\hat{\mathbb{P}}^{N_0}_{M_\varphi}}{\Phi}
\end{equation}
in dependence of the total number of discretization points 
$M_\varphi$.
As already mentioned, for $M_\varphi = 1$ the discretized projection 
operator~\eqref{eq:flamenko} is the unit operator.
This means that $\hat{\mathbb{P}}^{N_0}_{1}$ does not project at all
and attributes the unaltered original state to \textit{any} particle number 
$N_0$ compatible with that state's number parity, independent of that component 
being contained in the state's physical decomposition or not, cf.\ Fig.~\ref{fig:pnr1}.
From a different perspective, for 
$M_\varphi = 1$ the discretized projection operator attributes the complete 
sum of all physical components $\ket{\Psi^{N_1}}$ with their physical weight
$c^{N_1} = c^{N_1} (M_\varphi = \infty)$ to any particle number $N_0$ 
compatible with its number parity 
\begin{equation}
\label{eq:PTheta:M1}
\hat{\mathbb{P}}^{N_0}_{1} \ket{\Phi} 
= \ket{\Phi} 
= \sum_{N_1} c^{N_1} \ket{\Psi^{N_1}} \, .
\end{equation}
Each of the thus numerically ``projected'' states is equal, and its 
observables take the value of the sum rule for projection, even when the
respective component is absent from the original symmetry-breaking state.

These observations provide the starting point for the understanding of how the 
discretized projection operator~\eqref{eq:flamenko} generates projected 
states for finite values of $M_\varphi$ by eliminating non-targeted
components from the summation in Eq.~\eqref{eq:PTheta:M1}.
As demonstrated in Appendix~\ref{sec:discpnr}, for finite 
$M_{\varphi} > 1$ applying the discretized projection operator 
$\hat{\mathbb{P}}^{N_0}_{M_\varphi}$ on a state removes exactly 
all components that do not satisfy the condition
$N_1 = N_0 + 2 l {M_\varphi}$, $l \in \mathbb{Z}$,  from the original
state. The final result can be expressed in a compact way as the 
double sum
\begin{equation}
\label{eq:PTheta:M}
\hat{\mathbb{P}}^{N_0}_{M_{\varphi}} \ket{\Phi} = 
   \sum_{l \in \mathbb{Z}} \sum_{N_1 \ge 0} c^{N_1}  \ket{\Psi^{N_1}} \, \delta_{N_1 N_0 + 2lM_\varphi} \, . 
\end{equation}
The subset of non-targeted components contained in the original state
that are not eliminated by the discretized projection operator quickly 
becomes smaller with increasing $M_{\varphi}$. For a given $N_0$, the 
closest non-suppressed components are the ones at $N_0 \pm 2 M_{\varphi}$.

In theory, the non-vanishing irreps $N_0$ contained in a Bogoliubov quasiparticle state $\ket{\Phi}$
will fall into an interval bounded by $N_{\text{min}}$ and $N_{\text{max}}$.
The lower bound $N_{\text{min}}$ is given by the number of fully occupied single-particle states
in the canonical basis of $\ket{\Phi}$ while the upper bound $N_{\text{max}}$ is given by 
the total number of single-particle states with non-zero occupation in the same basis.
In practice, however, the wave function is generated numerically such that the bounds might be affected
by the numerical accuracy of the computation, which is ultimately limited by floating-point
arithmetic, and can only delimit an interval outside of which the respective 
components cannot be distinguished from numerical noise.

For a given irrep $N_0$ with non-zero weight, the discretized projection 
operator~\eqref{eq:flamenko} becomes exact when all other components
from $N_{\text{min}}$ to $N_{\text{max}}$ have been eliminated.
For a given $M_\varphi$, the interval for which the discretized numerical projection 
on particle number $N_0$ \eqref{eq:flamenko} becomes exact is 
\begin{equation}
\label{eq:Nproj:interval}
N_{\text{max}} - 2 M_\varphi + 2 
\leq N_0 
\leq N_{\text{min}} + 2 M_\varphi - 2 \, .
\end{equation}
Nevertheless, if one has a prior knowledge of the distribution of the
projected components, it is possible to adapt the discretization depending on the
localization of the targeted irrep within the distribution.
As it will be demonstrated later on, for Bogoliubov quasiparticle vacua one can assume 
up to a very good approximation a Gaussian shaped distribution centered around
the average particle-number of the state, see the discussion of Eq.~\eqref{eq:FloOni} in
what follows. In that case, for
$N_0$ next to the center of the distribution at 
$\big( N_{\text{min}} + N_{\text{max}} \big)/2 \approx \langle \op{N} \rangle$, this requires 
$M_{\varphi} > \big( N_{\text{max}} - N_{\text{min}} \big)/4 \approx 1.5 \, \langle \Delta \hat{N}^2 \rangle^{1/2}$
points, where we used Eq.~\eqref{eq:FloOni} for the estimate in terms of the dispersion
in particle number~\eqref{eq:N:dispersion}.
This represents the simplest case that sets the lower limit
for an acceptable value of $M_{\varphi}$. On the other hand, for targeted irreps $N_0$ at the boundary of the interval, 
all other components are eliminated for
$M_{\varphi} > \big( N_{\text{max}} - N_{\text{min}} \big)/2 \approx 3 \, \langle \Delta \hat{N}^2 \rangle^{1/2}$, see
Fig.~\ref{fig:pnr1}.

For components $N_0$ absent from the original state, however, the 
convergence towards the correct result $c^{N_0} = 0$ in general requires more
integration points.
Indeed, Eq.~\eqref{eq:PTheta:M} implies that the discretized projection 
operator has a periodicity of $2 M_\varphi$ and therefore
yields the same numerical result for particle numbers 
$N_0$ that differ by multiples of $2 M_\varphi$. As a consequence,
it generates mirror images of the results for $N_0$ that are 
repeated every $2 l M_\varphi$, as can be clearly seen in 
Fig.~\ref{fig:pnr1}. 
A special case is $M_\varphi = 1$, for which these mirror images superpose
in such a way that the numerically ``projected'' state is the sum of all 
physical components for all values of $N_0$ permitted by number parity.
When $M_\varphi > (N_{\text{max}} - N_{\text{min}})/2$, then some
components outside of this interval are also correctly identified as 
having weight zero, but not all of them.
A safe choice in that case would be to take 
$M_{\varphi} > | \big( N_{\text{max}} + N_{\text{min}} \big)/2 - N_0 | \approx | \langle \op{N} \rangle - N_0 |$,
although fewer points may already be sufficient.

As the periodic mirror images of the dominant components are pushed away
from the interval of physical components when increasing $M_{\varphi}$, 
the numerical results obtained for weights $|c^{N_0}(M_{\varphi})|^2$ of 
components $N_0$ far outside of the interval defined through 
Eq.~\eqref{eq:Nproj:interval} will oscillate between the values of various 
weights of physical components (or sums thereof) and zero, until they fall 
into the interval defined in Eq.~\eqref{eq:Nproj:interval}.
The effect can be seen for example in panel~(a) of Fig.~\ref{fig:pnr1}
for $|c^{14}|^2$ and $|c^{34}|^2$, which non-montonically jump around before
falling to zero. For components much further outside, such oscillation will 
repeat itself several times with increasing $M_{\varphi}$.

\begin{figure}[t!]
\centering  
  \includegraphics[width=8.0cm]{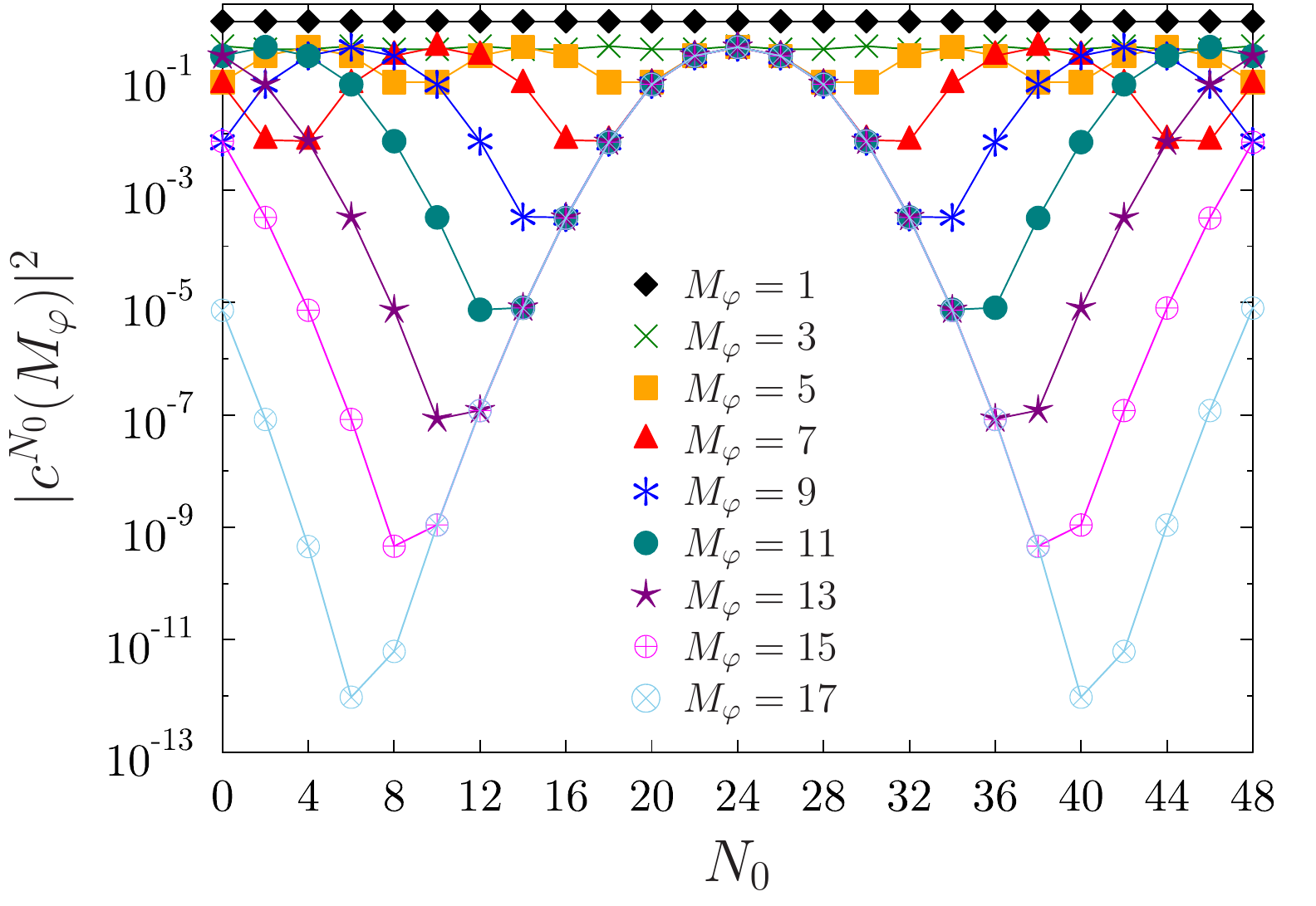}
\caption{
\label{fig:pnr1:log}
(Color online)
Same as panel~(a) of Fig.~\ref{fig:pnr1}, but in logarithmic scale and 
some additional values of $M_\varphi$.
}
\end{figure}

The queues of the distributions analyzed in Fig.~\ref{fig:pnr1} fall off
relatively slowly, which can be more clearly seen when plotting the same 
data on a logarithmic scale as done for the decomposition of one of the states 
in Fig.~\ref{fig:pnr1:log}. Still, the numerical convergence of these tiny 
components continues in the same way as the convergence of the dominant
ones until a level of $10^{-13}$ has been reached, beyond which the numerical 
noise from the calculation of many-body matrix elements sets in in our code. This figure
also shows even more clearly than Fig.~\ref{fig:pnr1} how the identical mirror
images at $N_0 + 2 M_\varphi$ of converged physical components at $N_0$ move 
outside with increasing $M_\varphi$ in the discretized projection operator.

\begin{figure}[t!]
\centering  
  \includegraphics[width=7.0cm]{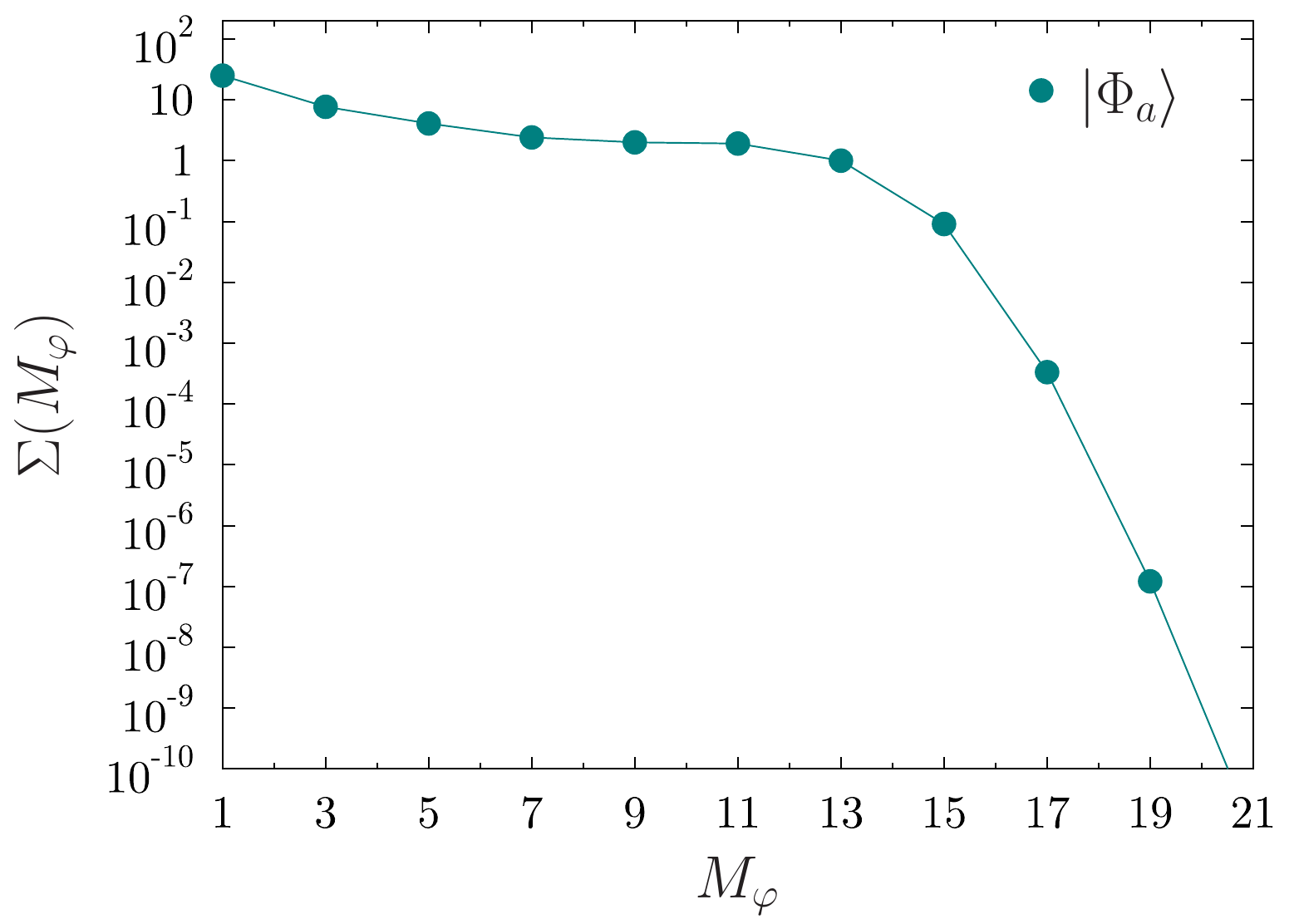}
\caption{
\label{fig:pnr1:sum}
(Color online)
Deviation~\eqref{eq:PNP:accuracy} in logarithmic scale of the sum rule for 
the components $|c^{N_0}(M_\varphi)|^2$ of the state $\ket{\Phi_a}$ from the
analytical value one as a function of $M_\varphi$. See text for details. 
to guide the eye. 
}
\end{figure}

The sum rule for the weights~\eqref{eq:sum:rule:1} establishes an additional
test of the internal consistency and the numerical accuracy of the projection 
of a given state. Focusing on state $\ket{\Phi_a}$, we display in 
Fig.~\ref{fig:pnr1:sum}, the sum rule for the components $|c^{N_0}(M_\varphi)|^2$
summed from $N_0=0$ to 48 and subtracted by one, i.e.\ the quantity
\begin{equation}
\label{eq:PNP:accuracy}
\Sigma (M_\varphi) \equiv \left| \sum_{N_0=0}^{48} |c^{N_0}(M_\varphi)|^2 - 1 \right| \, ,
\end{equation}
as a function of $M_\varphi$. 

As argued above, for $M_{\varphi} = 1$, the numerical projection operator
$\hat{\mathbb{P}}^{N_0}_{1}$
attributes the entire sumrule to any irrep $N_0$ compatible with the 
number parity of the original state~\eqref{eq:PTheta:M1}. Calculating
the sum rule in that case yields the mean-field expectation value of the 
operator in question times the number of irreps summed over. 
With increasing number of discretization points $M_{\varphi}$ in the numerical 
projector $\hat{\mathbb{P}}^{N_0}_{M_{\varphi}}$~\eqref{eq:flamenko}, the 
non-physical contributions to the matrix element for $N_0$ are eliminated 
as a result of relation~\eqref{eq:PTheta:M} until convergence to the 
physical value for the irrep $N_0$ is reached. For convergence of 
the sum rule, it is obviously necessary that the summation covers all 
irreps found in the original state $\ket{\Phi_a}$ and that the number 
of discretization point is sufficient to converge the calculation of
the projected matrix elements.

\begin{figure}[t!]
\centering  
  \includegraphics[width=7.2cm]{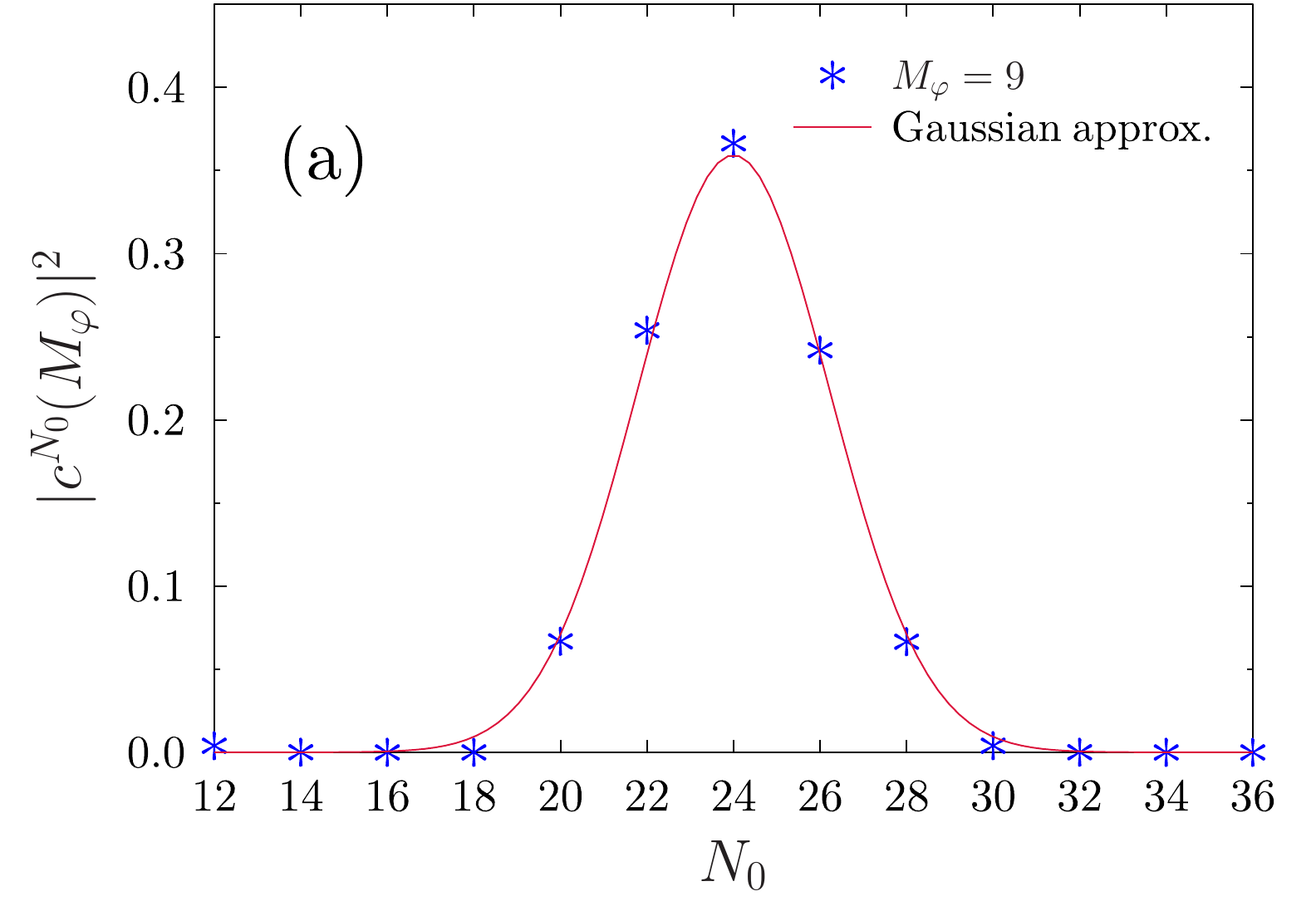}  \\ \vspace*{0.25cm} 
  \includegraphics[width=7.2cm]{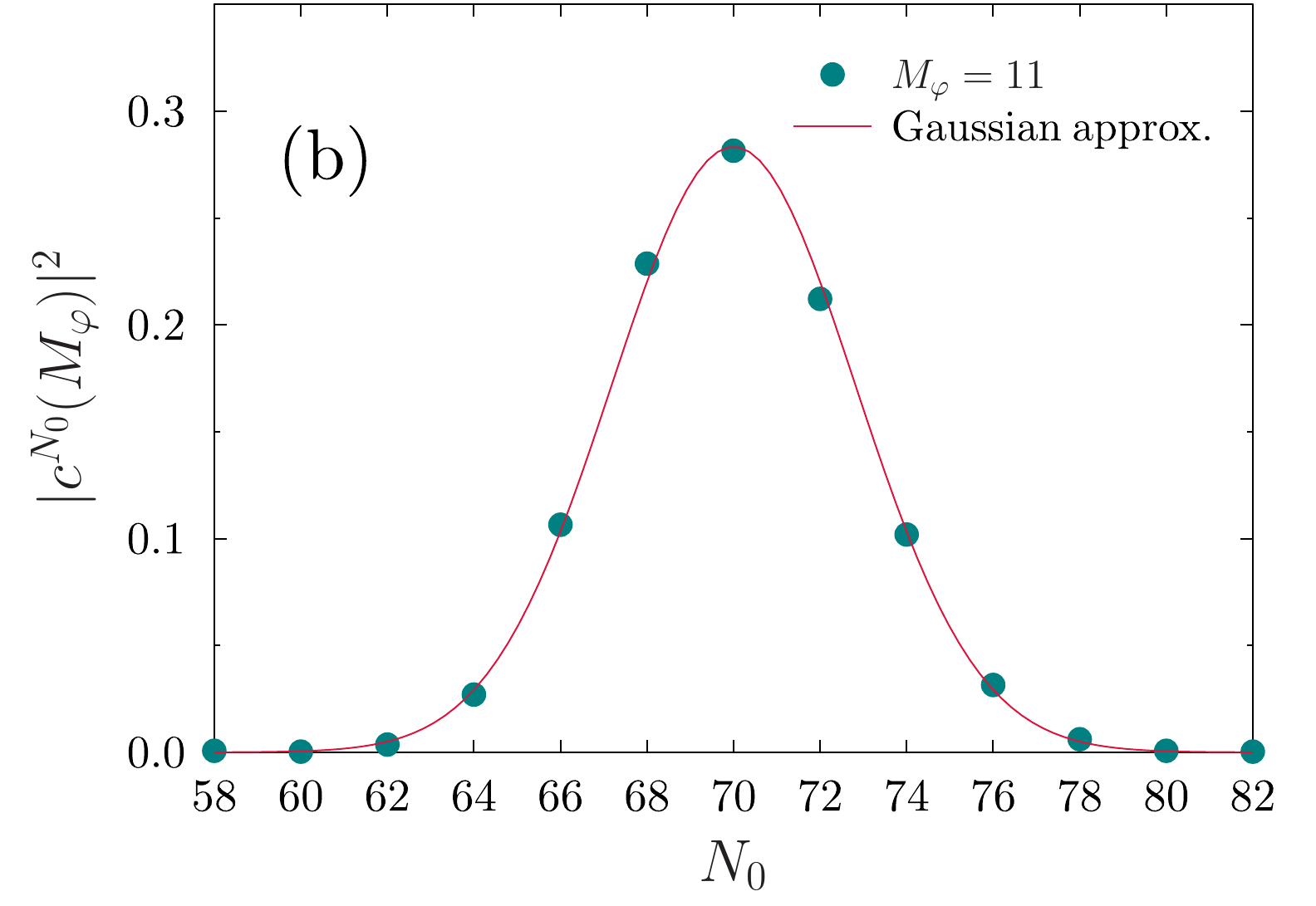} \\ \vspace*{0.25cm}
  \includegraphics[width=7.2cm]{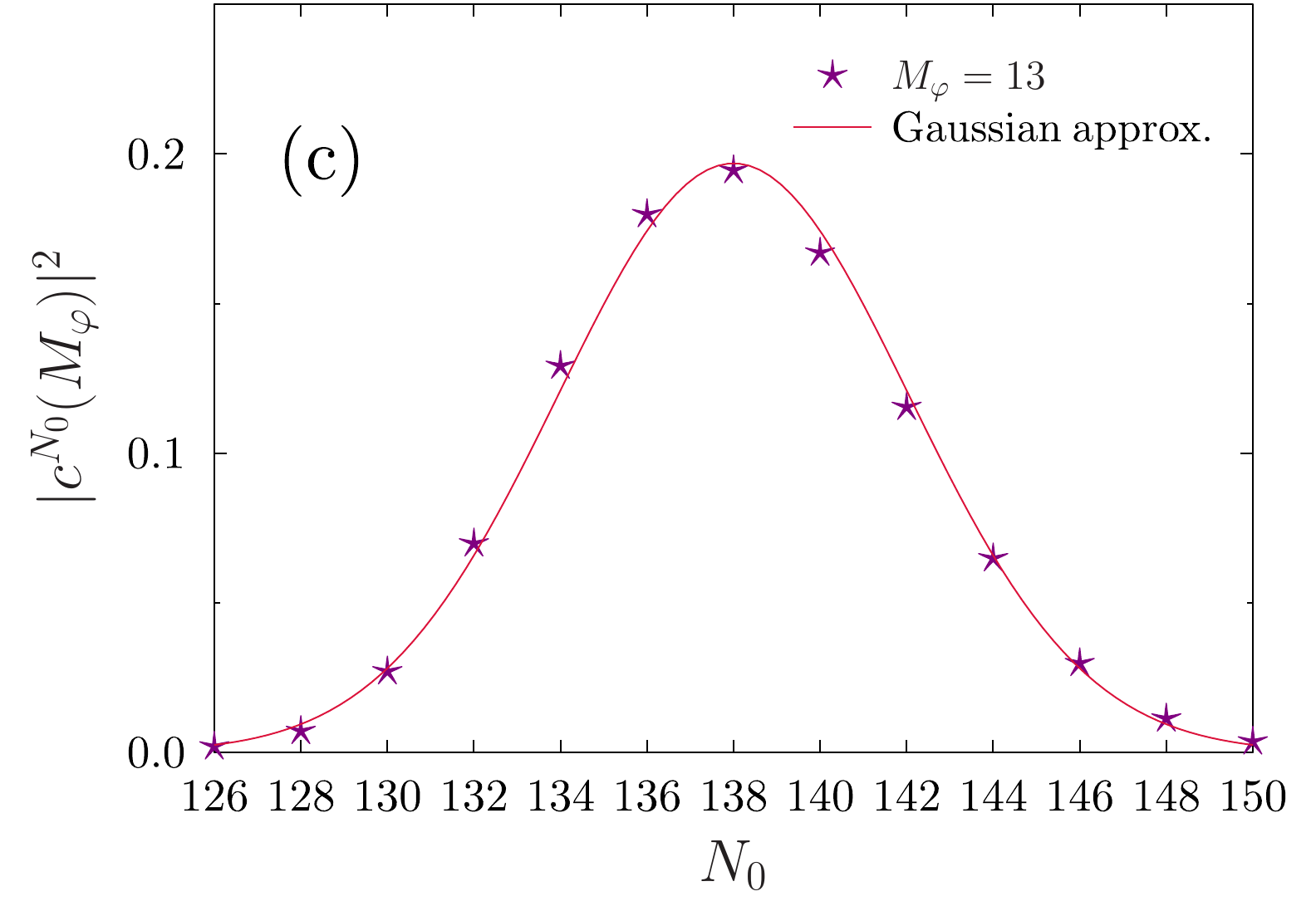}
\caption{
\label{fig:pnr1:gauss}
(Color online)
Comparison between the weight $|c^{N_0}(M_\varphi)|^2$ of the numerically projected 
states (markers) and its Gaussian approximation \eqref{eq:FloOni} 
(solid line) as a function of 
the number of neutrons $N_0$ on which one projects for 
the quasiparticle states $\ket{\Phi_a}$ (panel (a)), $\ket{\Phi_b}$ 
(panel (b)) and $\ket{\Phi_c}$  (panel (c)) as specified in the text.
On each panel, the number of points ${M_\varphi}$ used is the same as the
largest one displayed in Fig.\ \ref{fig:pnr1}. 
}
\end{figure}

To be sure about the numerical convergence of results without running test
calculations with different values of $M_{\varphi}$ requires the \textit{a
priori} knowledge of the boundaries $N_{\text{max}}$ and $N_{\text{min}}$
of the distribution of irreps in the original state. As has been demonstrated 
in Ref.~\cite{Flocard97a}, the distribution of the weights~\eqref{eq:cN2} of 
components with particle number $N$ in the decomposition of a fully paired 
Bogoliubov quasiparticle state $\ket{\Phi}$ can be estimated by a Gaussian 
centered around its average particle number $\elma{\Phi}{\hat{N}}{\Phi}$
\begin{align}
\label{eq:FloOni}
| c^{N_0} |^2 
& \approx  \frac{2}{\sqrt{2 \pi \sigma^2}}
      \exp \bigg[ - \frac{\big( \elma{\Phi}{\hat{N}}{\Phi} - N_0 \big)^2}
                         {2 \sigma^2} \bigg] \, ,
\end{align}
whose width is determined by that state's dispersion of particle number: 
$\sigma^2 = \elma{\Phi}{\Delta \hat{N}^2}{\Phi}$, see also Ref.~\cite{Samyn04a}.

For the fully paired quasiparticle vacua decomposed in Figs.~\ref{fig:pnr1}, 
the exact values of the $|c^{N_0}|^2$ are indeed very well approximated by 
the Gaussian of Eq.~\eqref{eq:FloOni}, even for very small components as
can be seen from Fig.~\ref{fig:pnr1:gauss}. However, it should be noted that the 
presence of such small components far from the center of the distribution
depends on choices made for cutoffs when solving the HFB equations.
A cutoff that limits pairing correlations to some valence space will
inevitably cut the tails from the distribution of irreps contained in the
symmetry-breaking state. In any event, the remaining differences between
the estimate and the calculated values are quite small, mainly in the form
of a slight asymmetry around the center. The latter is not too 
surprising as the estimate~\eqref{eq:FloOni} implies in one way or another 
equally distributed single-particle states and a state-independent pairing 
interaction, neither of which is the case in a realistic calculation. 
Nevertheless, the agreement is remarkable and the estimate~\eqref{eq:FloOni} 
can hence be used to determine an \textit{a priori} indication for the number 
of points ${M_\varphi}$ needed to converge the numerical projection, which 
in practice remains rather small ($M_\varphi \simeq 10$) for atomic nuclei.

It can be easily shown that for numerically projected 
matrix elements of any operator, for example a scalar operator $\hat{O}$,
the elimination of untargeted components follows the same rule as for
the plain norm overlap.
Indeed, the normalized expectation value of such operator can be written as
\begin{align}
\label{eq:PO:M}
\lefteqn{
\langle \hat{O} \rangle_{N_0} (M_\varphi) \equiv
\frac{\elma{\Phi}{\hat{O} \, \hat{\mathbb{P}}^{N_0}_{M_\varphi}}{\Phi}}
     {\elma{\Phi}{\hat{\mathbb{P}}^{N_0}_{M_\varphi}}{\Phi}}
}
      \nn \\
& =   \frac{
      \sum_{\substack{l_1\in\mathbb{Z} \\ N_1 \ge 0}} 
            |c^{N_1}|^2 \, \bra{\Psi^{N_1}} \hat{O} \ket{\Psi^{N_1}} \,
      \delta_{N_1,N_0 + 2 l_1 {M_\varphi}}
      }
      {\sum_{\substack{l_2 \in\mathbb{Z} \\ N_2 \ge 0}} |c^{N_2}|^2 \,
      \delta_{N_2, N_0 + 2 l_2 {M_\varphi}}}
\, .
\end{align}
For $M_\varphi = 1$, the normalized projected matrix element reduces to
the plain expectation value $\elma{\Phi}{\hat{O}}{\Phi}$.
Otherwise, for irreps with $|c^{N_0}|^2 \neq 0$ in $\ket{\Phi}$, 
all other contributions but the targeted one have been eliminated in
the numerator and the denominator when
$M_\varphi$ is large enough that $N_0$ falls into the interval defined by
Eq.~\eqref{eq:Nproj:interval}.
The rate of convergence, however, may depend also on the values of 
the \textit{exact} projected matrix elements in the numerator.

Note that, while theoretically Eq.~\eqref{eq:PO:M} can be
written only for irreps $N_0$ with non-zero weights \mbox{$|c^{N_0}|^2 \neq 0$}
in the original state, numerically neither the numerator nor the denominator
will ever fully vanish, such that numerically one ends up with the division 
of numerical noise representing $\elma{\Phi}{\hat{O} \, \hat{\mathbb{P}}^{N_0}_{M_\varphi}}{\Phi} = 0$ by
different numerical noise representing $\elma{\Phi}{\hat{\mathbb{P}}^{N_0}_{M_\varphi}}{\Phi} = 0$.
This is an artifact of the numerical treatment of projection, as formally operators can have 
non-zero expectation values only for irreps with non-zero weights, cf.\ Eq.~\eqref{eq:criterion}.
For that reason, the expectation value of operators for small components 
$\elma{\Phi}{\hat{\mathbb{P}}^{N_0}_{M_\varphi}}{\Phi}$ have to be considered with care.

\begin{figure}[t!]
\centering  
  \includegraphics[width=8.0cm]{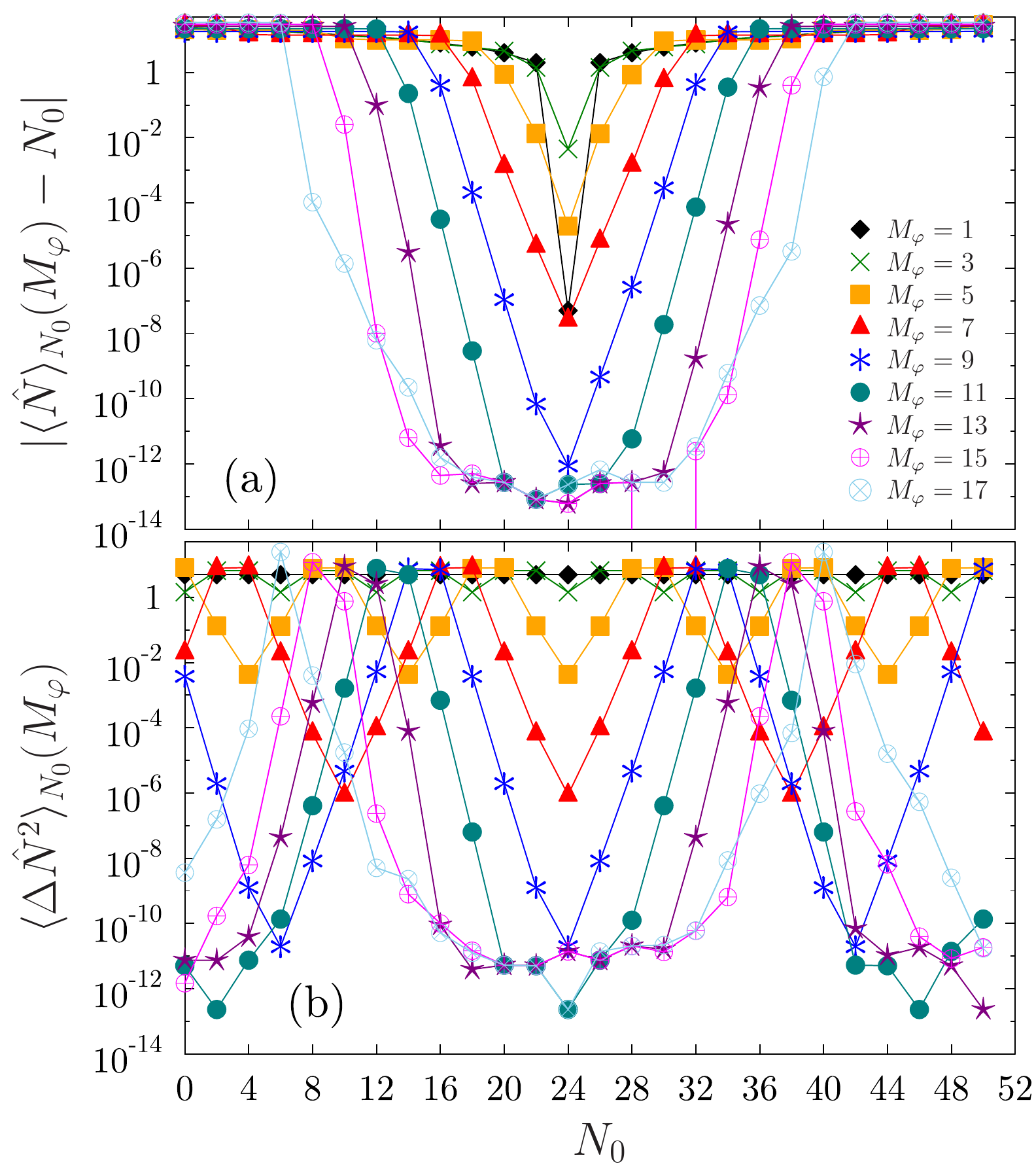} 
\caption{
\label{fig:pnr2:all}
(Color online)
Evolution of the deviation of the expectation value of the neutron number
from the value it is projected on (a) and dispersion of the neutron 
number (b) for states numerically projected on particle number $N_0$
from the state with $\bra{\Phi_a} \hat{N} \ket{\Phi_a} = 24$ 
for the number of discretization points ${M_\varphi}$ as indicated.
Points calculated with same $M_\varphi$ are connected by straight lines 
to guide the eye.
}
\end{figure}

As an example for the numerical convergence of the matrix elements projected 
on irreps in the tails of the distribution, Fig.~\ref{fig:pnr2:all} displays
the evolution of the deviation of the expectation value of the 
neutron number 
\begin{equation}
 \langle \hat{N} \rangle_{N_0} ({M_\varphi}) = 
 \frac{\elma{\Phi}{\hat{N} \, \hat{\mathbb{P}}^{N_0}_{M_\varphi}}{\Phi}}
           {\elma{\Phi}{\hat{\mathbb{P}}^{N_0}_{M_\varphi}}{\Phi}}
\end{equation}
from the value $N_0$ projected on, and also the dispersion 
\begin{equation}
\langle \Delta \hat{N}^2 \rangle_{N_0} ({M_\varphi}) 
= \langle \hat{N}^2 \rangle_{N_0} ({M_\varphi}) 
  - \langle \hat{N} \rangle^2_{N_0} ({M_\varphi})  \, ,
\end{equation}
in dependence of the number of 
discretization points ${M_\varphi}$ for a wide range of components 
projected from the state with $\bra{\Phi_a} \hat{N} \ket{\Phi_a} = 24$. 
Assuming again that the original state contains numerically significant 
irreps between $N_{\text{min}} \simeq 8$ and $N_{\text{max}} \simeq 40$, 
then Eq.~\eqref{eq:Nproj:interval} indicates that components can be expected 
to be converged for the maximum number of  \mbox{$M_\varphi = 17$}
points when they fall in the interval between about~8 and~40. In practice,
however, for the very small components below $N_0 \lesssim 12$ and above
$N_0 \gtrsim 32$, the precision of the numerical calculation of the matrix 
elements is visibly degraded compared to those in between. With our 
implementation that uses a double-precision floating-point format, 
further increasing the number of discretization points $M_\varphi$
does not significantly improve the quality of these components anymore.

\begin{figure}[t!]
\centering  
  \includegraphics[width=7.4cm]{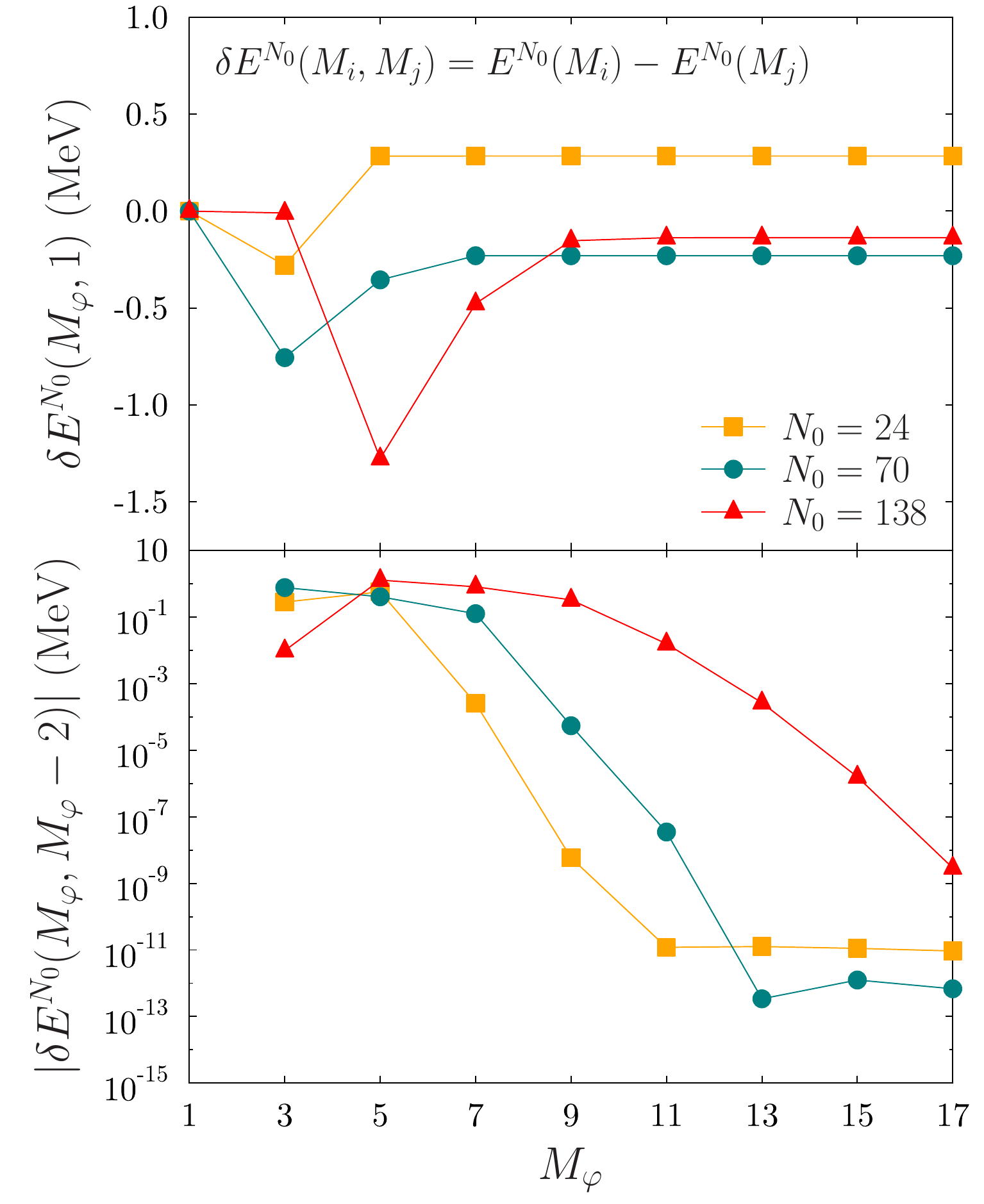}
\caption{
\label{fig:pnr3}
(Color online)
Evolution of the projected energy, for states with $N_0=24$, 70, and 138, 
projected from states with the same mean particle number as specified in 
Table~\ref{tab:qpstates:n}, as a function of  the total number of points 
${M_\varphi}$ used for the discretized projection operator.
}
\end{figure}

Another quantity that is sensitive to the number of particles is the projected binding energy
\begin{equation}
E^{N_0}({M_\varphi})  
=  \frac{\elma{\Phi}{\hat{H} \, \hat{\mathbb{P}}^{N_0}_{M_\varphi}}{\Phi}}
        {\elma{\Phi}{\hat{\mathbb{P}}^{N_0}_{M_\varphi}}{\Phi}}  \, .
\end{equation}
As we are interested only in the accuracy of the discretized projection operator, 
neither the precise form of the Hamiltonian, nor the exact values of the energies 
are relevant, except that we specify that a true Hamiltonian is used when evaluating 
the projected energies in order to avoid any influence of the possible problems
analyzed in Refs.~\cite{Donau98a,Anguiano01a,Dobaczewski07a,Robledo07a,Lacroix09a,Bender09a,Duguet09a,Robledo10a,Satula14a}
on our discussion.
The results are displayed in Fig.~\ref{fig:pnr3} for projection of the three Bogoliubov quasiparticle
vacua as specified in Table~\ref{tab:qpstates:n} on particle number $N=24$, 70, and 138, respectively
As we can see, as we increase the number of points in the discretization, the energy converges rapidly.
Beyond a certain value of $M_\varphi$, however, the numerical noise kicks in and increasing the number of
points does not improve anymore the projected energy. It is also interesting to note that for the 
state $\ket{\Phi_a}$,  the projected energy $E^{24}_a(M_\varphi)$ is several hundreds of keV higher than the expectation value
$\elma{\Phi_a}{\hat{H}}{\Phi_a}$ of the original state. 
Indeed, as previously explained in Sec.~\ref{subsect:projdiscussion}, if the projection method guarantees to
find at least one projected state of lower energy than the expectation value of the Hamiltonian 
of the original unprojected state, there is no reason that this
will be the case for the irrep one is interested in. The latter depends both on the Hamiltonian 
at hand and how the unprojected state has been obtained.

Finally, we note in passing that particle-number-projected overlaps $\bra{\Phi_a} \hat{\mathbb{P}}^{N_0}_{M_{\varphi}} \ket{\Phi_b}$
and non-diagonal matrix elements $\bra{\Phi_a} \hat{O} \, \hat{\mathbb{P}}^{N_0}_{M_{\varphi}} \ket{\Phi_b}$
will converge with increasing $M_{\varphi}$ according to the values of $N_{\text{min}}$ and $N_{\text{max}}$ found in the two states. 
In the typical case where the distributions of components for $\ket{\Phi_a}$ and $\ket{\Phi_b}$ have a large 
overlap, the extremal values $\text{max}(N_{\text{min},a}, N_{\text{min},b})$ and 
$\text{min}(N_{\text{max},a}, N_{\text{max},b})$ will govern the convergence of the numerical projection.
In more extreme cases, however, it is possible that the periodicity of the discretized projector induces
contamination coming from distant, but physical, components.

%
%=======================================================================
%
\section{Projection on angular momentum}
\label{sec:AMP}

\subsection{General Considerations}
\label{AMP:general}

The second projection that we will analyze in detail is the projection on total angular momentum $J$. 
Together with proton number, neutron number, and parity, angular momentum is the most
relevant quantum number for the analysis of spectroscopic data of atomic nuclei. As it provides 
selection rules for the existence of electromagnetic and other transitions and 
establishes rules for their relative strength, it provides the guideline to group states into 
characteristic level sequences \cite{Wigner59a}.  

As explained in Sect.~\ref{sec:PNP:general}, the angular-momentum projection of paired 
Bogoliubov-type quasiparticle vacua should be combined with particle-number projection.
Angular-momentum projection is nowadays a widely employed technique in the context 
of nuclear EDF methods \cite{Bender04a,Niksic06a,Kimura07a,Bender08a,Rodriguez10a,Guzman12a,Yao11a,Satula10a,Bally14a,Borrajo15a,Rodriguez15a,Egido16a,Egido16b,Robledo18a,Shimada15a,Ushitani19a,Bender19EDF3}.
And angular-momentum projected quasiparticle vacua can also used as building blocks for 
configuration-interaction methods. Prominent examples are the MONSTER/VAMPIR approach 
\cite{Schmid87a,Schmid04a}, the so-called projected shell model \cite{Hara95a,Sun16a} and 
the Monte-Carlo shell model \cite{Shimizu12a,Otsuka01a} that all present alternative numerical strategies 
to conventional shell-model calculations.

Only very few ground states of even-even nuclei will take the spherical symmetry of a $J=0$ state
when calculated in a (symmetry-unrestricted) self-consistent mean-field approach.
In fact, when calculated in HF approximation, only nuclei with subshells that are either 
completely filled or empty are even compatible with a  strict spherical symmetry of the wave function.\footnote{
In the zero-pairing limit of the HFB approximation, however, it is possible to obtain spherically
symmetric states also for open-shell systems. Nevertheless, as their resulting density matrices are not the ones of a single Slater
determinant \cite{Duguet20x,Duguet20y}, we will discard such possibility from the 
further discussion.}

Including pairing correlations in the modeling provides the means to describe spherical 
even-even open-shell systems with a single Bogoliubov quasiparticle vacuum, but such 
solutions are usually only found in the direct vicinity of major shell closures.
Similarly, because of self-consistent core-polarization effects, it is virtually impossible 
that the variationally determined states for odd and odd-odd nuclei will be eigenstates
of angular momentum. 

Indeed, for the vast majority of nuclei, correlations related to the spatial arrangement of 
the nucleons are important. This is evidenced for example by the observation of collective 
rotational excitations that are an ubiquitous feature of nuclei throughout the chart of nuclei 
\cite{Bohr76a,Rowe70a,BM98a,RS80a,Frauendorf01a}, or by the evolution of charge radii that cannot be 
explained by consecutive filling of spherical shells with increasing nucleon number
\cite{Angeli15a}. To efficiently describe such correlations, it is advantageous 
to work with deformed reference states. Doing so, however, rotational symmetry is broken, 
and the resulting reference states are not eigenstates of angular momentum anymore.
Angular momentum projection becomes mandatory to obtain a faithful and accurate description of 
related phenomena.

There is a noteworthy difference between the applications of angular-momentum and particle-number 
projection. In the literature, the latter is almost always exclusively used to extract one specific 
irrep from a symmetry-breaking state, which is very often the irrep with a particle number equal, 
or at least very close, to the one that the reference state has been constrained to, whereas angular-momentum
projection in the majority of cases is used to project out a spectrum of states with different $J$.
There are several reasons for this practice. The main reason is that nuclear structure changes often 
quickly with particle number, such that it is in most cases advantageous to construct reference states 
that are as close in particle number to the targeted one as possible. This is different for states
with different angular momentum in a given nucleus: experiment indicates that they often have almost
the same internal structure over a wide range of angular momenta, unless there is a backbending or 
other alignment phenomenon. And indeed, when being projected from a suitably chosen well-deformed 
quasiparticle vacuum, the angular-momentum restored states usually group into one or several rotational bands 
whose energies roughly scale with $J(J+1)$. Projecting on many irreps then even becomes a necessity 
for the interpretation of the internal structure of such symmetry-breaking state.

That collective rotational bands emerge naturally from an angular-momentum projected theory of 
quadrupole-deformed states can be shown analytically when expanding the Hamiltonian and 
norm kernels entering the projected energy in terms of rotation angles \cite{Peierls57a}, thereby 
justifying phenomenological collective models for deformed nuclei \cite{Rowe70a,BM98a,RS80a}.

The same kind of expansion for well-deformed nuclei at high rotational frequency also yields 
the self-consistently cranked HF and HFB methods as a first-order limit 
\cite{Kamlah68a,Beck70a,villars71a,Mang75a,Wong75a,islam79a}. In this approach, the energy is minimized with
an auxiliary condition on the expectation value on a component of the angular momentum 
vector, where the rotational frequency plays the role of the Lagrange parameter, which leads
to the so-called Thouless-Valatin moment of inertia \cite{Thouless62a}.

There are several possibilities how to embed angular-momentum projection into a given framework. 
When the framework is variational, then VAP calculations are in general preferable over PAV, but 
they are computationally so costly that up to  now such scheme has only been implemented for schematic 
bases \cite{Kanada98a} or small valence spaces  \cite{Schmid87a,Schmid04a,Gao15a}. 
Using a VAP scheme also has the drawback that in principle
one has to do one such calculation for each value of $J$ of interest.\footnote{Note that only the energy 
of the lowest irrep with given $J$ can be optimized in such calculation, but not the one of the other 
irreps of same $J$ that possibly can be projected out as well from the same reference state.} 
The vast majority of applications uses the PAV scheme instead. There are, however, several 
strategies to arrive at an intermediate scheme in the spirit of an RVAP or MAP. As far as the optimal 
deformation of the state to be projected is concerned, one can construct a set of deformation-constrained 
states first that are then in a second step all projected on $J$ in order to identify the one given the 
lowest energy. This is particularly important for light nuclei in general and also nuclei with ground states
at small deformation including spherical ones, for which the deformation of the reference state that gives
the lowest projected states is usually very different from the deformation of the self-consistent ground 
state \cite{Bender06a}. 
When combining angular-momentum projection with shape mixing in a GCM, 
such MAP is automatically done \cite{Bender06a}. When projecting from the lowest variational state
at a given deformation, such calculation is often characterized in terms of the Peierls-Yoccoz 
(or sometimes Yoccoz) moment of inertia \cite{Peierls57a} and systematically leads to excitation
spectra that are slightly too spread out. This can be compensated for when working with states
that are cranked to some rotational frequency, again either in a MAP scheme or by mixing states
at different rotational frequency in a GCM. The latter strategy is characterized as using 
Peierls-Thouless moments of inertia \cite{Peierls62a}. First applications along these lines using 
the Gogny EDF have recently become available \cite{Borrajo15a,Egido16b,Shimada15a,Ushitani19a}

We will now go into the details about the operation of projection on angular  momentum. 
Neither for its formal aspects nor for the analysis of the convergence of the discretized projection 
operator is it of importance in which framework they are used. As above, for the sake of clarity we 
drop all symmetry indices not related to angular momentum throughout this section.

%
%-----------------------------------------------------------------------
%
\subsection{Basic principles}

The rotations in space\footnote{Here, we consider the space 
that contains both spatial positions and spin degrees of freedom.}
of a single-particle state are characterized by the special unitary group 
$SU(2)$. Unlike global gauge rotations discussed above, $SU(2)$ is not 
Abelian. This will lead to a large number of formal and practical 
differences that make angular momentum projection more involved
than particle-number projection.

The rotational invariance of the many-body Hamiltonian ensures that its
eigenstates have good total angular momentum $J$. 
Contrarily to gauge invariance that is verified by protons and neutrons separately
-- and thus gives to two separate good quantum numbers $N$ and $Z$ -- 
the rotational invariance is a property of the total wave function made by all the nucleons.
Mathematically, this means that for a nucleus made of $A$ nucleons, the group of interest is the tensor product
$\text{diag}( SU(2)\times \ldots \times SU(2) )$,\footnote{Considering a group $G$, 
the group $\text{diag}(G\times G)$ is composed by the diagonal elements of the group direct product $G\times G$. 
Thus defined, $\text{diag}(G\times G)$ is trivially a subgroup of $G\times G$ and it is isomorphic to $G$.} 
where the direct product contains $A$ times the group $SU(2)$.
As $\text{diag}( SU(2)\times \ldots \times SU(2) )$ is isomorphic to $SU(2)$, for the sake of simplicity we will simply 
label it as $SU(2)_A$. The irreps of $SU(2)_A$ are labeled by the angular momentum quantum number $J$ and have 
each a dimension $d_J = 2J + 1$. This means that, but for the trivial representation ($J=0$), the irreps of $SU(2)_A$ are not one dimensional
as it was the case when dealing with the group $U(1)_N \times U(1)_Z$ for particle number. 
This is a consequence of the fact that the compact Lie group $SU(2)_A$ is not Abelian.

We choose to represent the unitary rotation operator $\hat{U}(g)$ of this
group in terms of a sequence of three subsequent rotations about the $z$, $y$
and again the $z$ axis of a fixed coordinate system parameterized through the three Euler angles 
$(\alpha, \beta, \gamma) \in [0, 2\pi] \times [0,\pi] \times [0, 4\pi]$
\cite{Varshalovich88a}
\begin{equation}
\label{eq:SU2:rot}
\hat{R}(\alpha,\beta,\gamma) 
= e^{-\iunit \alpha \hat{J}_z} \, 
  e^{-\iunit \beta  \hat{J}_y} \, 
  e^{-\iunit \gamma \hat{J}_z} \, ,
\end{equation}
where $\hat{J}_i$ ($i=x,y,z$) is the component along the $i$-axis of the total 
angular momentum vector $\hat{\vec{J}}$.
To lighten the notation, throughout this article we define the angular 
momentum operators without a factor $\hbar$.
We note that there exist also alternative representations of the rotation operator that
use a different sequences of rotations about different axes and angles 
\cite{Varshalovich88a}. 

As this will be of importance later on, we define next the three specific rotations 
of angle $\pi$ around each of the cartesian axis
\begin{subequations}
\begin{align}
\label{eq:Rx:def}
 \hat{R}_x &= e^{-\iunit \pi \hat{J_x}} =  e^{-\iunit \pi \hat{J_y}} \,  e^{-\iunit \pi \hat{J_z}} = \hat{R}(0,\pi,\pi) \, , \\
\label{eq:Ry:def}
 \hat{R}_y &= e^{-\iunit \pi \hat{J_y}} = \hat{R}(0,\pi, 0 ) \, , \\
\label{eq:Rz:def}
 \hat{R}_z &= e^{-\iunit \pi \hat{J_z}} = \hat{R}(0, 0 ,\pi) = \hat{R}(\pi, 0 , 0) \, , 
\end{align}
\end{subequations}
with the operator $\hat{R}_i$ ($i=x,y,z$) being usually denoted as the 
$i$-signature in the nuclear physics literature \cite{Doba00a}.

In the parameterization of Eq.~\eqref{eq:SU2:rot}, 
the volume of the group is given by
\begin{equation}
\label{es:SU2:vol}
v_{SU(2)_A} 
= \int_0^{2\pi} d \alpha 
  \int_0^{\pi} d \beta \, \sin (\beta) 
  \int_0^{4\pi} \! d \gamma 
= 16\pi^2 \, .
\end{equation}
Its value being $16 \pi^2$ is a consequence of looking at systems of 
Fermions, whose total wave functions can take integer or half integer
total angular momenta.
The different intervals covered by the Euler angles $\alpha$ 
and $\gamma$ can be interchanged without affecting the results.

The set of eigenstates $\ket{\Psi^{JM}_\epsilon}$ of $\hat{J}^2$ and $\hat{J}_z$
\begin{align}
\label{eq:SU2:basis}
\hat{J}^2 \ket{\Psi^{JM}_\epsilon} 
& = J(J+1) \ket{\Psi^{JM}_\epsilon} \, , \\
\hat{J}_z \ket{\Psi^{JM}_\epsilon} 
& = M \ket{\Psi^{JM}_\epsilon} \, , 
\end{align}
and their respective quantum numbers $J$ and $-J \leq M \leq +J$ provide the basis 
functions and labels for the irreps of $SU(2)_A$. The generic label $\epsilon$ 
in \eqref{eq:SU2:basis} represents all additional characteristics of a 
state not directly affected by spatial rotations. Under such rotation,
the basis functions transform as
\begin{equation}
\label{eq:RabgPsiJMK}
\hat{R}(\alpha,\beta,\gamma) \ket{\Psi^{J M}_\epsilon} 
= \sum_{M'=-J}^J D_{M'M}^{J} (\alpha,\beta,\gamma) \, \ket{\Psi^{JM'}_\epsilon} 
\, , 
\end{equation}
where the matrix representation $D_{M'M}^{J} \, (\alpha,\beta,\gamma)$
of the group in the space of the rotation angles is provided by the Wigner 
rotation matrices \cite{Varshalovich88a}
\begin{equation}
\begin{split}
\label{eq:DMpMJ=eMpdJeM}
D_{M'M}^{J} (\alpha,\beta,\gamma) 
&= \delta_{J'J} \, \elma{\Psi^{J'M'}_\epsilon}{\hat{R}(\alpha,\beta,\gamma)}{\Psi^{JM}_\epsilon} \\
&=  e^{-\iunit \alpha M'} \, d_{M'M}^{J}(\beta) \, e^{-\iunit \gamma M} 
\end{split}
\end{equation}
with $d_{M'M}^{J}(\beta)$ being a Wigner $d$-matrix\footnote{We note in passing that 
within our representation of rotations,  the Wigner $d$-matrices are always real.} 
\cite{Varshalovich88a}.

The operator that projects onto angular momentum $J$ for given $M$ and $K$
reads         
\begin{equation}
\begin{split}
\label{eq:PJMK}
\hat{P}^{J}_{MK} 
& = \frac{2J + 1}{16\pi^2} 
    \int_0^{2\pi} d \alpha 
    \int_0^{\pi} d \beta \, \sin (\beta) 
    \int_0^{4\pi} \! d \gamma \\
& \phantom{=} \times \,  D_{MK}^{J*}(\alpha,\beta,\gamma) \, 
    \hat{R}(\alpha,\beta,\gamma) \, , 
\end{split}
\end{equation}
and has the properties 
\begin{align}
\label{eq:PJMKPJpMpKp}
\hat{P}^{J}_{MK} \, \hat{P}^{J'}_{M' K'} 
& = \delta_{J J'} \, \delta_{K M'} \, \hat{P}^{J}_{M K'} \, , \\
\big( \hat{P}^{J}_{M K} \big)^\dagger 
& = \hat{P}^{J}_{K M} \, .
\end{align}
These two equations indicate that for $M \neq K$ the operator 
$\hat{P}^{J}_{M K}$ is a transfer operator as defined in 
Sect.~\ref{sec:proj}, which is again a consequence of $SU(2)$ being non-Abelian. 

While the integral representation~\eqref{eq:PJMK} is the most
transparent one from a group-theoretical point of view and the most
flexible one as far as its combination with other symmetry restorations
and configuration mixing is concerned, there are alternative representation 
of $\hat{P}^{J}_{MK}$ that have been sometimes used in specific contexts.
One is in terms of angular-momentum shift operators \cite{lowdin55c}, whose 
connection to the form~\eqref{eq:PJMK} is sketched in Ref.~\cite{MacDonald70a}.
Another one that is based on direct diagonalization of angular momentum has been
proposed in Ref.~\cite{morrison74a}. 
More recently, the projection by solving a system of linear equations 
was also proposed \cite{Johnson17a,Johnson18a}.
For a comparison of the computational
cost of the various possibilities see for example
Refs.~\cite{MacDonald70a,kelemen75a,kelemen76a,Johnson18a}.

Let us now consider the decomposition of a symmetry-breaking state 
$\ket{\Phi}$ into its content in orthogonal eigenstates of angular
momentum. 
As explained in Sec.~\ref{sec:proj}, from very general properties of
compact Lie groups it follows that the linear span 
$\text{span}(SU(2)_A \ket{\Phi})$ constructed from all rotated states, 
i.e.\ the set 
$\big\{ \hat{R}(\alpha,\beta,\gamma) \ket{\Phi} , \, 
(\alpha, \beta, \gamma) \in [0, 2\pi] \times [0,\pi] \times [0, 4\pi] \big\}$, 
can be decomposed as the direct sum of subspaces of dimension $d_J = 2J+1$ 
that each carry an irrep $J$ of $SU(2)_A$ 
\begin{equation}
\label{eq:SU2:span:decompo}
\text{span}(SU(2)_A \ket{\Phi}) 
= \bigoplus_{J} \bigoplus_{ \epsilon = 1}^{n_J} S^{J}_{\epsilon} \, .
\end{equation}
The number $n_J$ of subspaces $S^{J}_{\epsilon}$ carrying the same irrep $J$ that can be found
in the decomposition of $\text{span}(SU(2)_A \ket{\Phi})$ depends on the structure 
of $\ket{\Phi}$, and we use the label $\epsilon$ to distinguish between them.

The natural way to obtain a full set of orthonormal basis states
$\left\{ \ket{\Psi^{JM}_\epsilon}, J, M \in \llbracket -J, J \rrbracket, \epsilon \in \llbracket 1, n_J \rrbracket \right\}$
withtin $\text{span}(SU(2)_A \ket{\Phi})$ is to diagonalize the Hamiltonian
within this space.
To that end, 
in addition of the application of the operator it is necessary to solve 
the GEP
\begin{equation}
\label{eq:gepj}
\boldsymbol{H}^J \boldsymbol{f}^{J}_{\epsilon} 
= e^{J}_{\epsilon} \boldsymbol{N}^J \boldsymbol{f}^{J}_{\epsilon} \, ,
\end{equation}
where $\boldsymbol{H}^J$ and $\boldsymbol{N}^J$ are the Hamiltonian and 
norm matrices, respectively, whose matrix elements are
\begin{align}
\label{eq:projener}
H^{J}_{KK'} 
& =  \elma{\Phi}{\hat{H} \hat{P}^{J}_{KK'}}{\Phi}  \, ,\\
\label{eq:projover}
N^{J}_{KK'} 
& =  \elma{\Phi}{\hat{P}^{J}_{KK'}}{\Phi} \, . 
\end{align}
At the end of the procedure, we end up with symmetry-restored states of the form\footnote{The normalization of the state
is absorbed in the weights $f^{JK}_{\epsilon}$.}
\begin{equation}
\label{eq:PsiJMKeqSfPJMKPhi}
\ket{\Psi^{JM}_{\epsilon}} 
= \sum_{K=-J}^J f^{JK}_{\epsilon} \, \hat{P}^{J}_{MK}  \ket{\Phi} \, ,
\end{equation}
that are, from a variational point of view, the optimal states with good 
angular momentum in the vector space $\text{span}(SU(2)_A \ket{\Phi})$. 

Expanding $\ket{\Phi}$ on these orthonormal basis states, we can write
\begin{equation}
\label{eq:decompoJ}
\ket{\Phi} 
= \sum_J \sum_{\epsilon=1}^{n_J} \sum_{K=-J}^{J} 
  c^{JK}_{\epsilon} \ket{\Psi^{JK}_{\epsilon}} \, ,
\end{equation}
with the normalization of $\ket{\Phi}$ leading to the sum rule
\begin{equation}
\begin{split}
\label{eq:decompoJsumrule}
\norm{\Phi}{\Phi}  = 1 
&= \sum_J \sum_{\epsilon=1}^{n_J} \sum_{K=-J}^{J} 
  | c^{JK}_{\epsilon} |^2  \\
&= \sum_J \sum_{K=-J}^{J} N^{J}_{K K} \, .
\end{split}
\end{equation}
This sum rule can be used to probe the numerical accuracy of the AMP and determine the appropriate number of points 
required to converge the discretized integral over Euler angles (see Sec.\ \ref{sec:numAMP} for details) in 
numerical calculations. While both lines of Eq.~\eqref{eq:decompoJsumrule} are equivalent, the second one has the practical advantage
that it requires only the evaluation of projected matrix elements
and not the precise determination of the weights 
\begin{equation}
c^{JK}_{\epsilon} 
= \norm{\Psi^{JK}_{\epsilon}}{\Phi} = \sum_{K'=J}^{J} \big( f^{JK'}_{\epsilon} \big)^* \, N^{J}_{K'K} \, .
\end{equation}
that can be done only after solving the GEP of Eq.\ \eqref{eq:gepj}.
On the other hand, the advantage of calculating the coefficients $c^{JK}_{\epsilon}$ to make use of the first line of Eq.\ \eqref{eq:decompoJsumrule}
is that it allows to check the numerical accuracy of the resolution of the GEP. In the end, both equalities can be used at the same time to test
the numerical precision of at each step of the symmetry restoration.

It is interesting to remark that while the space $\text{span}(SU(2)_A \ket{\Phi})$ contains for a given irrep all components $M\in \llbracket -J, J \rrbracket$, 
the decomposition \eqref{eq:decompoJ} of the state $\ket{\Phi}$ may contain only part of them (but at least one). 
This is possible because $\text{span}(SU(2)_A \ket{\Phi})$ is by design invariant under rotation, i.e.\
it is generated by rotating $\ket{\Phi}$ in every possible way, and therefore contains
all the components $M$ of a given irrep, even those not present originally in the decomposition of $\ket{\Phi}$.
For that reason, when discussing angular-momentum projection, one often uses a notation which 
distinguishes the angular-momentum components $K$ that a symmetry-breaking 
state $\ket{\Phi}$ can be decomposed into from the angular-momentum 
components $M$ that enter the calculation of the matrix elements of
irreducible tensor operators between symmetry-restored states in the 
space $\text{span}(SU(2)_A \ket{\Phi})$.

In the literature, the $K$ are sometimes associated with the
$z$-component of angular momentum expressed in an ``intrinsic'' frame of reference,
whereas $M$ labels angular momenta in the ``laboratory'' frame of reference.
The notion of intrinsic and laboratory frame, however, has its ambiguities 
as it suggests that there are two different bases involved, one in an 
``intrinsic frame'' attached to the nucleus as chosen by the theoretician
when setting up the nucleus' wave function and another one corresponding 
to what an experimentalist is observing in the laboratory.
In the projection formalism described above, however,
the $z$ direction that the such interpreted $K$ and $M$ refer to is 
the same, i.e.\ the basis states $\ket{\Psi^{JM}_{\epsilon}}$
are part of the same set of eigenstates of $\hat{J}_z$.
The terminology of intrinsic and laboratory frame
is in fact used by analogy with the Unified Model of Bohr and Mottelson 
and other collective models \cite{BM98a,Rowe70a,RS80a}, 
where the total wave function is assumed to be a factorization of an intrinsic part, 
which captures the internal structure of the nucleus, times a Wigner $D$-matrix that
describes the orientation of the nucleus in the laboratory frame in terms of collective
angular momentum degrees of freedom. In that case, the Wigner rotation matrix does indeed
describe a transformation between two differently oriented frames of reference as mentioned 
above. The same relations are also used as an approximation when connecting self-consistent 
mean-field results for electric and magnetic moments with experimental data without
symmetry restoration \cite{SJ87a}.

By contrast, within the projection method,
the importance often given to the decomposition \eqref{eq:decompoJ} into $K$-components of a 
state $\ket{\Phi}$  has to be tempered by the the fact that this decomposition depends on the 
orientation chosen to represent the state within the reference frame. As far as the projection
method is concerned, this choice is completely arbitrary\footnote{Nevertheless, it may be interesting
to look at the $K$-decomposition of a state from a purely theoretical point of view.}
as from a formal point of view it has no effect on the final set of 
projected states $\ket{\Psi^{JM}_{\epsilon}} $. When the state to be projected has some intrinsic 
symmetries, as it is usually the case, some specific choices can be more advantageous 
than others from a computational point of view as they simplify the numerical construction of the 
reference state and the numerical solution of the GEP.

The more fundamental difference is between the states $\hat{P}^{J}_{MK} \ket{\Phi}$ 
simply obtained by the application of the projection operators 
and the states $\ket{\Psi^{JM}_{\epsilon}}$ that diagonalize the Hamiltonian
in $\text{span}(SU(2)_A \ket{\Phi})$. 
Indeed, the former keep a memory of the $K$-component they have been constructed from and therefore
of the particular orientation chosen for $\ket{\Phi}$. By contrast, the latter \eqref{eq:PsiJMKeqSfPJMKPhi}  are constructed
explicitly as a superposition of all $K$-components present in the decomposition of $\ket{\Phi}$
when diagonalizing the Hamiltonian, and as such they are independent
of the initial orientation chosen for $\ket{\Phi}$. All states 
$\ket{\Phi}$ and $\ket{\Phi'} = \hat{R}(\alpha,\beta,\gamma) \ket{\Phi}$ 
that differ only by a rotation will generate the same subspace $\text{span}(SU(2)_A \ket{\Phi})$
and therefore the same symmetry-restored states 
after solving the GEP \eqref{eq:gepj}.
Nevertheless, it is important to stress that wether we consider the states $\hat{P}^{J}_{MK} \ket{\Phi}$ 
or $\ket{\Psi^{JM}_{\epsilon}}$, the projection cannot be reduced to a mere change 
of reference frame. As a matter of fact, this interpretation would be contradictory to the fact 
that the Hamiltonian is rotationally invariant as it implies that one obtains different values 
for the energy in different reference frames.

%
%-----------------------------------------------------------------------
%
\subsection{Distinguishing odd and even nuclei}

We have seen in Sect.~\ref{subsect:numberparity} that Bogoliubov
quasiparticle vacua $\ket{\Phi}$ are always eigenstates of a number parity 
operator~\eqref{eq:nparNop}, which permits to distinguish the states 
$\ket{\Phi}$ that only decompose on irreps with even particle number from 
those that decompose exclusively on irreps with odd particle number. The
symmetry under this operator can then be used to reduce the integration
interval for numerical particle number projection, provided that the number
parity of a given Bogoliubov state is known a priori.

Given our choice of parametrization for the Euler angles, 
the operator that takes the same role for angular-momentum corresponds
to a rotation about $2\pi$ around the $z$-axis
\begin{equation}
 \hat{\Pi}_J = e^{-2 \iunit \pi \hat{J}_z} \, ,
\end{equation}
although, in practice any rotation about $2\pi$ around an arbitrary
axis yields the same result.
This operator is linked to the point-group operators discussed in 
Sect.~\ref{sec:simple} below by being equal to the square of any
signature operator $\hat{\Pi}_J = \big( \hat{R}_x \big)^2
 = \big( \hat{R}_y \big)^2
 = \big( \hat{R}_z \big)^2$,
which is why we will call it \textit{squared signature}.
Together with the usual binary operation, 
the set 
$\{ \hat{\eins}, \hat{\Pi}_J \}$ defines a cyclic group of order two with two 
distinct irreps labeled by $\pi_J = \pm 1$, and which is a finite
subgroup of $SU(2)_A$.

Because of the rules of angular-momentum coupling, 
the wave function of 
a system composed of an even number of fermions is such that 
$\pi_J = 1$, whereas a state with an odd number of fermions has 
$\pi_J = -1$. This implies that for a fermionic wave function number parity
and squared signature are equal
\begin{equation}
\pi_J = \pi_n \, .
\end{equation}
As a consequence, the Bogoliubov quasiparticle state $\ket{\Phi}$, which 
is a superposition of projected states with either an even or an odd number 
of fermions, is such that 
\begin{equation}
 \hat{\Pi}_J \ket{\Phi} = \pi_{J} \ket{\Phi} \, .
\end{equation}
which leads also to the equality 
\begin{equation}
 c^{JK}_{\epsilon} = \pi_{J} \, (-)^{2J} \, c^{JK}_{\epsilon}
\end{equation}
for the weights $c^{JK}_{\epsilon}$ entering the decomposition~\eqref{eq:decompoJ}.

As a result, a quasiparticle state with squared signature $\pi_{J} = +1$
is a superposition of basis states $\ket{\Psi_{\epsilon}^{JK}}$ with an integer angular momentum,
whereas a quasiparticle state with $\pi_{J} = -1$ is a superposition of basis states with 
an half-integer angular momentum.

Analogously to the simplification of particle-number projection for states
with good number parity, we can simplify angular-momentum projection for
eigenstates of squared signature. Indeed, as can be easily 
shown, in this case the integration interval for the angle $\gamma$ 
in the projection operator can be reduced from $[0,4\pi]$ to 
$[0,2\pi]$ \cite{BallyPHD,Varshalovich88a}, leading to the numerically
less costly reduced projection operator 
\begin{equation}
\label{eq:redamr}
\begin{split}
\hat{\mathcal{P}}^{J}_{MK} 
&= \frac{2J + 1}{8\pi^2} \int_0^{2\pi} d \alpha \int_0^{\pi} d \beta \sin (\beta) \int_0^{2\pi} \! d \gamma \\
  &\phantom{=} \times \:   D_{MK}^{J*} (\alpha,\beta,\gamma) \, \hat{R}(\alpha,\beta,\gamma) \, . 
\end{split}
\end{equation}
Its form is the same for both possible eigenvalues of squared signature. 
However, the reduced projection operator~\eqref{eq:redamr} cannot distinguish 
anymore between states having different eigenvalues of $\hat{\Pi}_J$. 
Indeed, the operator of 
Eq.~\eqref{eq:redamr} is now a projection operator for a specific class of 
states, where the irrep $J$ one projects out has to be chosen according to the 
squared signature of the states it acts on, i.e.\ $\pi_{J} = (-)^{2J}$. 

%
%=======================================================================
%
\subsection{Evaluation of observables}
\label{sec:obsJ}

A large number of the observables of interest in nuclear structure are either irreducible tensor
operators with respect to $SU(2)_A$ or can be decomposed in terms of sums and/or products of
such operators. This is in particular the case for the observables of interest in
nuclear spectroscopy such as the energy, radii, or electromagnetic moments and 
transition moments.

Labeling generically $\hat{T}^{\lambda}_{m}$ an irreducible tensor of rank $\lambda$,
with $m \in \llbracket -\lambda, \lambda \rrbracket$, 
for any angle $(\alpha, \beta, \gamma)$ we have the relation
\begin{equation}
\label{eq:OPtransfo}
 \hat{R} (\alpha, \beta, \gamma) \, \hat{T}^{\lambda}_{m} \, \hat{R}^\dagger (\alpha, \beta, \gamma) 
 = \sum_{m'=-\lambda}^{\lambda} D^{\lambda}_{m' m} (\alpha, \beta, \gamma) \, \hat{T}^{\lambda}_{m'} \, .
\end{equation}
In the case of a scalar operator $\hat{T}^{0}_{0}$, e.g.\ the energy or the charge radius, 
the above relation reduces to the simple commutation relation
\begin{equation}
\label{eq:OPcommu}
 \big[ \hat{R}(\alpha, \beta, \gamma), \hat{T}^{0}_{0} \big] = 0 \, .
\end{equation}
from which directly follows
the commutation relation of the scalar operator with the projection operators
\begin{equation}
\label{eq:OPcommu2}
 \forall J, \, \forall M,K \in \llbracket -J,J \rrbracket^2 \, , \quad \big[ \hat{P}^{J}_{M K}, \hat{T}^{0}_{0} \big] = 0 \, .
\end{equation}
From a practical point of view, relations~\eqref{eq:PJMKPJpMpKp}
and~\eqref{eq:OPcommu2} imply that the evaluation 
of projected matrix elements of scalar operators requires only  one application of the 
projection operator, either on the bra on the ket, which substantially reduces the 
numerical cost. We have already used this relation above when discussing the 
derivation of the GEP~\eqref{eq:gepj}.

For irreducible tensor operators of higher rank, the commutator 
$\big[ \hat{P}^{J}_{MK}, \hat{T}^{\lambda}_{m} \big]$ is non-zero, but can 
be evaluated using relation \eqref{eq:OPtransfo} and using the Clebsch-Gordan
series for the product of two Wigner $D$ matrices \cite{Varshalovich88a}.
With this, one finds for the reduced matrix elements of the tensor operator
$\hat{T}^{\lambda}_{m}$, as defined through the Wigner-Eckart theorem \cite{BT97a,Drake06a}, 
between two projected states
\begin{equation}
\begin{split}
\label{eq:OPredu}
 &\redu{\Psi^{J'}_{\epsilon'}}{\hat{T}^{\lambda}}{\Psi^{J}_{\epsilon}} = \sqrt{2J'+1}
  \sum_{K'=-J'}^{J'} \sum_{K=-J}^{J} \big(f^{J'K'}_{\epsilon'} \big)^* \, f^{JK}_{\epsilon} \\
  &\qquad \times \sum_{m=-\lambda}^{\lambda} \sum_{M=-J}^{J} 
  \left( J M \lambda m \vert J' K' \right)
  \elma{\Phi}{\hat{T}^{\lambda}_{m}\hat{P}^{J}_{MK}}{\Phi} \, 
\end{split}
\end{equation}
where the $\left( J M \lambda m \vert J' K' \right)$ are Clebsch-Gordan coefficients 
\cite{Varshalovich88a}. The reduced matrix elements are
independent of the $z$-component of the angular momenta of the bra and the ket.
Matrix elements between specific states from the two irreps can then be obtained from
\begin{equation}
\label{eq:OPreconstru}
 \elma{\Psi^{J'M'}_{\epsilon'}}{\hat{T}^{\lambda}_{m}}{\Psi^{JM}_{\epsilon}} = 
 \frac{\left( J M \lambda m \vert J' M' \right)}{\sqrt{2J'+1}} 
 \redu{\Psi^{J'}_{\epsilon'}}{\hat{T}^{\lambda}}{\Psi^{J}_{\epsilon}} \, .
\end{equation}
Proceeding in such a way, one formally reduces the calculation of doubly-projected 
matrix elements $ \elma{\Phi}{\hat{P}^{J'}_{M'K'} \, \hat{T}^{\lambda}_{m} \, \hat{P}^{J}_{MK}}{\Phi}$
between all non-trivial combinations of $K$, $K'$, $M$, $M'$, and $m$ to the evaluation
of  singly-projected matrix elements $\elma{\Phi}{\hat{T}^{\lambda}_{m} \, \hat{P}^{J}_{MK}}{\Phi}$
between all combinations in the much smaller set of $K$, $M$, and $m$.
This brings a substantial reduction of the numerical cost of the calculation. The state on which 
the projection operator acts can either be the bra or, as chosen here, the ket.

From a different perspective, it is the Clebsch-Gordan coefficient by which the projected matrix 
element $\elma{\Phi}{\hat{T}^{\lambda}_{m}\hat{P}^{J}_{MK}}{\Phi}$ is multiplied
in Eq.~\eqref{eq:OPredu} that directly selects, through the orthogonality relation 
\begin{equation}
\begin{split}
 \sum_{m=-\lambda}^{\lambda} 
 \sum_{M=-J}^{J} & \left( J M \lambda m \vert J' K' \right) \, \left( J M \lambda m \vert J'' K'' \right) \\
  &= \delta_{J' J''} \, \delta_{K' K''} \, ,
\end{split}
\end{equation}
the projected angular momentum $J'$ and component $K'$ of the bra that gives a non-zero
contribution to the reduced matrix element without the need to explicitly project it out.

The discussion above can be trivially generalized to the calculation of matrix elements 
between states $\ket{\Psi^{J'M'}_{a \epsilon'}}$ and $\ket{\Psi^{J'M'}_{b \epsilon}}$ projected 
out from different reference states $\ket{\Phi_a}$ and $\ket{\Phi_b}$, respectively. 
In that context, and using that $\hat{T}^{\lambda\dagger}_{m} = (-1)^{m} \, \hat{T}^{\lambda}_{-m}$, 
one can derive an additional useful symmetry for the reduced matrix elements \cite{Suhonen07}
\begin{equation}
\label{eq:Tlambda:radmat:symm}
\redu{\Psi^{J'}_{a \epsilon'}}{\hat{T}^{\lambda}}{\Psi^{J}_{b \epsilon}} 
=  (-1)^{J'-J} \redu{\Psi^{J}_{b \epsilon}}{\hat{T}^{\lambda}}{\Psi^{J'}_{a \epsilon'}}^* \, .
\end{equation}
This relation implies that having calculated the reduced matrix elements 
$\redu{\Psi^{J'}_{a \epsilon'}}{\hat{T}^{\lambda}}{\Psi^{J}_{b \epsilon}}$ 
for all relevant combinations of ($J$,  $\epsilon$) and ($J'$, $\epsilon'$) 
for given $\ket{\Phi_a}$ and $\ket{\Phi_b}$ along the lines of Eq.~\eqref{eq:OPreconstru}, 
one can use relation~\eqref{eq:Tlambda:radmat:symm} to reconstruct all required
reduced matrix elements $\redu{\Psi^{J}_{b \epsilon}}{\hat{T}^{\lambda}}{\Psi^{J'}_{a \epsilon'}}$
without the need to numerically calculate matrix elements where 
the projection operator acts on $\ket{\Phi_a}$.

Examples of observables of interest whose operators cannot be written in 
terms of irreducible tensor operators are the spatial densities and transition densities 
between symmetry-restored states \cite{Yao12a,Yao15a}. 
As a consequence, their calculation requires the projection (hence, rotation) of both states
entering the matrix elements.
There are, however, tricks to simplify the application of one of the 
rotation operators, see \cite{Yao12a,Yao15a}. 
Note, however, that the resulting densities take the quantum numbers of the projected
states for which they are calculated.

%
%-----------------------------------------------------------------------
%
\subsection{Numerical implementation}
\label{sec:numAMP}

For several reasons, the numerical evaluation of the angular-momentum projection 
of Bogoliubov-type quasiparticle vacua is much more involved than 
their particle-number projection. 
First, in the general case one has to deal with the evaluation of a three-dimensional
integral over Euler angles instead of two separate one-dimensional integrals
over proton and neutron gauge angles, respectively.
Second, as it will be explained later on, the number of points required to accurately discretize 
each of the three integrals  over Euler angles is usually larger than what is typically needed 
for gauge-space integrals as discussed in Sect.~\ref{subsect:PN:num:imp}, such 
that the total number of integrands to be evaluated can be several orders of magnitude larger.
Also, unless when working in a spherical basis, which becomes problematic at large deformation,  
the numerical representation of rotation in space about the Euler angle $\beta$ always mixes 
all states of same parity in the single-particle basis with each other.

Nevertheless, each combination of angles $(\alpha,\beta,\gamma)$ being independent from the others, 
one deals with an embarrassingly parallel problem that can be parallelized with almost a perfect
linear scaling. In addition, profiting from the structure of the integrands, 
it is possible to use efficient discretization of the three integrals. 

As the integrals over $\alpha$ and $\gamma$ have a structure alike the one of the integral over gauge angles, we use for
both of them a discretization similar to the one of Fomenko for particle-number projection~\eqref{eq:flamenko}
\begin{equation}
 \label{eq:momenko}
  \hat{\mathbb{P}}^{K_0}_{z,M_\gamma} 
  = \frac{1}{M_\gamma} \, \sum_{n=1}^{M_\gamma} e^{-\iunit  2 \pi \frac{n-\frac12}{M_\gamma} (\hat{J}_z - K_0)} \, ,
\end{equation}
explicated here for the angle $\gamma$. There are a few comments that we can make about 
the motivation of this choice
\begin{enumerate}[(i)]
\item
The use of a midpoint rule \eqref{eq:momenko} instead of the repeated rectangular rule used to discretize the
particle-number projection operator~\eqref{eq:flamenko} will be of advantage for the efficient exploitation of symmetries 
of the integrand that result from intrinsic symmetries of the states to be projected and that will be discussed
in Sect.~\ref{sec:simple}. As a consequence, however, the first point of the discretization is at 
$\frac{\pi}{M_\gamma}$ instead of zero. Therefore, the operator~\eqref{eq:momenko} does not reduce
to the identity for $M_\gamma=1$, and the case of ``no projection'' has thus to be treated separately. 

\item
Note that because of the identical choice of quadrature for the integrals over $\alpha$ and $\gamma$, 
the action of the discretized projection operator is trivially symmetric under the exchange of the bra and ket as long as the 
same number of discretization points is used for both of these angles.

\item 
The operator $\hat{\mathbb{P}}^{K_0}_{z,M_\gamma}$ 
is a discretization of the reduced projection operator of Eq.~\eqref{eq:redamr}. Therefore we can use 
$\hat{\mathbb{P}}^{K_0}_{z,M_\gamma}$ only to project states that have a squared signature equal \mbox{to $(-)^{2K_0}$}.
A more general operator that works whatever the squared signature of the reference state, meaning that its
application to a state with squared signature $\pi_{Ja} \neq (-)^{2K_0}$ yields matrix elements that are numerically zero,
could be defined by replacing, either for $\alpha$ or $\gamma$, the factor $2\pi$ in the exponential by a factor $4\pi$. 
Like in the case of particle number projection, applying such operator would be twice as costly.
\end{enumerate}
By applying the discretized operator $\hat{\mathbb{P}}^{K_0}_{z,M_\gamma}$ on a reference state $\ket{\Phi_a}$ with squared signature 
$\pi_{Ja} = (-)^{2K_0}$, we remove exactly\footnote{The rationale used to demonstrate this is identical to the one use in Appendix
\ref{sec:discpnr} for particle-number projection.}
all $K$ components such that $K_1 \neq K_0 + lM_\gamma$, $l \in \mathbb{Z}$
from the wave function 
\begin{equation}
\label{eq:projeffectK}
 \hat{\mathbb{P}}^{K_0}_{z,M_\gamma} \ket{\Phi_a} = \sum_{J_1} \sum_{\epsilon=1}^{n_{J_1}}
 \sum_{l \in \mathbb{Z}} (-)^l \, c^{J_1 K_0 + lM_\gamma}_{a \epsilon} \ket{\Psi^{J_1 K_0 + lM_\gamma}_{a \epsilon}} \, .
\end{equation}
Unfortunately, there is no discretization for the integral over $\beta$ that would lead to a similar removal 
of directly identifiable $J$ components from the non-projected state.
But the structure of the volume element suggests to use a $M_\beta$-point Gauss-Legendre
quadrature rule
\begin{equation}
  \label{eq:gaussleg}
  \hat{\mathbb{P}}^{J_0 M_0 K_0}_{y,M_\beta} =      
  \frac{2J_0 +1}{2}  \sum_{i=1}^{M_\beta} \, \omega_i \, d_{M_0 K_0}^{J_0} (\beta_i) \, e^{-\iunit  \beta_i \hat{J}_y} \, ,
\end{equation}
where $\beta_i = \arccos(x_i)$, $x_i$ being the abscissa and $w_i$ the weight of the $i$-th point in the quadrature.
One of the advantageous features of this quadrature is that it is exact for any polynomial 
in $x = \cos(\beta)$ that has an order lower or equal to $(2 M_\beta - 1)$. Given that the 
transformation of a state $\ket{\Psi^{J_1 K_0}_{a \epsilon}}$ under rotation involves Wigner $d$-matrices 
of rank $J_1$, Eqs.~\eqref{eq:decompoJ}, \eqref{eq:RabgPsiJMK}, and~\eqref{eq:DMpMJ=eMpdJeM} imply that
each component $J_1$ present in the state $\ket{\Phi_a}$ contributes as a polynomial of order $(J_1 + J_0)$ 
in $\cos(\beta)$ to the expectation value 
$d_{M_0 K_0}^{J_0} (\beta_i) \, \elma{\Phi_a}{e^{-\iunit  \beta_i \hat{J}_y}}{\Phi_a}$
of the l.h.s.\ of Eq.~\eqref{eq:gaussleg}. Gauss-Legendre quadrature of the integral over this polynomial 
becomes exact if $M_\beta \geq (J_0 + J_1 + 1)/2$. This means that when dealing with a wave function 
$\ket{\Phi_a}$ that is a superposition of irreps up to some $J_{\text{max}}$, the projection of the 
component with $J_0$ using $\hat{\mathbb{P}}^{J_0 M_0 K_0}_{y,M_\beta}$ becomes formally exact for

\begin{equation}
\label{eq:betaconv}
M_\beta \geq \frac{J_0 + J_{\text{max}} + 1}{2} \, .
\end{equation}
Depending on the weight of the highest $J$ components in the state $\ket{\Phi_a}$, smaller
values of $M_\beta$ might well be sufficient to calculate projected matrix elements with a 
numerically acceptable precision.

The full discretized projection operator on angular momentum thus reads
\begin{equation}
\label{eq:PJKM:discretized:full}
\hat{\mathscr{P}}^{J_0 M_0 K_0}_{M_\alpha M_\beta  M_\gamma} 
 = \hat{\mathbb{P}}^{M_0}_{z,M_\alpha}  \, 
 \hat{\mathbb{P}}^{J_0 M_0 K_0}_{y,M_\beta} \, \hat{\mathbb{P}}^{K_0}_{z,M_\gamma} \, ,
\end{equation}
with $M_\alpha$ points for $\alpha$, $M_\beta$ points for $\beta$, and $M_\gamma$ points for $\gamma$. 
Note first that, as the rotations around axes $z$ and $y$ do not commute it is 
not possible to change the order in which the discretized projection operators are applied on a state.
Also, it is important to remark that because the orthogonality relation between two Wigner $d$-matrices is realized only for matrix elements of the same row ($M_0$)
and column ($K_0$) \cite{Varshalovich88a}, the accuracy of the operator that involves rotations around the $y$ axis by $\beta$, and which is always
applied as the second one,
 is directly impacted by the accuracy of the operators
acting on $\alpha$ and $\gamma$. An accurate numerical projection thus requires the simultaneous convergence of the three parts of the operator 
$\hat{\mathscr{P}}^{J_0 M_0 K_0}_{M_\alpha M_\beta M_\gamma}$.

Eventually, considering a number of points $M_\alpha$, $M_\beta$, $M_\gamma$ large enough, the full discretized projection operator 
will select only the desired component out of the reference state $\ket{\Phi_a}$, i.e.\
\begin{equation}
 \hat{\mathscr{P}}^{J_0 M_0 K_0}_{M_\alpha M_\beta M_\gamma}  \ket{\Phi_a} 
\: \underset{M_\alpha, M_\beta, M_\gamma \to \infty}{\xrightarrow{\hspace*{1.40cm}}} \: 
\sum_{\epsilon=1}^{n_{J_0}} c^{J_0 K_0}_{a \epsilon} \ket{\Psi^{J_0 M_0}_{\epsilon  a}} \, .
\end{equation}
To illustrate our discussion on the behavior of the discretized projection operators, we will 
study the numerical convergence of the angular-momentum projection of several Slater determinants 
constructed in the $sd$-shell valence space using the numerical suite \textsf{TAURUS} \cite{Bally20a,Bally19a}.
The use of Slater determinants for such analysis has the advantage that no 
particle-number projection is required, thus simplifying the calculations and removing a source 
of numerical inaccuracies (in particular as different $[N,Z]$ components
will in general have different angular-momentum decompositions).
Similarly, the $sd$-shell being solely made of positive parity single-particle states, the Slater determinants 
are automatically eigenstates of parity with eigenvalue $+1$.
Performing the calculations with only a few particles in a restricted model space further simplifies the calculations, reduces 
the numerical noise (in particular as rotations are more accurately represented in a spherical harmonic oscillator basis than 
on a coordinate-space mesh)
and provides a natural cutoff for the highest angular momentum than can appear in the decomposition of a given state.

To start with, we consider four Slater determinants built in the $sd$-shell using the USDB interaction \cite{Brown06a}.
The states were obtained each starting from a randomly-generated and symmetry-unrestricted seed wave function but were constrained to have a
certain value of the triaxial parameters $(\beta,\gamma)$.
Their main characteristics are summarized in Tab.~\ref{tab:qpamr}.
To cover all cases of major interest, we consider for states for even-even 
[case (a) and (c)]  and odd [case (b) and (b)] systems that at convergence either 
adopt an axial [cases (c) and (d)] or non-axial [case (a) and (b)] shape.
We have to stress, however, that in a fully symmetry-unrestricted code the
symmetries adopted by the self-consistent solutions are never perfect, such
that there might also appear tiny contributions in the decomposition that
result from the slight breaking of these intrinsic symmetries.

\begin{table}
\begin{ruledtabular}
\begin{tabular}{cccccccc}
Label & $Z_{\text{val}}$ & $N_{\text{val}}$ & $(\beta, \gamma)$ & $\langle \hat{J}_z \rangle$ & $\langle \hat{J}^2 \rangle$ &
$P$ & $T$ \\
\hline
a & 2 & 2 & $(0.075,30^\circ)$ & 0   &  20   & $+$ & $+$ \\ 
b & 1 & 2 & $(0.025,30^\circ)$ & -2.58 &  26.2 & $+$ & $-$ \\
c & 2 & 2 & $(0.130,0^\circ)$ & 0   &  16.1 & $+$ & $+$ \\ 
d & 2 & 3 & $(0.130,0^\circ)$ & 1.5 &  18.9 & $+$ & $-$
\end{tabular}
\caption{\label{tab:qpstates:j}
Characteristics of the four Bogoliubov quasiparticle states considered for
the convergence analyses displayed in Figs.~\ref{fig:amr1},
\ref{fig:amr2}, \ref{fig:amr3}, and \ref{fig:amr4} (see text). The values of $Z_{\text{val}}$ and
$N_{\text{val}}$ indicate the number of valence protons and neutrons, respectively, whereas
$\beta$ and $\gamma$ indicate the intrinsic quadrupole deformation. Finally, 
$\langle \hat{J}_z \rangle$ and $\langle \hat{J}^2 \rangle$ denote the mean values
of the angular-momentum operators as indicated. The columns $P$ and $T$ indicate if the 
symmetry-unrestricted self-consistent procedure converged towards a state that adopted 
the symmetry to be even or odd under space inversion and time reversal, 
either exactly or to a good numerical approximation, respectively (see text).
}
\label{tab:qpamr}
\end{ruledtabular}
\end{table}

\begin{figure}[t!]
\centering  
  \includegraphics[width=8.0cm]{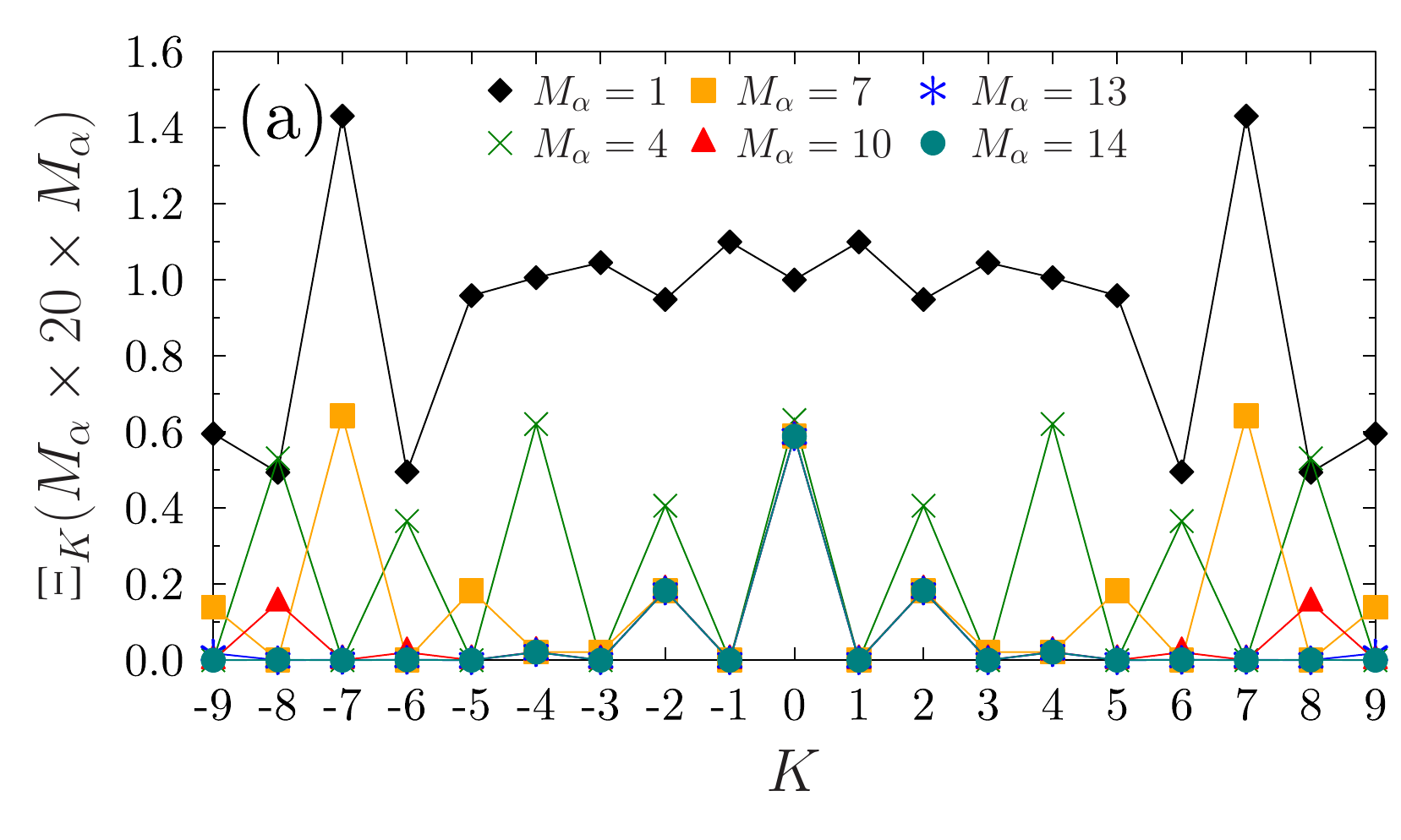}  \\
  \includegraphics[width=8.0cm]{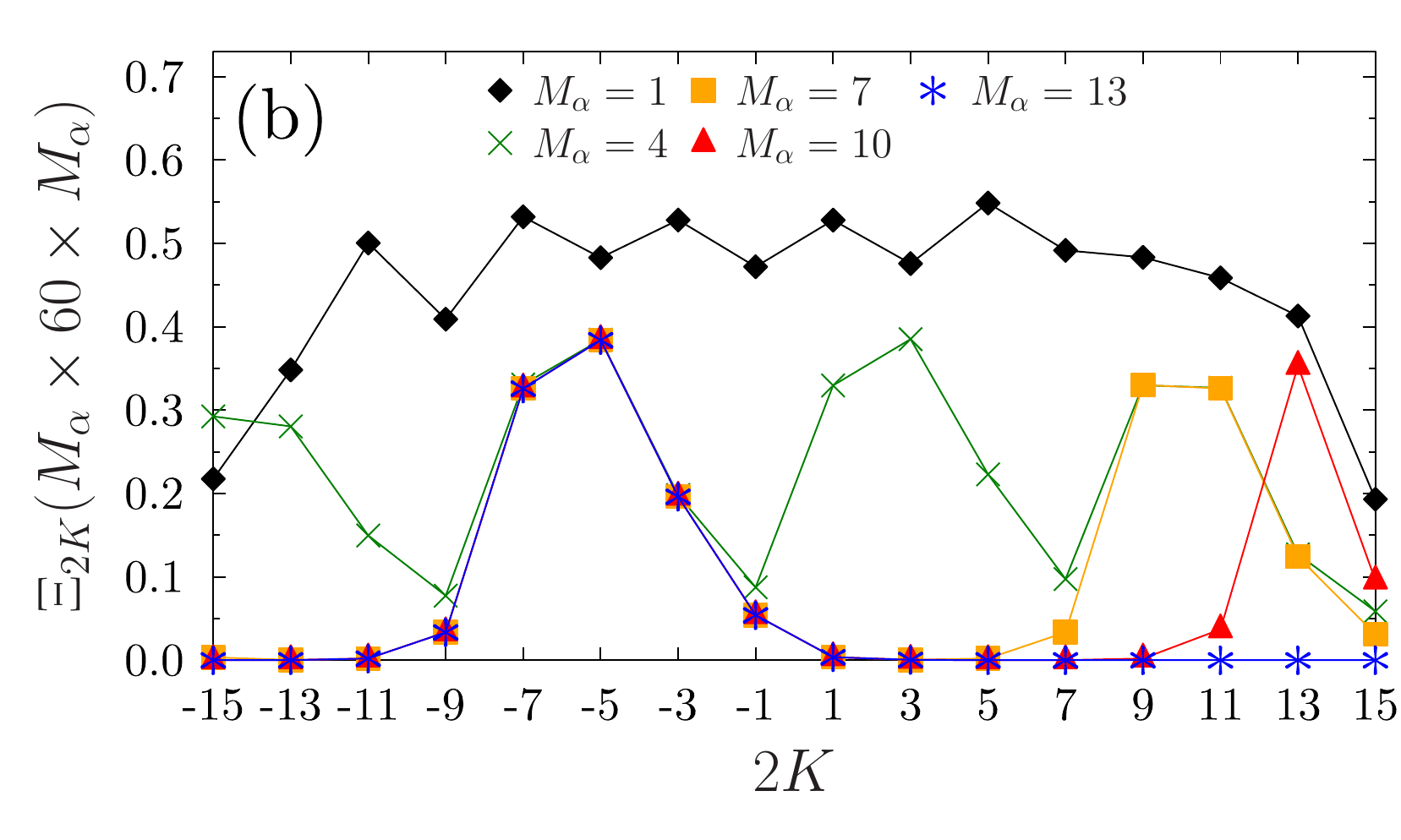} 
\caption{
\label{fig:amr1}
(Color online) Evolution of $\Xi_{K} (M_{\alpha} \times M_{\beta} \times M_{\gamma})$ 
as a function of the value of $K$ on which one projects for 
different choices for the number of discretization points $M_\alpha=M_\gamma$ for
the quasiparticle states $\ket{\Phi_a}$ (panel (a)), $\ket{\Phi_b}$ (panel (b)),  
as specified in Table~\ref{tab:qpstates:j}.
Points calculated with same $M_\alpha=M_\gamma$ are connected by straight lines 
to guide the eye. 
On each panel, the results obtained with the largest value of $M_\alpha=M_\gamma$
cannot be distinguished from exact results for the values of $J_0$ shown.
}
\end{figure}

To investigate the projection on the third-component of the angular momentum,
in Fig.~\ref{fig:amr1} we plot the sum of $J_0$-components for a given $K_0$, i.e.\
\begin{equation} 
 \label{eq:sumofJ}
 \Xi_{K_0} (M_\alpha \times M_\beta \times M_\gamma) = \sum_{J_0} \elma{\Phi}{\hat{\mathscr{P}}^{J_0 K_0 K_0}_{M_\alpha M_\beta M_\gamma}}{\Phi} \, ,
\end{equation}
for different choices for the number of discretization points $M_\alpha=M_\gamma$, while we fix a large value of $M_\beta = 20$.
As will be shown later on, with the latter choice, we can assume that the numerical evaluation of 
$\hat{\mathbb{P}}^{J_0 M_0 K_0}_{y,M_\beta}$ is converged for the states considered for our analysis.

We first notice that the two decompositions converge towards the physical values as one increases the number of discretization points 
in a manner that is very similar to what we observed for the  projection on particle number in Sect.~\ref{subsect:PN:num:imp}.
In particular, the discretized projection operator creates mirror images that are more and more pushed away to higher angular momenta
until only the true distribution remains. Nevertheless, there are some key differences with the particle-number case.
First of all, here the copies appear with the smaller periodicity $M_\alpha$ (instead of $2M_\phi$), which is to be expected given the 
factor $2\pi$ (instead of $\pi$) in the discretized operator of Eq.~\eqref{eq:momenko}. Also, the mirror images are not 
perfect copies of each other but present some noticeable asymmetries. 
As pointed out above, an imperfect projection on $M_0$ or $K_0$ will 
spoil the application of the intermediate operator $\hat{\mathbb{P}}^{J_0 M_0 K_0}_{y,M_\beta}$ that is in between
the two others, which might be the cause of those differences.

It is also interesting to remark that the two states display very different decompositions. The even-even state $\ket{\Phi_a}$ has a
perfectly symmetric decomposition centered around $K_0=0$ and with only even values of $K_0$. 
On the other hand, the odd-even state $\ket{\Phi_b}$ has non-vanishing components 
only for negative values of $K_0$ but do not display any selection rules for the values of $K_0$.
Typically, the distribution in terms of $K$ can be very different depending on the nature of the
unprojected state (even or odd system, degree of non-axiality, etc) and the orientation of its major
axes and therefore, contrarily to the particle-number case, there is no simple relation
that could help us determine the minimal number of points sufficient to project out the targeted 
states with a good accuracy. However, symmetries of the unprojected many-body states, either
imposed on them or adopted by them when solving the self-consistent HFB equations,
will affect the components  that can exist in their decompositions, as we will discuss in the 
Sec.~\ref{sec:simple}.

Another complication comes from the fact that an accurate projection of all $K_0$-components for a given $J_0$ is 
necessary to perform the subsequent mixing of $K$-components, which is necessary to fully diagonalize the 
Hamiltonian in the subspace of the projected states $\big\{ \hat{P}^{J_0}_{MK} \ket{\Phi} \big\}$, Eq.~\eqref{eq:gepj}.
This fact tends to increase the number of discretization points required compared to the particle-number projection as this means that one needs to
use enough points to converge even the extreme values of $K_0$ at the tail of the distribution.
In the valence-space calculations employed here to illustrate convergence, however, we can easily calculate the 
extremal values that $K_0$ can take and that are $\pm 8$ for state $\ket{\Phi_a}$ and $\pm13/2$ for 
state $\ket{\Phi_b}$. When the number of discretiztion points is insufficient, however, we can see that even 
higher values of $K_0$  numerically give non-vanishing results as they are part of mirror images
of the physical distribution.

\begin{figure}[t!]
\centering  
  \includegraphics[width=7.0cm]{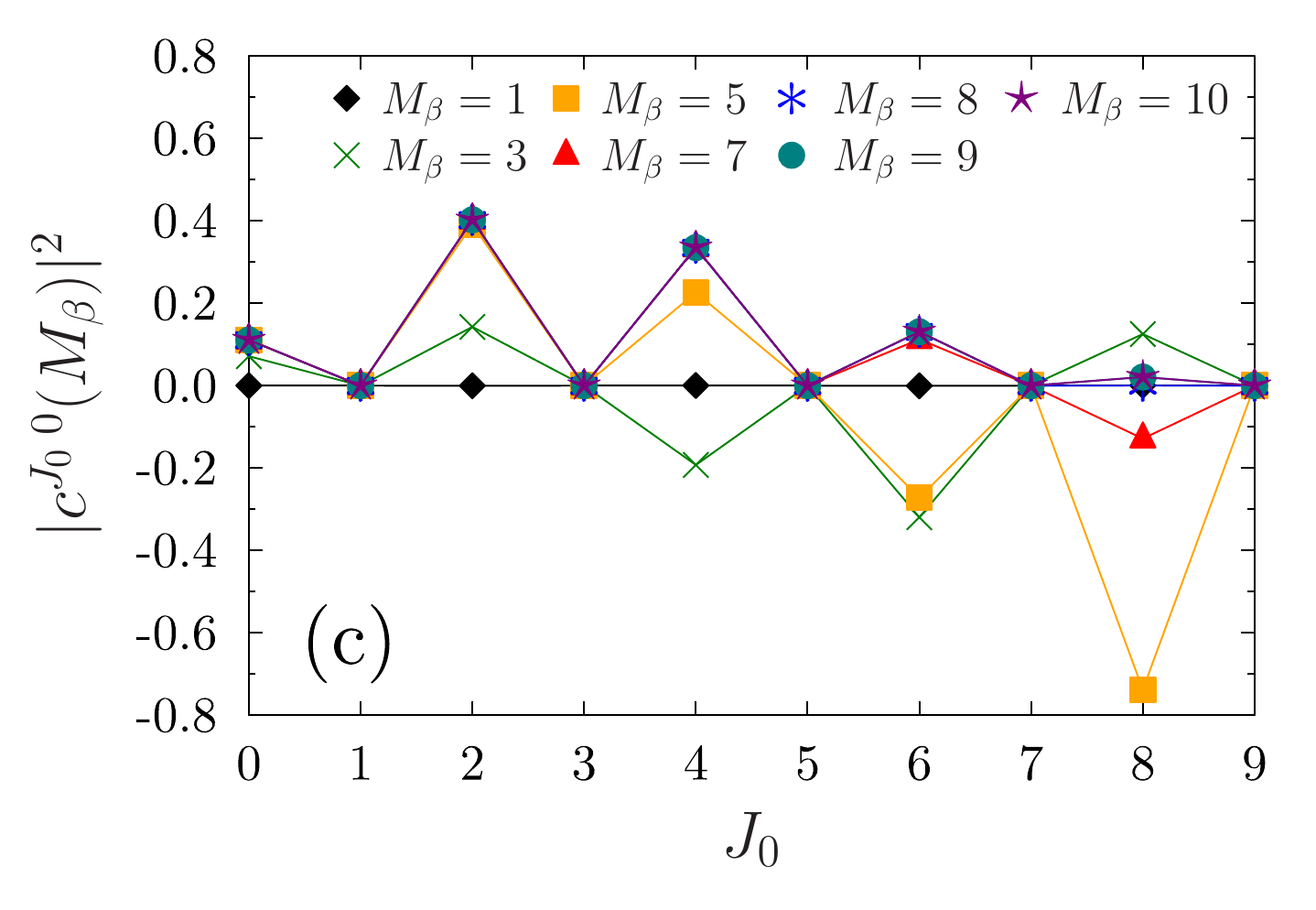} \\
  \includegraphics[width=7.0cm]{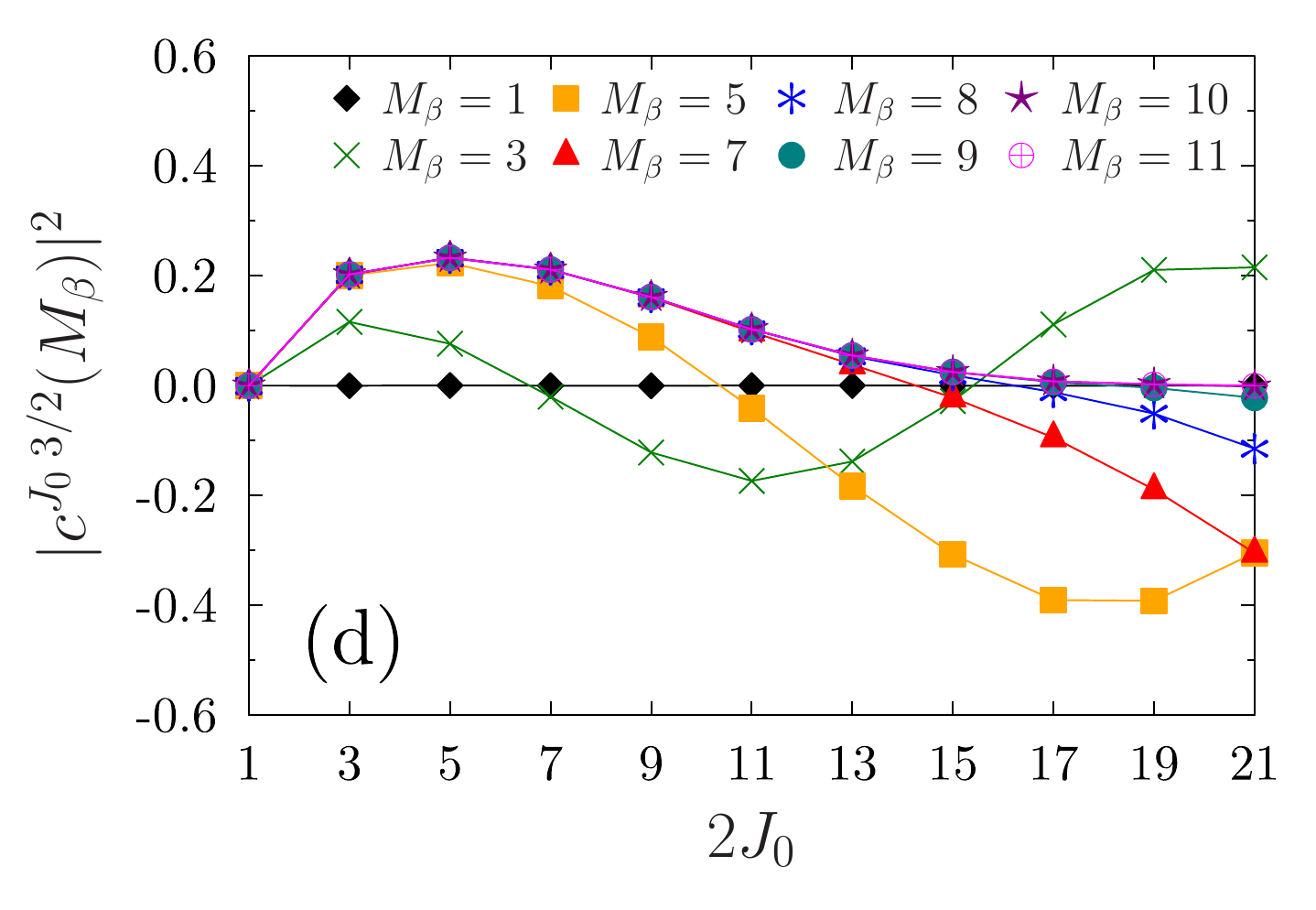}  
\caption{
\label{fig:amr2}
(Color online)
Evolution of the weight $|c^{J_0 K_0}(M_\beta)|^2$ of the numerically projected 
states as a function of the angular momentum $J_0$ on which one projects for 
different choices for the number of discretization points $M_\beta$ for
the quasiparticle states $\ket{\Phi_c}$ (panel (c)), $\ket{\Phi_d}$ (panel (d)),  
as specified in Table~\ref{tab:qpstates:j}.
Points calculated with same $M_\beta$ are connected by straight lines 
to guide the eye. 
On each panel, the results obtained with the largest value of ${M_\beta}$
is the first one for which the projection is exact for all values of $J_0$ plotted.
}
\end{figure}

As mentioned before, the operator $\hat{\mathbb{P}}^{J_0 M_0 K_0}_{y,M_\beta}$ 
has a different numerical structure than the other two that enter the full discretized 
projection operator of Eq.~\eqref{eq:PJKM:discretized:full} and that behave very similar
to the discretized particle-number operator. To analyze the convergence of
the operator $\hat{\mathbb{P}}^{J_0 M_0 K_0}_{y,M_\beta}$ more specifically, we consider
now the two states $\ket{\Phi_c}$ and $\ket{\Phi_d}$ as specified in Tabke~\ref{tab:qpstates:j}
that adopted axial symmetry and therefore can be treated as
eigenstates of $\hat{J}_z$ with eigenvalue $K$, such that only the integral over
$\beta$ has to be calculated numerically.
In Fig.~\ref{fig:amr2}, we display the of weight of the projected component $K_0$
\begin{equation} 
 \label{eq:weiJKex}
|c^{J_0 K_0} (M_\beta)|^2 = \elma{\Phi}{\hat{\mathbb{P}}^{J_0 K_0 K_0}_{y,M_\beta}}{\Phi} \, ,
\end{equation}
for different choices for the number of discretization points $M_\beta$.
Analogously to what is observed in Fig.~\ref{fig:amr1}, the discretized projection operator converges progressively as one increases the number 
of discretization points. The convergence pattern is, however, quite different. For a given value of $M_\beta$, one observes an oscillatory behavior
as a function of $J_0$ that numerically can even take negative values whereas the exact projected weights $|c^{J_0 K_0}|^2$
considered here are necessarily a positive quantity that can only take values between zero and one.
When increasing the number of discretization points $M_\beta$, the oscillatory behavior of the distribution of the 
$|c^{J_0 K_0}(M_\beta)|^2$ disappears and converges towards the physical values for the weights.

Again, we remark that the distributions of the weights of the irreps that can be projected out from the two states 
$\ket{\Phi_c}$ and $\ket{\Phi_d}$ are very different. The even-even state $\ket{\Phi_c}$ contains only even 
values $J_0$ and the projection vanishes identically for odd values of $J_0$ for all values of $M_\beta$, even the 
smallest ones.
That the state $\ket{\Phi_c}$ can only be decomposed onto irreps with even $J_0$ follows from the relations 
for the decomposition  of states with intrinsic symmetries that will be elaborated in Sect.~\ref{sec:simple}, and is
a consequence of this state having positive parity by construction and taking an axial shape. As the Wigner matrices
$d^J_{00}(\cos(\beta))$ are even (odd) functions of $\cos(\beta)$ for even (odd) values of $J$, the 
kernel $\elma{\Phi_c}{\hat{R}_{y}(0,\beta,0)}{\Phi_c}$ of an axial state whose physical decomposition 
only contains even values of $J$ will also be an even function of $\cos(\beta)$. When choosing a 
discretization of $\cos(\beta)$ that is symmetric around zero as done with Eq.~\eqref{eq:gaussleg},
then the numerical integrals over such kernel times an odd Wigner matrix $d^J_{00}(\cos(\beta))$
are automatically zero, irrespective of the number of discretization points $M_\beta$.
Such selection rule is specific to $K=0$ states and has no analogue
for the state $\ket{\Phi_d}$ that has $K = 3/2$. For the latter, the only component that 
necessarily has to vanish is $J_0 = 1/2$, which numerically is achieved
already for small values of $M_\beta$.

\begin{figure}[t!]
\centering  
  \includegraphics[width=7.0cm]{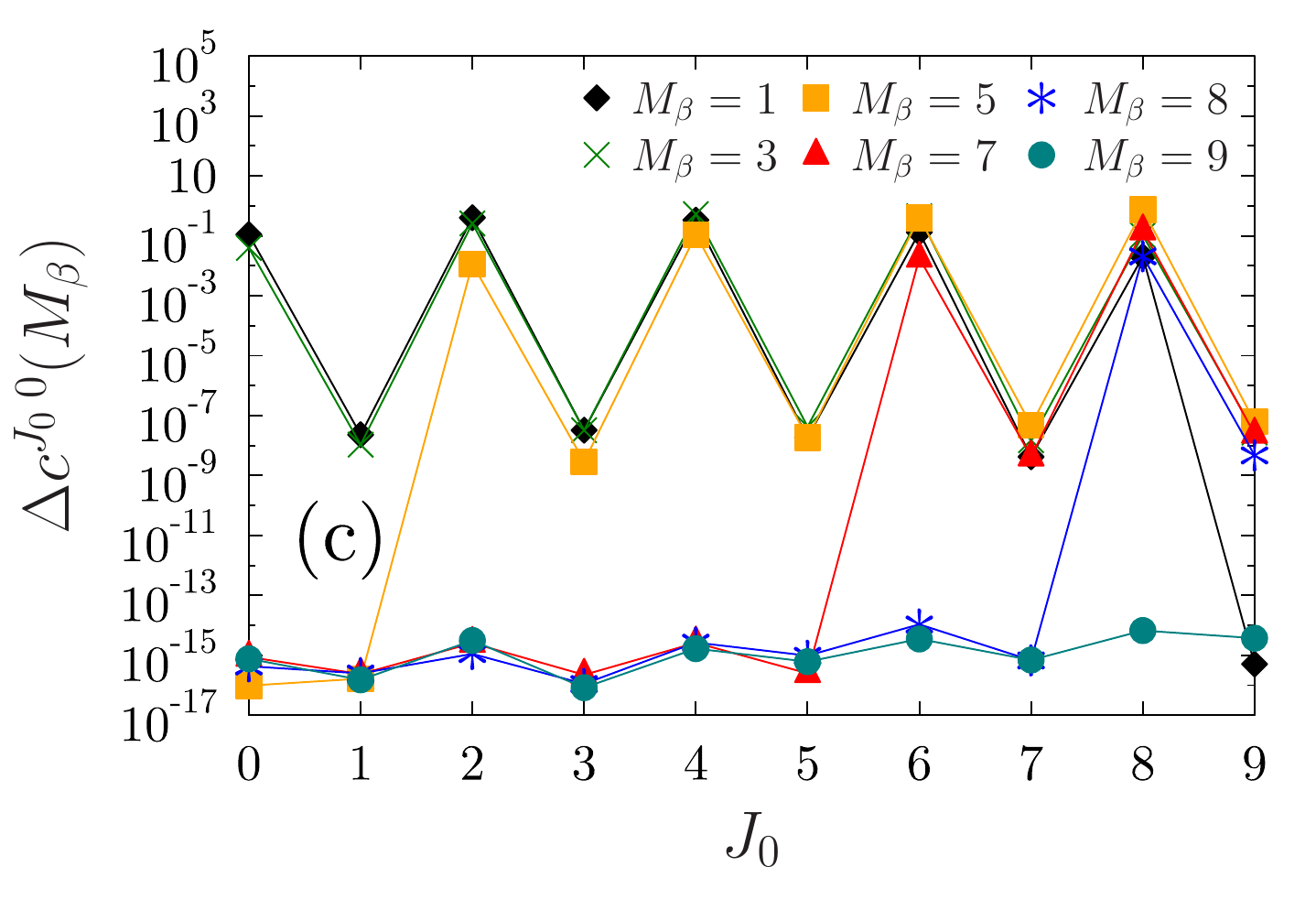} \\
  \includegraphics[width=7.0cm]{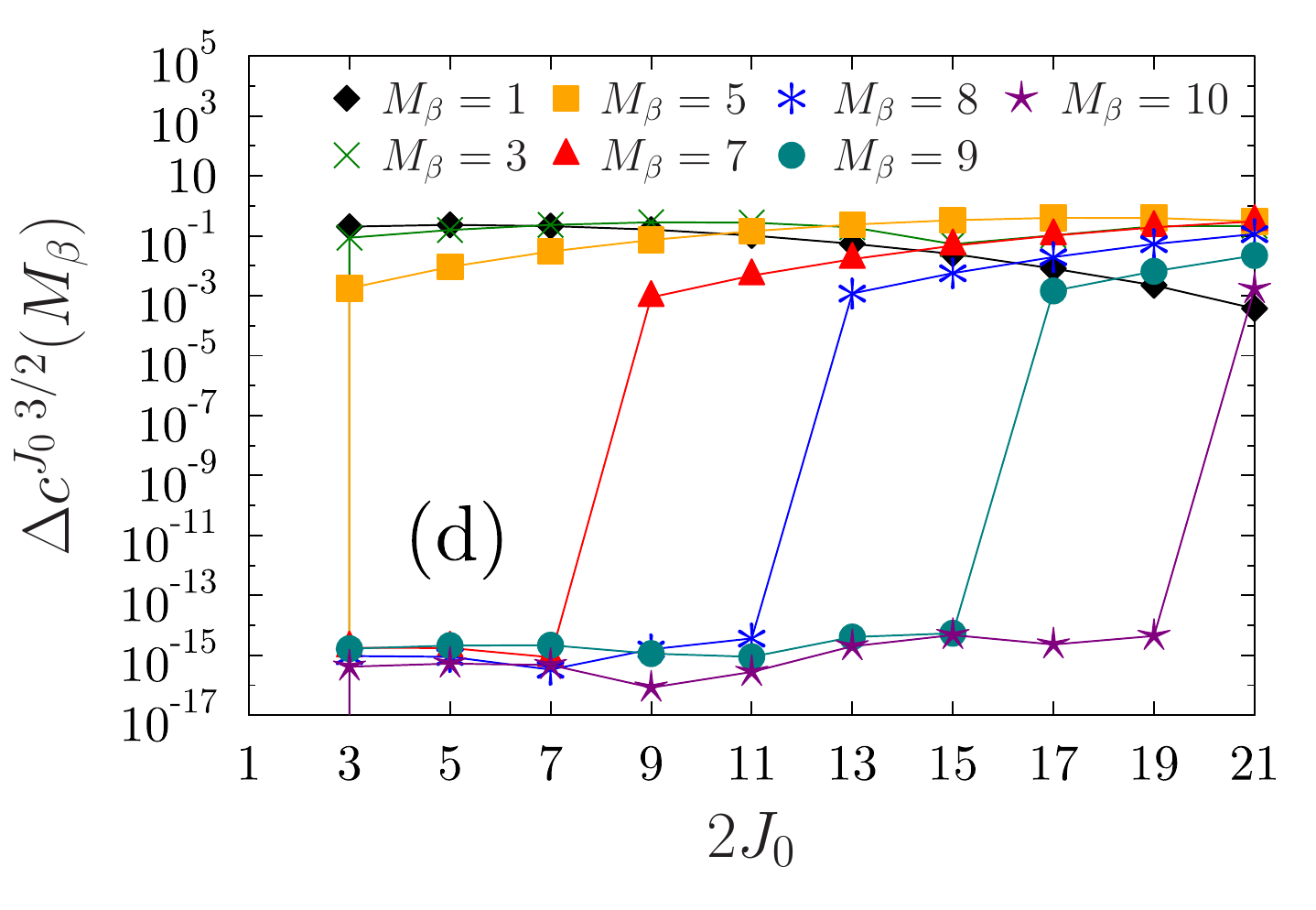}  
\caption{\label{fig:amr3} (Color online)
Evolution of the difference between the 
numerical values for the weight $|c^{J_0 K_0}(M_\beta)|^2$ at $M_\beta$ as indicated
and the converged values $|c^{J_0 K_0}(M^{c}_\beta)|^2$ obtained with $M^{c}_\beta = 11$
in logarithmic scale 
as a function of angular momentum $J_0$ on which one projects for
the quasiparticle states $\ket{\Phi_c}$ (panel (c)), $\ket{\Phi_d}$ (panel (d)),  
as specified in Table~\ref{tab:qpstates:j}.
Points calculated with same $M_\beta$ are connected by straight lines 
to guide the eye. 
}
\end{figure}

Given that the state $\ket{\Phi_c}$ is made of two neutrons and two protons in the $sd$-shell, the largest possible
value of $J_0$ is 8, which is also what is found at convergence in the numerical decomposition.
Having an additional neutron, the state $\ket{\Phi_d}$ can contain components having up to $J_0 =  19/2$. In this case, only 
$M_\beta = 11$ reproduces exactly this result.

The convergence pattern explicated in Eq.~\eqref{eq:betaconv} can be more explicitly seen in Fig.~\ref{fig:amr3}, 
where we display (in logarithmic scale) the same projected weights but subtracted from the converged values 
obtained with the discretization $M^{c}_\beta$, which is equal to the largest value of $M_\beta$ used in each panel of 
Fig.~\ref{fig:amr2}, i.e.\
\begin{equation} 
 \label{eq:weiJKexprec}
 \Delta c^{J_0 K_0} (M_\beta) = \left| \elma{\Phi}{\hat{\mathbb{P}}^{J_0 K_0 K_0}_{y,M_\beta}}{\Phi}  -
                                \elma{\Phi}{\hat{\mathbb{P}}^{J_0 K_0 K_0}_{y,M^{c}_\beta}}{\Phi} \right| \, .
\end{equation}
For example, in panel~(d), the discretization $M_\beta = 8$ gives exact results for components up to 
$J_0 = 11/2$: the difference to the
converged value obtained with $M^{c}_\beta = 11$ is of the order of the numerical noise.
This can be easily understood using Eq.~\eqref{eq:betaconv} as the largest physical
value of $J_1$ in the distribution is $19/2$ and therefore $11/4 + 19/4 + 1/2 = 8$. 
Note that this rule applies even if the value $J_0$ onto which one projects is not in the distribution
of components contained in the state to be projected.
Indeed, $M_\beta = 11$ is required to get the appropriate vanishing result for $J_0 = 21/2$, 
a component that cannot be contained in the Slater determinant $\ket{\Phi_d}$.

We note that the components with odd $J_0$ are not exactly zero on panel~(c). This is a
consequence of the state $\ket{\Phi_c}$ not having adopted a perfectly axially-symmetric 
 shape when stopping the HFB iterations in the symmetry-unrestricted code. 
Even though the $\Delta c^{J_0 K_0} (M_\beta)$ are all tiny for odd values of $J_0$, at small 
values of $M_\beta$  they are nevertheless approximately eight order of magnitudes larger than 
the fully converged results.

\begin{figure}[t!]
\centering  
  \includegraphics[width=7.0cm]{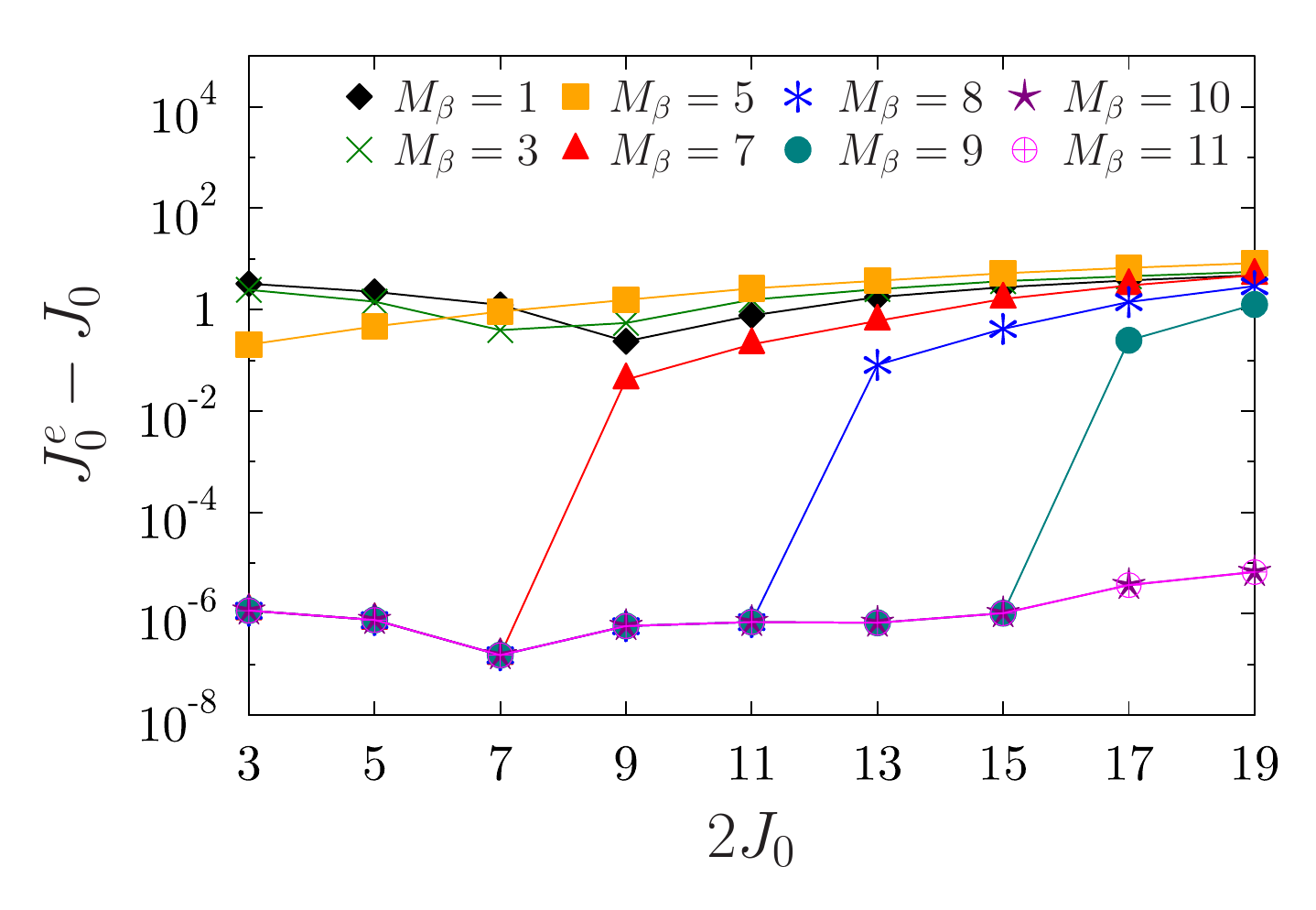}  
\caption{
\label{fig:amr4}
(Color online)
Evolution in logarithmic scale of the difference $J_0^e -J_0$, where $J_0^e$ is extracted from 
Eq.~\eqref{eq:convj}, for the state $\ket{\Phi_d}$
as a function of the angular momentum $J_0$ onto which one projects for 
different choices for the number of discretization points $M_\beta$.
Points calculated with same $M_\beta$ are connected by straight lines 
to guide the eye. 
}
\end{figure}

Finally, in Fig.~\ref{fig:amr4} we plot, for the state $\ket{\Phi_d}$, the difference between the value of
angular momentum $J_0^e$ extracted from the numerically projected matrix elements of $\hat{J^2}$ 
through the equation
\begin{equation}
\label{eq:convj}
\elma{\Phi}{ \hat{J^2} \, \hat{\mathbb{P}}^{J_0 K_0 K_0}_{y,M_\beta}}{\Phi} 
 =  J_0^e (J_0^e + 1) 
\end{equation}
and the targeted value $J_0$ as put into the projection operator.
As one can see, the convergence of $J_0^e$ follows the same pattern as for the overlap. 
At the same discretization of the projection operator, however, the numerical accuracy 
of $J_0$ value of is general inferior.

The possible maximal angular momentum, and thereby the possible maximal width
of the distribution of angular-momentum components, increases with
the number of active particles in the model space. As a consequence, the angular-momentum 
projection of a state represented in the full space of occupied single-particle states can become 
much costlier than projecting a state represented in a small valence space not only because of the larger
size of the single-particle basis, but also because of the larger number of discretization points that 
is needed for the projection operator.

\begin{figure}[t!]
\centering  
\includegraphics[width=8.0cm]{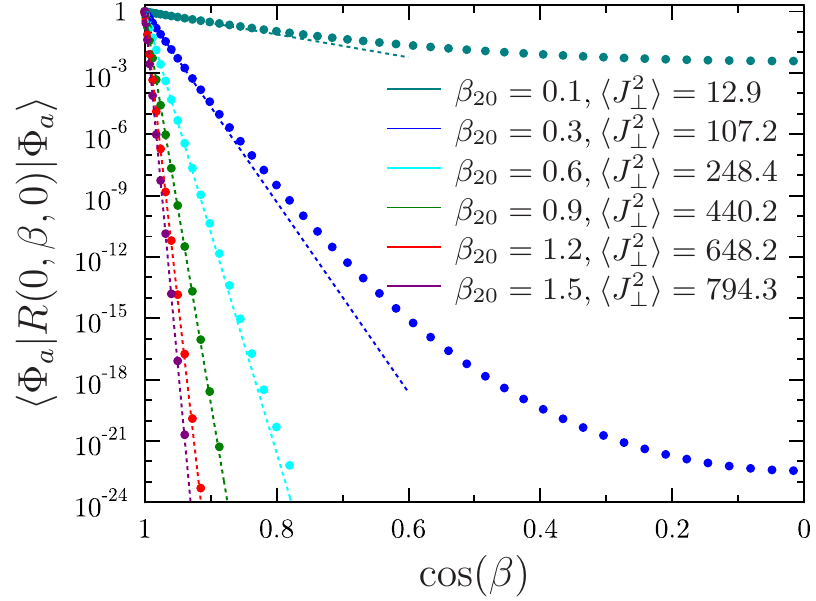}  
\caption{\label{fig:amrPu}
(Color online)
Norm kernel as a function of the Euler angle $\beta$ on a logarithmic scale 
for axial reflection-symmetric states of \nuc{240}{Pu} 
with dimensionless quadrupole moment $\beta_{20}$ and dispersion of angular momentum 
$\langle \Phi_a | \hat{J}^2_\perp | \Phi_a \rangle$ perpendicular to the symmetry axis as 
indicated. Dots indicate calculated values for the kernel using $M_{\beta} = 96$ points, whereas 
the dotted lines represent the estimate defined in Eq.~\eqref{eq:overlap:estimate}. For axial reflection-symmetric 
states, the norm kernel is a symmetric function around $\beta = \pi/2$ $\Leftrightarrow$ $\cos(\beta) = 0$, 
$\langle \Phi_a | \hat{R}(0,\beta,0) | \Phi_a \rangle = \langle \Phi_a | \hat{R}(0,\pi-\beta,0) | \Phi_a \rangle$,
cf.\ Eqns.~\eqref{eq:sym2} and~\eqref{eq:sym3}, such that only half the integration interval 
for $\beta$ is shown.
}
\end{figure}

As a rule of thumb, in a given model space the distribution of components contained 
in a state stretches further out with increasing deformation and increasing collective
angular momentum \cite{grummer78a}, thereby requiring an increasing number of discretization 
points for $\hat{\mathbb{P}}^{J_0 K_0 K_0}_{y,M_\beta}$ that are needed to attain the
same level of convergence for a component with given $J_0$. This, however, is partially 
counterbalanced by the norm and operator kernels becoming increasingly sharply peaked 
as a function of $\beta$ \cite{Bender06a}, such that the kernels might not have to be numerically 
evaluated for all combinations of Euler angles in the discretized projection operator. For the sake
of clarity, we will exemplify this for the simple case of axial and reflection-symmetric 
time-reversal-invariant quasiparticle vacua of \nuc{240}{Pu}
calculated with a Skyrme EDF in a space that comprises all single-particle states with non-zero
occupation. We choose states with a dimensionless prolate quadrupole moment of $\beta_{20} = 0.1$, 0.3,
0.6, 0.9, 1.2 and 1.5. The state at 0.3 is close to the normal-deformed ground state, the state at 
0.6 in the region of the inner fission barrier (whose saddle point corresponds to a non-axial state
not considered here), the state at 0.9 close to the deformation of the excited
superdeformed fission isomer, and the states at 1.2 and 1.5 in the outer barrier (whose actual 
saddle point corresponds to a non-axial reflection-asymmetric state not considered here).

\begin{figure}[t!]
\centering  
\includegraphics[width=8.0cm]{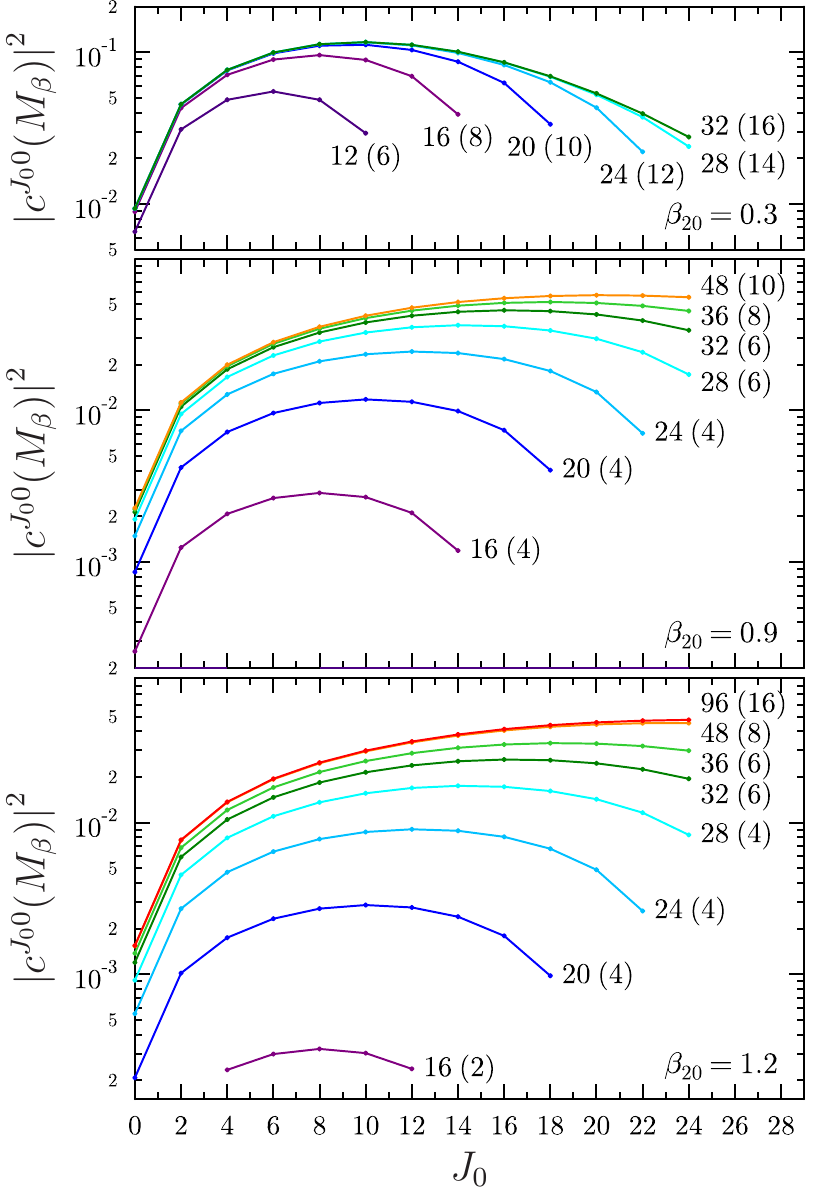}  
\caption{\label{fig:amrPu:cJ0}
(Color online)
Decomposition of the weights $\big| c^{J_0 0}\big|^2$ of the projected components as 
defined in Eq.~\eqref{eq:weiJKex} on a logarithmic scale for axial reflection-symmetric 
states of \nuc{240}{Pu} with dimensionless quadrupole moment $\beta_{20}$  
and number of discretization points $M_\beta$ as indicated. The number in parentheses
indicates the number of Euler angles whose contribution is at most ten orders of magnitude
smaller than the largest one. Because of the reflection symmetry imposed on the 
quasiparticle vacua, the $\big| c^{J_0 0}\big|^2$ are exactly zero for odd angular momenta
and therefore not plotted.
}
\end{figure}

To focus on angular-momentum projection, the states $| \Phi_a \rangle$ used to prepare 
Figs.~\ref{fig:amrPu} and~\ref{fig:amrPu:cJ0} are normalized particle-number projected quasiparticle vacua.
Recalling that the weights $\big| c^{J_0 K_0}\big|^2$ of the numerically projected components 
are the sum over the product of  the norm kernel  $\langle \Phi_a | \hat{R}(\alpha,\beta,\gamma) | \Phi_a \rangle$ 
times a factor that is at most on the order one, it follows immediately that in a numerical representation that 
uses double-precision floating point arithmetic that Euler angles for which of the norm kernel between normalized 
states is smaller than $10^{-16}$ cannot meaningfully contribute to the sum. As the level of numerical noise is 
usually at a larger level, the limit for significant contributions can already be drawn at about $10^{-10}$.

Figure~\ref{fig:amrPu} shows the norm kernel as a function of the Euler angle $\beta$ for these 
states.  For the smallest deformation, $\beta_{20} = 0.1$, all Euler angles $\beta$ contribute significantly to 
the projected matrix elements. But already at the typical ground-state deformation of heavy nuclei 
$\beta_{20} \simeq 0.3$, the norm kernel falls  off to values much smaller than $10^{-10}$ over a wide 
range of Euler angles, such that only those with $|\cos(\beta)| \gtrsim 0.7$ will contribute to the projected matrix 
elements. With increasing deformation the peak of the norm kernel narrows further, which reflects that for 
simply geometrical reasons an ever increasing part of the wave function that is at the tip of an elongated 
shape is rotated outside of the nucleus. This behavior of the norm kernel indicates that there is an additional
condition for the numerical convergence of angular momentum projection: there has to be a sufficient number
of integration points in the region where the overlap is peaked.

The norm kernels between two different states $| \Phi_a \rangle$ and $| \Phi_b \rangle$ might have 
a different angular dependence, depending on differences in elongation of the shapes and differences 
in orientation. For example, when calculating matrix elements between prolate and oblate states, 
the norm kernel is peaked at $\beta = \pi/2$ instead \cite{Bender04b}. Although they are of course 
not strictly proportional to each other, for very deformed states all operator kernels 
$\langle \Phi_a | \hat{T}^{\lambda}_{\mu} \, \hat{R}(\alpha,\beta,\gamma) | \Phi_a \rangle$
have in general the same overall dependence on the Euler angles as the norm kernel, which for tensor
operators of non-zero rank $\lambda$ might be superposed with an oscillatory behavior.

Figure~\ref{fig:amrPu:cJ0} shows the convergence of the numerical calculation of the
weights $\big| c^{J_0 0}\big|^2$ of the components at three different deformations as a 
function of $M_\beta$. The number of points in the full interval $ 0 \leq \beta \leq \pi$
that is necessary to reach a sufficient level of convergence increases dramatically with 
deformation, which is again a direct consequence of the rule established in 
Eq.~\eqref{eq:betaconv}. It is, however, not necessary to evaluate the matrix elements
at all of these integration points, as is indicated on Fig.~\ref{fig:amrPu:cJ0} by the
number of significant integration points given in the parentheses. In fact, to reach a similar
level of convergence, the number of necessary significant integration points is similar 
in all three cases, but they do not cover the same range of Euler angles.

The Wigner $d$-matrix $d^{J_0}_{00}(\beta)$ equals the Legendre polynomial of order $J_0$
in $\cos(\beta)$ \cite{Varshalovich88a}. As a consequence, when using the 
discretization~\eqref{eq:gaussleg} the weights $\big| c^{J_0 0}\big|^2$ are numerically 
exactly zero for $J_0 = M_\beta$, as in that case all abscissas coincide with the zeros 
of $d^{M_\beta}_{00}(\beta)$. Therefore, for discretizations with $M_\beta \leq 24$ 
the curves on Fig.~\ref{fig:amrPu:cJ0} end at $J_0 = M_\beta-2$. The weights of some components 
with $J_0 \gg M_\beta$ are again numerically non-zero as in the examples discussed 
above, but have been omitted on Fig.~\ref{fig:amrPu:cJ0} for the sake of clarity. From the
discussion leading to Eq.~\eqref{eq:betaconv} it is clear that the numerically calculated 
values of these components cannot be correct.

Assuming that $| \Phi_a \rangle$ is a normalized axial reflection-symmetric state, 
at small Euler angles $\beta$ the norm kernel for projection of that state can be 
estimated to be \cite{Baye84a}
\begin{equation}
\label{eq:overlap:estimate}
\langle \Phi_a | \hat{R}(0,\beta,0) | \Phi_a \rangle
\simeq \exp \big\{ - [1 - \cos(\beta)] \, \langle \Phi_a | \hat{J}^2_\perp | \Phi_a \rangle \big\} \, .
\end{equation}
where $\hat{J}^2_\perp$ is the square of the expectation value of angular momentum in a direction 
perpendicular to the symmetry axis. As can be seen from Fig.~\ref{fig:amrPu}, this estimate closely 
follows the calculated points for values of $\cos(\beta)$ between 1.0 and about 0.85, which is sufficient 
to estimate the range of Euler angles $\beta$ that will contribute significantly to the projected overlap 
and operator matrix elements. This information can be used to adapt $M_\beta$ such that one can expect to
have a certain number of significant integration points, and also to optimize approximate schemes for 
angular momentum projection that rely on interpolation between a small number of exactly calculated 
kernels \cite{Bender04b,Bender06a}. If needed, higher-order corrections to Eq.~\eqref{eq:overlap:estimate} 
can also be derived. Generalizations of Eq.~\eqref{eq:overlap:estimate} to full 3d rotations of non-axial 
states as the ones given in \cite{islam79a,Baye84a} can also be used to anticipate the number of discretization 
points needed to project more complex states.

To conclude, the number of discretization points for each Euler angle required to represent accurately
the three-dimensional  integral over Euler angles depends very sensitively on the nature and orientation
of the state one starts with. The large variety of possible decompositions that can be encountered prevents 
the existence of a simple recipe for the selection of integration points that can be expected to be near-optimal for 
all situations. To achieve a numerically precise projection of irreps up to $J_0$ one needs an estimate
of the largest values of $J$ and $K$ contained in decomposition of a given states, as well as an estimate
of the width of the norm kernel as a function of the Euler angles. In certain cases, the number of 
significant discretization points that have to be evaluated numerically in order to reliably perform 
the angular-momentum projection can be as large as several times $10^4$.
%
%=======================================================================
%
\section{Projection of cranked triaxial quasiparticle states}
\label{sec:simple}

%---------------------------------------------------------------------------------------------------------------------------

\subsection{General considerations}
\label{sec:simple:general}

Even considering the simple structure of a Bogoliubov-type quasiparticle vacua, numerical calculations 
for completely symmetry-unrestricted Bogoliubov reference states are far from being trivial.
Besides their numerical cost, there are also practical issues such as the necessity to fix
the position of the center-of-mass and also the orientation of the nucleus when self-consistently 
constructing the reference states, which can become complicated to implement and delicate to 
converge. In addition, there is a large body of empirical evidence that for many phenomena 
of interest the self-consistent solution of the HFB equation adopts one or several symmetries,
which can be used to simplify the numerical treatment by imposing these as intrinsic symmetries 
throughout the calculation.

All such conserved intrinsic space-time symmetries have in common 
that in one way or another they introduce linear dependences between
the states in the vector space $\text{span}(SU(2)_A \ket{\Phi})$ that can be 
constructed through rotations of a given Bogoliubov state $ \ket{\Phi}$.
This has two important consequences. 
First, these linear dependences can be used to reduce the number of rotated states that actually 
have to be constructed through application of the rotation operator. Second,  these 
linear dependences reduce the number of different irreps $\epsilon$ for a given value of $J$  
that  a given Bogoliubov state $ \ket{\Phi}$ can be decomposed into in Eq.~\eqref{eq:SU2:span:decompo}, 
from $2J+1$ to a smaller number, sometimes even zero. The latter manifests itself either through
vanishing matrix elements in the Hamiltonian and norm kernels of Eqns.~\eqref{eq:projener} 
and~\eqref{eq:projover}, or through linear dependences between their respective matrix 
elements, which in either case reduces the rank of these matrices. 
We have seen already some examples for such reduction in Sect.~\ref{sec:numAMP}.
Ultimately, the linear dependences between rotated states  
can be exploited to reduce the number of actual matrix elements 
$\elma{\Phi_a}{\hat{T}^{\lambda}_{m} \hat{P}^{J}_{MK}}{\Phi_b}$
that have to be calculated numerically. Even more importantly,
the reduction of the rank of the norm and Hamiltonian matrices 
might jeopardize the numerical solution of the the GEP~\eqref{eq:gepj} when not being taken care of.

As an example for the practical  consequences of intrinsic symmetries, we will discuss
here the case of time-reversal breaking triaxial HFB states, which offer the flexibility to cover a large number of cases of interest
in nuclear spectroscopy \cite{Bender08a,Rodriguez10a,Yao11a,Bally14a,Borrajo15a,Egido16b,Egido16a},
such as arbitrary quadrupole deformation and collective rotation about a principal axis. There are several possibilities 
how triaxial symmetry can be realized by imposing different subgroups of the double point symmetry group 
$D_{2h}^{TD}$ as defined in \cite{Doba00a,Doba00b}.
To exemplify the procedure, we will use here the subgroup consisting of parity $\hat{P}$,
the $x$-signature  $\hat{R}_x$, and the $y$-time-simplex $\hat{S}^{\mathcal{T}}_y \equiv \hat{R}_y \hat{P} \hat{\mathcal{T}}$,
with $\hat{\mathcal{T}}$ being the time-reversal operator. We will label the group generated by these 
three operators as $\langle  \hat{P}, \hat{R}_x , \hat{S}^{\mathcal{T}}_y  \rangle$.

This choice is not unique, and there are others that offer the same physical degrees of freedom 
\cite{Doba00b}, but differ by an exchange of coordinate axes. The actual
specific choice of symmetry group has some consequences for the convenience of numerical 
implementations.  The choice made here allows for simplifications during the mixing of $K$-components  
when solving the GEP that will be discussed in Sect.~\ref{sec:kmixing}. 
For single-reference calculations, however,  it is of advantage to impose $z$-signature $\hat{R}_z$ 
instead of $x$-signature as described in \cite{Bonche87a}. The transformation between both 
representations \cite{BallyPHD} can be achieved with an axis permutation operator as proposed 
in Ref.~\cite{Bonche91a}.

The symmetries of the quasiparticle vacua have actually to be implemented 
at the level of the one-quasiparticle states, meaning that one has to use a basis of single-particle states 
that respect this set of intrinsic symmetries, and construct Bogoliubov matrices that have the 
appropriate symmetry-preserving block structure. This has, however, no importance for our sidcussion.
For further details we refer to Refs.~\cite{Doba00a,Doba00b}.

Let us now specify the properties of a quasiparticle vacuum $\ket{\Phi_a}$ that has
the symmetries of the group $\langle  \hat{P}, \hat{R}_x , \hat{S}^{\mathcal{T}}_y  \rangle$. This implies
that
\begin{subequations}
\label{eq:triax:symm}
\begin{align}
\hat{P} \ket{\Phi_a}       & = p_a    \, \ket{\Phi_a} \, ,\\
\hat{R}_x \ket{\Phi_a}   & = \eta_a  \,   \ket{\Phi_a} \, ,\\ 
\hat{S}^{\mathcal{T}}_y \ket{\Phi_a}  &= \phantom{\pi_a} \, \ket{\Phi_a} \, ,
\end{align}
\end{subequations}
where only the first two relations are eigenvalue equations, whereas the
third one for the anti-linear operator $\hat{S}^{\mathcal{T}}_y $ fixes a phase  
through a symmetry transformation \cite{Doba00a,Doba00b}. From  
$\ket{\Phi_a}$ being a fermionic product state one also automatically has the relations
\begin{subequations}
\label{eq:fermionic:symm}
\begin{align}
 \hat{\Pi}_{A} \ket{\Phi_a} &= \pi_a \, \ket{\Phi_a} \, ,\\
\hat{R}_\mu^2 \ket{\Phi_a} &= \pi_{a} \, \ket{\Phi_a} \quad \text{for $\mu=x$, $y$, $z$,}\\
 \hat{\mathcal{T}}^2 \ket{\Phi_a}   &= \pi_{a} \, \ket{\Phi_a}  \, .
\end{align}
\end{subequations}
for the number parity $\hat{\Pi}_{A} $ \eqref{eq:nparNop}, squared signature 
$\hat{R}_\mu^2$ and the square of the time-reversal operator.

As we assume that $\ket{\Phi_a}$ is a direct product of a neutron and a proton 
many-body wave functions, and as our symmetry operators factorize
in the corresponding tensor product space, for example
$\hat{\Pi}_{A} \equiv \hat{\Pi}_{N} \otimes \hat{\Pi}_{Z}$,
the total number parity $\pi_{a}$, total parity $p_a$,  and total signature $\eta_a$, 
are then the products of the proton and neutron number parities, parities, 
and signatures, respectively,
\begin{subequations}
\begin{alignat}{2}
  \pi_{a} &= \pi_{na} \, \pi_{za}  \, &&,\\
    p_{a} &= p_{na} \, p_{za} \, &&,\\
 \eta_{a} &= \eta_{na} \, \eta_{za} \, &&,
\end{alignat}
\end{subequations}
where if the particle species $q = n$, $p$, has even number parity we have 
\begin{equation}
\pi_{qa} = +1 \, , \quad
p_{qa} = \pm 1 \, , \quad  
\eta_{qa} = \pm 1 \, ,
\end{equation}
whereas if the particle species $q$ has an odd number parity we have
\begin{equation}
 \pi_{qa} =-1 \, , \quad
   p_{qa} = \pm 1 \, , \quad
\eta_{qa} = \pm i \, . 
\end{equation}
Finally, defining a short-hand notation for the time-reversed state
\begin{equation}
 \label{eq:ta}
  \ket{\overline{\Phi}_a} =  \hat{\mathcal{T}} \, \ket{\Phi_a} \, ,
\end{equation}
we can derive an additional useful relation between $\ket{\Phi_a}$ and its 
time-reversed $\ket{\overline{\Phi}_a}$ from~\eqref{eq:triax:symm}
\begin{equation}
 \label{eq:trz}
 \hat{R}_z \ket{\Phi_a} = \pi_{a} \, p_a \, \eta_a^* \, \ket{\overline{\Phi}_a} \, ,
\end{equation}
which follows from
$\hat{S}^T_y \equiv \hat{R}_y \, \hat{P} \, \hat{\mathcal{T}} = \hat{R}_z \, \hat{R}_x \, \hat{P} \, \hat{\mathcal{T}}$.
Further details about the derivation of these relations can be found in Ref.~\cite{BallyPHD}.
As said above, quasiparticle vacua respecting these symmetries cover the vast 
majority of cases of interest for single-reference applications to nuclear spectroscopy.
And we will see that these symmetries allow also for great simplifications at the 
multi-reference level. 

We note that the symmetries of the $\langle  \hat{P}, \hat{R}_x , \hat{S}^{\mathcal{T}}_y  \rangle$
point-group can only be imposed on the unrotated states. The noncommutativity of rotations 
about different axes implies that the symmetry relations under $\hat{R}_x$ and $\hat{S}^{\mathcal{T}}_y$
transformations get lost when rotating a state to an arbitrary angle $\hat{R}(\alpha,\beta,\gamma) \, \ket{\Phi_a}$,
such that only parity remains as a common symmetry of the rotated and unrotated states.

%
%-----------------------------------------------------------------------
%
\subsection{Symmetries of the rotated matrix elements}

The numerical cost of the evaluation of an angular-momentum projected matrix element 
scales with the number $M_\alpha \times M_\beta \times M_\gamma$ of discretization points.
As we have seen in Sect.~\ref{sec:numAMP}, to numerically converge projected matrix elements 
might require to consider several tens of thousands of combinations of rotation angles, which 
can become a bottleneck in MR calculations that involve angular-momentum projection.

Therefore, to keep the computational time under control it is advantageous to find ways to reduce the 
number of angles that have to be calculated without altering the accuracy of the 
discretized integrals. By using the symmetries defined in Sect.~\ref{sec:simple:general},
we can derive a set of helpful relations between rotations by certain angles.

We will consider here the general case where we want to evaluate projected matrix elements of
the $2\lambda + 1 $ components $m$ of a spherical tensor operator $\hat{T}^{\lambda}_{m}$ 
between two different reference states $\ket{\Phi_b}$ and $\ket{\Phi_a}$, which  are assumed to have 
the same number parity for both protons and neutrons. Defining
$\hat{\overline{T}}{}^{\lambda}_{m} \equiv \hat{\mathcal{T}} \, \hat{T}^{\lambda}_{m} \, \hat{\mathcal{T}}^\dagger$
as a shorthand for the operator obtained by a time-reversal transformation of $\hat{T}^{\lambda}_{m}$,
we obtain the following seven relations \cite{BallyPHD}
\begin{subequations}
\begin{widetext}
\begin{alignat}{3}
 \label{eq:sym1}
 \bra{\Phi_a} \hat{R}(\pi-\alpha,\beta,\pi-\gamma) \, \hat{T}^{\lambda}_{m} \ket{\Phi_b}  
 & =  (-)^{\lambda} \, \pi_a \, \eta_{a} \, \eta_{b}^* \, && \bra{\Phi_a} \hat{R}(\alpha,\beta,\gamma) \, \hat{T}^{\lambda}_{-m} \ket{\Phi_b} \, &, \\ 
 \hspace{-1.0cm}
 \label{eq:sym2}
 \bra{\Phi_a} \hat{R}(\pi+\alpha,\pi-\beta,2\pi-\gamma) \, \hat{T}^{\lambda}_{m} \ket{\Phi_b} 
 & =  (-)^{\lambda} \, \pi_a \, \eta_{b}^* \, &&  \bra{\Phi_a} \hat{R}(\alpha,\beta,\gamma) \hat{T}^{\lambda}_{-m} \ket{\Phi_b} \, &, \\ 
 \hspace{-1.0cm}
 \label{eq:sym3}
 \bra{\Phi_a} \hat{R}(2\pi-\alpha,\pi-\beta,\pi+\gamma) \, \hat{T}^{\lambda}_{m} \ket{\Phi_b} 
 & =  \pi_a \, \eta_{a} \,  && \bra{\Phi_a} \hat{R}(\alpha,\beta,\gamma) \, \hat{T}^{\lambda}_{m} \ket{\Phi_b} \, &, \\ 
 \hspace{-1.0cm}
 \label{eq:sym4}
 \bra{\Phi_a} \hat{R}(\alpha,\pi-\beta,\pi-\gamma) \, \hat{T}^{\lambda}_{m} \ket{\Phi_b} 
 & =  (-)^{\lambda+m} \, p_{a} \, p_{b} \, \eta_{a} \, && \bra{\Phi_a} \hat{R}(\alpha,\beta,\gamma) \, \hat{\overline{T}}{}^{\lambda}_{-m} \ket{\Phi_b}^* \, &, \\ 
 \hspace{-1.0cm}
 \label{eq:sym5}
 \bra{\Phi_a} \hat{R}(\pi-\alpha,\pi-\beta,\gamma) \, \hat{T}^{\lambda}_{m} \ket{\Phi_b} 
 & =  (-)^{m} \, p_{a} \, p_{b} \, \eta_{b}^* \, && \bra{\Phi_a} \hat{R}(\alpha,\beta,\gamma) \, \hat{\overline{T}}{}^{\lambda}_{m} \ket{\Phi_b}^* \, &, \\ 
 \hspace{-1.0cm}
 \label{eq:sym6}
 \bra{\Phi_a} \hat{R}(\pi+\alpha,\beta,\pi+\gamma) \, \hat{T}^{\lambda}_{m} \ket{\Phi_b} 
 & =  (-)^{m} \, \pi_a \, p_{a} \, p_{b} \, \eta_{a} \, \eta_{b}^* \, && \bra{\Phi_a} \hat{R}(\alpha,\beta,\gamma) \, \hat{\overline{T}}{}^{\lambda}_{m}\ket{\Phi_b}^* \, &, \\ 
 \hspace{-1.0cm}
 \label{eq:sym7}
 \bra{\Phi_a} \hat{R}(2\pi-\alpha,\beta,2\pi-\gamma) \, \hat{T}^{\lambda}_{m} \ket{\Phi_b} 
 & =  (-)^{\lambda+m} \, p_{a} \, p_{b} \, && \bra{\Phi_a} \hat{R}(\alpha,\beta,\gamma) \, \hat{\overline{T}}{}^{\lambda}_{-m} \ket{\Phi_b}^* \, &, 
 \end{alignat}
\end{widetext} 
Each of these relations
is obtained by replacing either the left and/or right quasiparticle vacuum entering the matrix element with a suitable symmetry-transformed
state from one of the relations in Eq.~\eqref{eq:triax:symm} or~\eqref{eq:fermionic:symm}, followed by commutation of the rotation 
operator contained in the symmetry transformation with $\hat{T}^{\lambda}_{m}$ according to Eq.~\eqref{eq:OPtransfo}, using the 
special values of the Wigner rotation matrices at rotation angles that are multiples of $\pi$ \cite{Varshalovich88a} and regrouping 
of the remaining rotation operators into a single one \cite{BallyPHD}. The above relations connect the matrix elements of any 
irreducible tensor operator at Euler angles $(\alpha, \beta, \gamma)$ with those at seven other combinations of Euler angles in the 
full integration interval of the discretized projection operator~\eqref{eq:PJKM:discretized:full}, which reduces the 
necessary computational cost of the projected matrix element by a factor of eight.

In addition to these relations, there is another one that relates matrix elements involving the time-reversed of one of the 
states with matrix elements at a different rotation angle
\begin{alignat}{1}
\label{eq:sym8}
\bra{\Phi_a} & \hat{R}(\alpha,\beta,\pi+\gamma) \, \hat{T}^{\lambda}_{m} \ket{\Phi_b} 
\nn \\
 & =  (-)^{m} \, p_{b} \, \eta_{b}^* \, \bra{\overline{\Phi}_a} \hat{R}(\alpha,\beta,\gamma) \, \hat{\overline{T}}{}^{\lambda}_{m}  \ket{\Phi_b}^*  \, .
\end{alignat}
\end{subequations}
This relation is different in several respects from those of Eqns.~(\ref{eq:sym1}-\ref{eq:sym7}). First, it only is a symmetry 
relation when time-reversal is conserved, i.e.\ when $\ket{\overline{\Phi}_a}  = \ket{\Phi_a}$. In the more general
case that we address here the matrix element on the r.h.s.\ of Eq.~\eqref{eq:sym8} still has to be numerically evaluated in 
addition to the one on the r.h.s.\ of Eqns.~(\ref{eq:sym1}-\ref{eq:sym7}).
However, constructing the time-reversed of a product state is usually much less costly, and numerically more precise,
than rotating it. Second, relation \eqref{eq:sym8} can be combined with any of the seven relations  (\ref{eq:sym1}-\ref{eq:sym7})
above, such that the number of necessary applications of the rotation  operator is reduced by a factor of 16, from the 
full interval $[0,2\pi] \times [0,\pi] \times [0,2\pi]$ to the much smaller one 
$\left[0,\frac\pi2\right] \times \left[0,\frac\pi2\right] \times \left[0,\pi\right]$. But for 
each of the remaining  combinations of angles ($\alpha,\beta,\gamma$), one has to construct two matrix elements, 
one involving the original state, the other its time-reversed.  Omitting global pre-factors, the application of the 
reduced angular momentum projection operator~\eqref{eq:redamr} becomes with this
\begin{widetext}
\begin{align}
\label{eq:monster}
 & \int^{2\pi}_{0} d\alpha  \, \int^{\pi}_{0} d\beta \, \sin(\beta) \, \int^{2\pi}_{0} d\gamma \, 
       D_{MK}^{J*} (\alpha,\beta,\gamma) \,  \bra{\Phi_a} \op{R}(\alpha,\beta,\gamma)  \hat{T}^{\lambda}_{m} \ket{\Phi_b} \nn \\
 & = \int^{\frac{\pi}{2}}_{0} d\alpha  \, \int^{\frac{\pi}{2}}_{0} d\beta \, \sin(\beta) \, \int^{\pi}_{0} d\gamma \, 
     \Bigg[ \bigg\{ 
      \Big[ D_{MK}^{J*} (\alpha,\beta,\gamma) \, 
       +  \pi_a \eta_{a} D_{MK}^{J*} (2\pi-\alpha,\pi-\beta,\pi+\gamma)
      \Big] \bra{\Phi_a}\op{R}(\alpha,\beta,\gamma)  \hat{T}^{\lambda}_{m} \ket{\Phi_b} \nn \\
 &   +  (-)^{\lambda} \Big[  \pi_a \eta_{a} \eta_{b}^* D_{MK}^{J*} (\pi-\alpha,\beta,\pi-\gamma) 
       +  \pi_a \eta_{b}^*  D_{MK}^{J*} (\pi+\alpha,\pi-\beta,2\pi-\gamma) \,
       \Big]  \bra{\Phi_a} \op{R}(\alpha,\beta,\gamma) \hat{T}^{\lambda}_{-m} \ket{\Phi_b} \nn \\
 &   + (-)^{m} \Big[ p_{a} p_{b} \eta_{b}^* D_{MK}^{J*} (\pi-\alpha,\pi-\beta,\gamma) 
       +   \pi_a p_{a} p_{b} \eta_a \eta_{b}^* D_{MK}^{J*} (\pi+\alpha,\beta,\pi+\gamma)
       \Big] \bra{\Phi_a} \op{R}(\alpha,\beta,\gamma) \hat{\overline{T}}{}^{\lambda}_{m} \ket{\Phi_b}^* \nn \\
 &   + (-)^{\lambda+m} \Big[  p_{a} p_{b} \eta_{a} D_{MK}^{J*} (\alpha,\pi-\beta,\pi-\gamma)
       +   p_{a} p_{b}  D_{MK}^{J*} (2\pi-\alpha,\beta,2\pi-\gamma)
       \Big] \bra{\Phi_a} \op{R}(\alpha,\beta,\gamma) \hat{\overline{T}}{}^{\lambda}_{-m} \ket{\Phi_b}^*
     \bigg\} \nn \\ 
  & +  (-)^{m} p_{b} \eta_{b}^*  \bigg\{
       \Big[ D_{MK}^{J*} (\alpha,\beta,\pi+\gamma) \, 
        +  \pi_a \eta_{a} D_{MK}^{J*} (2\pi-\alpha,\pi-\beta,2\pi+\gamma)
       \Big] \bra{\overline{\Phi}_a}\op{R}(\alpha,\beta,\gamma)  \hat{\overline{T}}{}^{\lambda}_{m} \ket{\Phi_b}^* \nn \\
  &   + (-)^{\lambda} \Big[ \pi_a \eta_{a} \eta_{b}^* D_{MK}^{J*} (\pi-\alpha,\beta,-\gamma) 
        + \pi_a \eta_{b}^*  D_{MK}^{J*} (\pi+\alpha,\pi-\beta,\pi-\gamma) \,
        \Big]  \bra{\overline{\Phi}_a} \op{R}(\alpha,\beta,\gamma) \hat{\overline{T}}{}^{\lambda}_{-m} \ket{\Phi_b}^* \bigg\} \nn \\
 & +  (-)^{m} p_{b} \eta_{b}  \bigg\{
        (-)^{m} \Big[ p_{a} p_{b} \eta_{b}^* D_{MK}^{J*} (\pi-\alpha,\pi-\beta,\pi+\gamma) 
        +  \pi_a p_{a} p_{b} \eta_a \eta_{b}^* D_{MK}^{J*} (\pi+\alpha,\beta,2\pi+\gamma)
        \Big] \bra{\overline{\Phi}_a} \op{R}(\alpha,\beta,\gamma) \hat{T}^{\lambda}_{m} \ket{\Phi_b} \nn \\
  &   + (-)^{\lambda+m} \Big[ p_{a} p_{b} \eta_{a} D_{MK}^{J*} (\alpha,\pi-\beta,-\gamma)
        +  p_{a} p_{b}  D_{MK}^{J*} (2\pi-\alpha,\beta,\pi-\gamma)
        \Big] \bra{\overline{\Phi}_a} \op{R}(\alpha,\beta,\gamma) \hat{T}^{\lambda}_{-m} \ket{\Phi_b}
      \bigg\} \Bigg] \, .
\end{align}
\end{widetext}
The above equation is valid whatever the eigenvalue of the $x$-signature and whatever 
the number parity of the considered quasiparticle vacua. Thereby, this equation constitutes 
a generalization of the expression given in Ref.~\cite{Enami99a} that is limited to quasiparticle vacua with 
a $x$-signature $\eta_a = \eta_{na} =  \eta_{pa} = 1$ describing even-even nuclei.

As already mentioned when defining them through Eqns.~\eqref{eq:momenko} 
and~\eqref{eq:gaussleg}, respectively, the use of the symmetries (\ref{eq:sym1}-\ref{eq:sym8}) of the 
rotated matrix elements imposes some conditions on the form of the discretized 
operators $\hat{\mathbb{P}}^{K_0}_{z,M_\gamma}$ and $\hat{\mathbb{P}}^{J_0 M_0 K_0}_{y,M_\beta}$.
First, the angles connected by the symmetries have all to be contained in a given discretization of the 
complete intervals. Second, it should be avoided that discretization points are located on the borders of the 
reduced intervals $\left[0,\frac\pi2\right] \times \left[0,\frac\pi2\right] \times \left[0,\pi\right]$. Indeed, 
when the quasiparticle vacua have eigenvalues of parity or signature that are different from $+1$, or when 
the two states entering the matrix element have different signature,\footnote{We note that although states of 
opposite signature are orthogonal, they have in general non-zero angular-momentum projected matrix elements 
if their parity is the same.}
then (\ref{eq:sym1}-\ref{eq:sym8}) imply 
that for symmetry reasons the norm overlap and some other operator matrix elements are zero on some of the 
boundaries of the reduced integration intervals. Using a discretization with points on the boundaries would 
then implicitly reduce the order of the quadrature rules. In addition, the same technical problems already 
mentioned in Sect.~\ref{subsect:PN:num:imp} would present themselves for the evaluation of those operator 
kernels that remain non-zero at those angles. 

The definitions Eqns.~\eqref{eq:momenko} and~\eqref{eq:gaussleg} of the discretized projection
operators respect these requirements as long as the values of $M_\alpha$, $M_\beta$, and $M_\gamma$
are chosen to be of the form
\begin{equation}
 \begin{split}
 M_\alpha & =  4i \, ,\\ 
 M_\beta  & =  2j  \qquad i,j,k \in \mathbb{N} \, ,\\
 M_\gamma & =  2k  \, ,
 \end{split}
\end{equation}
because we reduced by a factor of four, two, and again two the three intervals on which 
we carry out the rotations over $\alpha$, $\beta$, and $\gamma$, respectively.

%
%-----------------------------------------------------------------------
%
\subsection{Mixing of $K$-components}
\label{sec:kmixing}

As discussed in Sec.\ \ref{sec:proj} on the general principles of symmetry projection, 
linear dependences among the rotated states as discussed in the previous subsection
result in a reduction of the dimensionality of the subspace they span.
While for some intrinsic symmetries this eliminates specific values of $J$ and/or $K$  
directly from the spectrum of components a given quasiparticle vacuum can be decomposed
into, working with eigenstates of the $x$-signature introduces a symmetry in the norm and 
operator kernels that has to be specifically taken care of.

Indeed, considering a quasiparticle state $\ket{\Phi_a}$ that is an irrep of
$\langle  \hat{P}, \hat{R}_x , \hat{S}^{\mathcal{T}}_y  \rangle$,
a rotation by an angle $\pi$ around the $x$-axis transforms its projected component 
$K$ into its projected component $-K$
\begin{equation}
\label{eq:RxPsiJK}
 \op{R}_x \ket{\Psi^{J K}_{a \epsilon}} = e^{-\iunit \pi J} \, \ket{\Psi^{J -K}_{a \epsilon}} \, ,
\end{equation}
which can be easily shown inserting Eq.~\eqref{eq:Rx:def} into Eq.~\eqref{eq:RabgPsiJMK}
and then using the special values of the Wigner function $D_{M K}^{J} (0,\pi,\pi)$
\cite{Varshalovich88a} for given $J$, $M$, and $K$. This relation
implies that the coefficients $c^{J \pm K}_{a \epsilon}$ in the decomposition \eqref{eq:decompoJ} 
are related through
\begin{equation}
\label{eq:kandmk}
 c^{J -K}_{a \epsilon} = \eta_a \, e^{+\iunit \pi J} \, c^{JK}_{a \epsilon} \, .
\end{equation}
This relation among the weights of the components that can be projected out from a state
with given signature translates into a symmetry relation among elements \eqref{eq:projover} 
of the projected norm matrix $\boldsymbol{N}^J$
\begin{subequations}
\label{eq:norm:signature:symmetry}
\begin{alignat}{2}
N_{K K'}^J 
 & = \eta_{a}   \, e^{+\iunit \pi J} \, && N_{K -K'}^J  \, , \\ 
 & = \eta_{a}^* \, e^{-\iunit \pi J} \, && N_{-K K'}^J  \, , \\ 
 & =       && N_{-K -K'}^J \, ,
\end{alignat}
\end{subequations}
These relations imply that the norm matrix has null eigenvalues whenever decomposing 
a state with good $x$-signature. More precisely, there are $d_{\text{null}} = \frac{2J+1}{2}$
of them for half-integer angular momenta, and $d_{\text{null}} = \frac{2J+1}{2} \pm \frac12$ 
of them for integer angular momenta 
(the sign for $\pm 1/2$ depends on whether ${N}^J_{00}$ is null for reason of other intrinsic 
symmetries or not).
Notice that this does not necessarily exhaust the number of zero norm eigenvalues of these 
matrices as others can be null as well , for example, for reason of additional intrinsic symmetries that are 
not imposed by the numerical representation, but that are self-consistently taken by the state.

As the Hamiltonian is rotationally invariant, it commutes with $\op{R}_x$ and we find analogous 
symmetry relations for the elements~\eqref{eq:projener} of the Hamiltonian matrix $\boldsymbol{H}^J$
\begin{subequations}
\begin{alignat}{2}
 H_{K K'}^J 
 &= \eta_{a}    \, e^{+\iunit \pi J} \, && H_{K -K'}^J  \, ,\\ 
 &= \eta_{a}^*  \, e^{-\iunit \pi J} \, && H_{-K K'}^J \, , \\
 &=                                     && H_{-K -K'}^J \, .   
\end{alignat}
\end{subequations}
To solve the GEP \eqref{eq:gepj} numerically, we have to first remove all the null eigenvalues of 
the norm matrix that arise because of the intrinsic $x$-signature symmetry. In order to 
do so, we extend on an idea proposed by Enami \etal~\cite{Enami99a}, that is, 
we realize a similarity transformation
that partitions the norm, Hamiltonian, and other operator matrices into a 
$d_{\text{null}} \times d_{\text{null}}$ block that acts in the null space arising 
from the intrinsic $x$-signature symmetry, and a
$(2J + 1 - d_{\text{null}}) \times (2J + 1 - d_{\text{null}})$ block in the remaining space
of vectors that are not connected by a signature transformation. For the simplest 
case of states $\ket{\Phi_a}$ with even number parity of protons and neutrons 
and positive $x$-signature $\eta_a = +1$, a possible choice for $\boldsymbol{W}$ 
has been given in Ref.~\cite{Enami99a}. The generalization to arbitrary number parities 
and eigenvalues of signature is straightforward.

For a Bogoliubov quasiparticle vacuum $\ket{\Phi_a}$ with even number parity and 
$x$-signature $\eta_a = \pm 1$, which implies projection onto integer angular 
momentum $J$, the transformation $\boldsymbol{W}$ can be taken as the 
$(2J+1) \times (2J+1)$ orthogonal matrix
\begin{small}
\begin{equation}
\label{eq:WJeta:even}
 \boldsymbol{W}^J_{\eta_a} 
 \equiv
 \left( 
 \begin{array}{ccccccc}
  \tfrac{1}{\sqrt{2}} & & & \larzero  & & & \tfrac{\eta_a (-)^{J}}{\sqrt{2}}\phantom{a} \\
   & \ddots & & & & \iddots & \\
   & & \tfrac{1}{\sqrt{2}} & & \tfrac{\eta_a (-)^{J}}{\sqrt{2}}\phantom{a} & & \\
   \larzero & & & 1 & & & \larzero \\
   & & \tfrac{\eta_a (-)^{J+1}}{\sqrt{2}} & & \tfrac{1}{\sqrt{2}} & & \\
   & \iddots & & & & \ddots & \\
   \tfrac{\eta_a (-)^{J+1}}{\sqrt{2}}  & & & \larzero  & & & \tfrac{1}{\sqrt{2}}
   \end{array} 
 \right) \, .
\end{equation}\end{small}%
Using relations~\eqref{eq:norm:signature:symmetry} along the way, the
norm matrix resulting from a similarity transformation with this matrix
takes the structure
\begin{small}
\begin{align}
\big({\boldsymbol{W}^{J}_{\eta_a}} \big)^{-1} & \boldsymbol{N}^J \, \boldsymbol{W}_{\eta_a}^J
\nonumber \\
= &
\left(
\begin{array}{@{}c|c|c@{}}
  \phantom{\larzero} & & \phantom{\larzero} \\
  \larzero & \larzero  &\larzero \\
  \phantom{\larzero} & & \phantom{\larzero} \\
   \hline
   \phantom{\Big|}\larzero \phantom{\Big|}  & N^J_{0 0}  & \sqrt{2} N^J_{0 1} \cdots  \sqrt{2} N^J_{0 J} \\
  \hline
  \phantom{yolo} \larzero \phantom{yolo} & 
  \begin{matrix}
   \sqrt{2} N^J_{1 0} \\
   \vdots \\
   \sqrt{2} N^J_{J 0}
  \end{matrix} &
  \begin{matrix}
    2 N^J_{1 1}   &  \cdots  &  2 N^J_{1 J}  \\
    \vdots &  \ddots &  \vdots \\
    2 N^J_{J 1}   &  \cdots &  2 N^J_{J J}  \\
  \end{matrix}
 \end{array}
\right) 
\nonumber \\
\equiv &
\left(
  \begin{matrix}
    \norzero   &  \norzero \\
    \norzero   &  \tilde{\boldsymbol{N}}^J 
  \end{matrix}
\right) 
\, ,
\end{align}\end{small}%
and similar for the transformed Hamiltonian matrix and the matrices representing other 
operators.

For a Bogoliubov quasiparticle vacuum $\ket{\Phi_a}$ with odd number parity and 
$x$-signature $\eta_a = \pm i$, which implies projection onto half-integer angular 
momentum $J$, $\boldsymbol{W}^J_{\eta_a}$ can be taken as the $(2J+1) \times (2J+1)$ orthogonal 
matrix
\begin{small}
\begin{equation}
\label{eq:WJeta:odd}
 \boldsymbol{W}^J_{\eta_a} 
 \equiv
 \left( 
 \begin{array}{ccccccc}
  \tfrac{1}{\sqrt{2}} & & & \larzero  & & & \tfrac{\eta_a e^{+\iunit \pi J}}{\sqrt{2}} \\
   & \ddots & & & & \iddots & \\
   & & \tfrac{1}{\sqrt{2}} & & \tfrac{\eta_a e^{+\iunit \pi J}}{\sqrt{2}} & & \\
   \larzero & & & & & & \larzero \\
   & & \tfrac{\eta_a e^{-\iunit \pi J}}{\sqrt{2}} & & \tfrac{1}{\sqrt{2}} & & \\
   & \iddots & & & & \ddots & \\
   \tfrac{\eta_a e^{-\iunit \pi J}}{\sqrt{2}}  & & & \larzero  & & & \tfrac{1}{\sqrt{2}} \\
   \end{array} 
 \right) \, .
\end{equation}\end{small}%
In that case, the similarity-transformed norm matrix takes the form
\begin{align}
\big({\boldsymbol{W}^{J}_{\eta_a}} \big)^{-1} \, \boldsymbol{N}^J \, \boldsymbol{W}_{\eta_a}^J
= &
\left(
\begin{array}{@{}c|c@{}}
  \phantom{\larzero} & \phantom{\larzero} \\
  \larzero & \larzero \\
  \phantom{\larzero} & \phantom{\larzero} \\
  \hline
  \phantom{yolo} \larzero \phantom{yolo} & 
  \begin{matrix}
    2 N^J_{\frac12 \frac12}  &  \cdots & 2 N^J_{\frac12 J} \\
    \vdots & \ddots & \vdots \\
    2 N^J_{J \frac12}  &  \cdots & 2 N^J_{J J} \\
  \end{matrix}
 \end{array}
\right)
\nonumber \\
\equiv &
\left(
  \begin{matrix}
    \norzero  &  \norzero \\
    \norzero  &  \tilde{\boldsymbol{N}}^J 
  \end{matrix}
\right)
 \, .
\end{align}
We remark that, in the general case, the transformation matrix $\boldsymbol{W}^J_{\eta_a}$ does not only depend on the angular momentum onto 
which one projects, but also on the $x$-signature $\eta_a$ of the state $\ket{\Phi_a}$.

The advantage of performing such transformation is that it replaces the GEP \eqref{eq:gepj} that has
$d_{\text{null}}$ zero eigenvalues for symmetry reasons by a GEP that is reduced to the 
$(2J + 1 - d_{\text{null}}) \times (2J + 1 - d_{\text{null}})$ 
non-trivial blocks $\boldsymbol{\tilde{N}}^J$ and $\boldsymbol{\tilde{H}}^J$ of the transformed norm and Hamiltonian
matrices
\begin{equation}
\label{eq:gepsim}
\boldsymbol{\tilde{H}}^J \boldsymbol{\tilde{f}}^{J}_{\epsilon} 
= e^{J}_{\epsilon} \boldsymbol{\tilde{N}}^J \boldsymbol{\tilde{f}}^{J}_{\epsilon} \, .
\end{equation}
The transformed generalized eigenvectors $\boldsymbol{\tilde{f}}^{J}_{\epsilon}$ are related 
to the original ones through
\begin{equation}
\left(
\begin{matrix}
\boldsymbol{0} \\
\boldsymbol{\tilde{f}}^{J}_{\epsilon} 
\end{matrix}
\right)
= \boldsymbol{W}^{-1}_{\eta_a} \, \boldsymbol{f}^{J}_{\epsilon} 
\end{equation}
and are a linear combination of projected components with $K$ and $-K$ components 
induced by the $x$-signature symmetry. The possibility to eliminate the redundant
components this way is what has motivated us to choose the $x$-signature 
in the angular momentum projection over the $z$-signature used
in single-reference calculations.

The new GEP of Eq.~\eqref{eq:gepsim} obtained after similarity transformation gives same 
generalized eigenvalues $e^{J}_{\epsilon}$ as the ones of the original GEP. 
As for projected states, they can be shown to be equivalent to the original ones. 
Therefore we obtain same results by solving the transformed problem.

The definition of the transformation $\boldsymbol{W}^J_{\eta_a}$ is not unique; there are 
possible alternative forms that differ by global phases or that result in a different placement 
of the non-trivial block in the transformed matrix. But all of these lead to the same projected 
states after solution of the GEP.

The above relations can be easily generalized to the transformation of the norm and energy
kernels between two different states as they enter the a GCM calculation based on angular-momentum
projected states, and which are simply obtained as
$\big({\boldsymbol{W}^{J}_{\eta_a}} \big)^{-1} \, \boldsymbol{N}^{aJbJ} \, \boldsymbol{W}_{\eta_b}^J$
and similar for the energy kernel.
The same relation can also be used to transform the matrices representing the matrix elements
of any other scalar operator to the reduced space.

For higher-rank tensor operators that connect different irreps, the transformation of the
initial and final state will in general be different as their angular momenta and signatures
can be different. For a set of tensor operators $\hat{T}^{\lambda}_{\mu}$, and defining the 
reduced matrix elements before $K$-mixing as\footnote{We recall that $K$ and $K'$ are mere 
labels of components that have to be mixed by the GEP~\eqref{eq:gepj} in order to obtain
the projected states. They should not be confused with the labels $M$ and $M'$ of components 
within the irreps of the angular-momentum projected states after $K$ mixing.
}
\begin{align}
\boldsymbol{T}^{aJ'bJ}_{K'K} 
\, \equiv \, & \sqrt{2J'+1}
       \sum_{\mu = -\lambda}^{\lambda} \sum_{M=-J}^{J} \left( J M \lambda \mu \vert J' K' \right) \nn \\
& \times \elma{\Phi_a}{\hat{T}^{\lambda}_{\mu}\hat{P}^{J}_{MK}}{\Phi_b} \, ,
\end{align}
one then has to compute in the general case the transformed matrix
\begin{align}
\left(
\begin{array}{@{}c|c@{}}
 \phantom{\boldsymbol{\tilde{T}^{aJ}}} & \phantom{\norzero} \\
 \hline
 \phantom{\norzero} & \boldsymbol{\tilde{T}}^{aJ'bJ} 
\end{array}
\right)
\equiv \big({\boldsymbol{W}^{J'}_{\eta_a}} \big)^{-1} \, \boldsymbol{T}^{aJ'bJ} \, \boldsymbol{W}_{\eta_b}^{J} \, ,
\end{align}
where $\boldsymbol{\tilde{T}}^{aJ'bJ}$ is now a $(2J' + 1 - d'_{\text{null}}) \times (2J + 1 - d_{\text{null}})$
rectangular matrix. 

%
%-----------------------------------------------------------------------
%
\subsection{Time-reversed partners}
\label{subsec:timerev}

Let us consider a state $\ket{\Phi_a}$ and its time-reversed partner 
$\ket{\overline{\Phi}_a}$ as defined in Eq.~\eqref{eq:ta}. 
Using the symmetry relation~\eqref{eq:trz}, one can derive \cite{BallyPHD} that their components $c^{JK}_{a \epsilon}$ and 
$\bar{c}^{JK}_{a \epsilon}$, respectively, in the decomposition on the basis states of $\text{span}(SU(2)\ket{\Phi_a})$
are related through the equation
\begin{equation}
\label{eq:tandmt}
\bar{c}^{JK}_{a \epsilon} = p_a \, \eta_a  \, e^{+\iunit \pi K} \, c^{JK}_{a \epsilon} \, . 
\end{equation}
As a consequence, the elements of the norm matrix $\boldsymbol{\bar{N}}^J$ for the 
time-reversed state $\ket{\overline{\Phi}_a}$ 
are directly related to those of $\boldsymbol{N}^J$ through the relation
\begin{equation}
\label{eq:tsim}
 \bar{N}^J_{KK'} = (-)^{K-K'} \, N^J_{KK'} \, . 
\end{equation}
As the Hamiltonian commutes with $\hat{\mathcal{T}}$, we also have an equivalent relation for the Hamiltonian matrix $\boldsymbol{\bar{H}}^J$ 
of the time-reversed state $\ket{\overline{\Phi}_a}$
\begin{equation}
\label{eq:tsim2}
 \bar{H}^J_{KK'} = (-)^{K-K'} \, H^J_{KK'} \, .
\end{equation}
Equations~\eqref{eq:tsim} and~\eqref{eq:tsim2} can be seen as a similarity transformation 
through the $(2J+1) \times (2J+1)$ diagonal unitary matrix $\boldsymbol{S}$ whose matrix elements are 
\begin{equation} 
 S_{KK'} = e^{-\iunit \pi K} \, \delta_{KK'} \, .
\end{equation}
With the same rationale as the one used in the previous section,
it is then straightforward to understand that a pair of time-reversed states 
gives the same projected states after $K$-mixing, up to a phase factor.    
As example, we mention the case of two one-quasiparticle states 
built by blocking in a Bogoliubov even-even vacuum the two time-reversed single-particle states of opposite signatures. 
Even if they are orthogonal, those two one-quasiparticle states will give, up to linear dependence, the same projected states.
As a result, when building projected states, it is sufficient to consider only one out of the two states  
$\ket{\Phi_a}$  and $\ket{\overline{\Phi}_a}$, but not both of them. 

%
%-----------------------------------------------------------------------
%
\subsection{Reduction to a real symmetric GEP}

Another relation can be obtained by noticing that, within basis states of an irrep of $SU(2)$,
the time-reversal operation links a state with $K$ and a state with $-K$. 
With our choice of representation for the rotation and time-reversal operators, we obtain
\begin{equation}
\label{eq:tandmk}
\bar{c}^{JK}_{a \epsilon} = (-)^{J+K} \, \big( {c}^{J -K}_{a \epsilon} \big)^* \, .
\end{equation}
Combining this relation with Eqs.~\eqref{eq:kandmk} and~\eqref{eq:tandmt}, we identify that 
\begin{equation}
\label{eq:realc}
 {c}^{JK}_{a \epsilon} = p_a \, (-)^{2J} \, \big( {c}^{JK}_{a \epsilon} \big)^* \, .
\end{equation}
As a result, depending on the parity of the state $\ket{\Phi_a}$ and on the value of $J$, 
the weights ${c}^{JK}_{a \epsilon}$ are either real or purely imaginary numbers. 
Therefore, any product of the form $\big({c}^{JK'}_{a \epsilon}\big)^* {c}^{JK}_{a \epsilon}$ is real, 
and consequently so are the norm matrix $\boldsymbol{N}^J$ and the Hamiltonian matrix $\boldsymbol{H}^J$. 
The final consequence is that our problem reduces to a real symmetric GEP.

%
%-----------------------------------------------------------------------
%
\subsection{Representative applications}
\label{subsect:applications:esperance}

%
%-----------------------------------------------------------------------
%
\subsubsection{Examples of decomposition}
\label{subsubsect:applications:esperance:decomposition}

\begin{table}[t!]
\begin{ruledtabular}
\begin{tabular}{cccccccc}
label & $\langle \hat{Z} \rangle$ & $\langle \hat{N} \rangle$ & $(\beta_2, \gamma)$ & $\eta$ & $p$ & $\langle \hat{d}_z \rangle$ & $\langle \hat{d}_z \rangle_{\text{axial}}$ \\
\hline
``spherical''  & 12 & 13 & $(0.0,0^\circ)$   & $-i$ & 1 &  0.50  & 0.50 \\ 
``axial''      & 12 & 13 & $(0.52,0^\circ)$  & $-i$ & 1 &  0.50  & 0.50 \\
``triaxial 1'' & 12 & 13 & $(0.60,14^\circ)$ & $-i$ & 1 &  0.65  & 0.50 \\
``triaxial 2'' & 12 & 13 & $(0.77,30^\circ)$ & $-i$ & 1 &  1.19  & 2.50 \\
%``axial''      & 12 & 13 & $(0.40,0^\circ)$  & $-i$ & 1 &  0.50  \\
%``triaxial 1'' & 12 & 13 & $(0.45,14^\circ)$ & $-i$ & 1 &  0.65  \\
%``triaxial 2'' & 12 & 13 & $(0.52,30^\circ)$ & $-i$ & 1 &  1.19   
\end{tabular}
\caption{Characteristics of the quasiparticle vacua discussed in Fig.~\ref{fig:phy1}, where
$\beta_2$ and $\gamma$ are defined as in Ref.~\cite{Bender08a} characterize the absolute size and non-axiality
of quadrupole deformation, while $\eta$ is the eigenvalue of $z$-signature $\hat{R}_z$, $p$ the eigenvalue 
of parity, and $\langle \hat{d}_z \rangle$ 
the $z$ component of the decoupling vector as defined in 
\cite{Olbratowski06a} that for states with $\langle \hat{P}, \hat{R}_x , \hat{S}^{\mathcal{T}}_y \rangle$ 
plays the role of measuring the $z$ component of total angular momentum in the non-projected state (see text).
We also give the value of the decoupling vector $\langle \hat{d}_z \rangle_{\text{axial}}$ of the equivalent single-particle state
when considering a pure axial deformation.
All states are constructed as blocked one-quasiparticle excitations of \nuc{25}{Mg} 
and therefore are odd under time-reversal.
}
\label{tab:qpphy}
\end{ruledtabular}
\end{table}

Let us first study the decompositions in terms of $J$- and $K$-components of some quasiparticle vacua
that have the intrinsic symmetries of the group $\langle  \hat{P}, \hat{R}_x , \hat{S}^{\mathcal{T}}_y  \rangle$.
For a product state having such point-group symmetry all local densities and currents have three 
plane symmetries \cite{Doba00a,Doba00b,Bonche87a}.
Examples of the decomposition of time-reversal-invariant quasiparticle vacua with even number parities 
for protons and neutrons that have this symmetry, which implies $p = \eta = +1$, have been discussed earlier
in Refs.~\cite{Bender08a,Rodriguez10a}. We consider here instead four different one-quasiparticle states 
of \nuc{25}{Mg} constructed by (self-consistently) blocking different selected single-particle states and 
imposing constraints on their average quadrupole deformation during the minimization procedure. 
These states are taken from the set of reference states entering the projected GCM calculation
presented in Ref.~\cite{Bally14a} and have been selected because they are representative examples
for the effects of projection. The states were generated using 
the cartesian 3d coordinate-space HFB code \textsf{CR8} that is based on the principles outlined 
in Ref.~\cite{Bonche87a} and which has been updated to handle the Skyrme Hamiltonian SLyMR0 of 
Ref.~\cite{Sadoudi13a} in connection with exact Coulomb exchange and Coulomb pairing, and which 
was used for the calculation described below. The HFB and the projection code \cite{esperance} use 
the same accurate Lagrange-mesh technique for derivatives \cite{Ryssens15a}, which can also be used to 
define rotation operators. While the projection code assumes a 
$\langle  \hat{P}, \hat{R}_x , \hat{S}^{\mathcal{T}}_y  \rangle$ point-group symmetry, the mean-field code 
imposes a $\langle  \hat{P}, \hat{R}_z , \hat{S}^{\mathcal{T}}_y  \rangle$ symmetry instead, which is 
of advantage when generating cranked HFB states as those that will be discussed in Sect.~\ref{subsec:cranked}.
The transformation between both representations is achieved with an axis permutation  operator as defined 
in Ref.~\cite{Bonche91a}, see Ref.~\cite{BallyPHD} for details. 
The main characteristics of the states can be found in Tab.~\ref{tab:qpphy}. For the sake of simplicity 
we label the four states as ``spherical'', ``axial'' and ``triaxial'', which refers to the shape 
of their matter density distribution. In all cases, the shape has been constrained
with quadrupole constraints that lead to the values for the quadrupole moments $\beta_2$ and $\gamma$. 
This does, however not mean that the ``spherical'' state truly takes spherical symmetry. While it has 
zero quadrupole moments, the distributions of spin and current cannot be spherically symmetric 
for a state with odd-number parity that in general has a non-zero expectation value of angular momentum. 
Similarly, the ``axial'' state has been constrained to an axial shape of its matter density distribution, 
which does not automatically mean that currents and spin densities are axially symmetric.

Working with eigenstates of the $x$-signature implies that only the component of their angular momentum
vector pointing in $x$ direction is measurable; hence, can have a non-zero expectation value.\footnote{This
does not mean that the angular-momentum vector of these states is aligned with the $x$-axis. An eigenstate
of $\hat{J}_x$ that automatically is an eigenstate of $x$-signature still can have an expectation value
of $\hat{J}^2$ that is larger than $J_x(J_x+1)$.
} 
This has direct consequences for the decomposition of such states into eigenstates of $\hat{J}_z$ by applying
the projection operator $\hat{P}^{J}_{K K}$. As has been worked out in detail in Ref.~\cite{Olbratowski06a},
eigenstates of a signature with respect to some cartesian coordinate axis can always be expressed 
as a superposition of an eigenstate of a signature with respect to a different cartesian coordinate axis
and its time-reversed state, where both have equal weight and some relative phase. As they are eigenstates 
of $\hat{J}_z$, the $K$ components projected out with $\hat{P}^{J}_{K K}$ are by construction also 
eigenstates of the $z$-signature. From this follows that a quasiparticle vacuum that is eigenstate of the
$x$-signature will always decompose into pairs of components $\pm K$ with equal weight, which 
is the foundation of Eq.~\eqref{eq:kandmk} derived earlier. The transformation $\boldsymbol{W}^J_{\eta}$ 
as defined in Eqns.~\eqref{eq:WJeta:even} or~\eqref{eq:WJeta:odd} then recombines each pair of 
components $\pm K$ into a single state whose measurable angular momentum points again into 
$x$ direction. This choice of the $x$-axis for the direction of measurable angular momenta might not be 
the most intuitive, but has the advantage that redundant states in the decomposition can be eliminated
with symmetry arguments based on conserved signature of the reference state.\footnote{Note that 
what is discussed in the literature as the signature of eigenstates of angular momentum usually has 
a different meaning: it is the signature with respect to a rotation by $\pi$ around the axis into which 
points angular momentum. This property is for example used to characterize experimentally observed
rotational bands \cite{BM98a,RS80a,Szymanski84a,Ejiri89a}. We will come back to that in Sect.~\ref{subsec:cranked}.
} 
We also recall that the physical states
are those obtained from $K$ mixing solving the GEP~\eqref{eq:gepsim}, such that the decomposition 
into $K$ components is merely a diagnostic tool.
The following analysis of decomposition addresses $J$ and $K$ components as obtained from the 
application of $\hat{P}^{J}_{K K}$ before applying the similarity transformation 
$\boldsymbol{W}^J_\eta$.

The same arguments apply to the characterization of angular momenta of the quasiparticle
vacua that serve as reference states. The shape of the states is oriented such that prolate states 
($\gamma = 0^\circ$) are symmetric around the $z$-axis in the projection code. The size of the 
angular momentum of a single-particle state along the symmetry axis of the ``axial'' state is 
obtained as the length of the ``decoupling vector'''\footnote{Throughout this subsection, we use
lower case symbols when referring to operators acting in the single-particle space and their matrix
elements in order to distinguish them from the upper case symbols that refer to operators in Fock space.}
$d_z \equiv \langle \hat{j}_z \, \hat{\mathcal{T}} \rangle$ 
as defined in Ref.~\cite{Olbratowski06a}. The same orientation has been chosen also for the 
``spherical'' state.

\begin{figure}[t!]
\centering  
  \includegraphics[width=8.5cm]{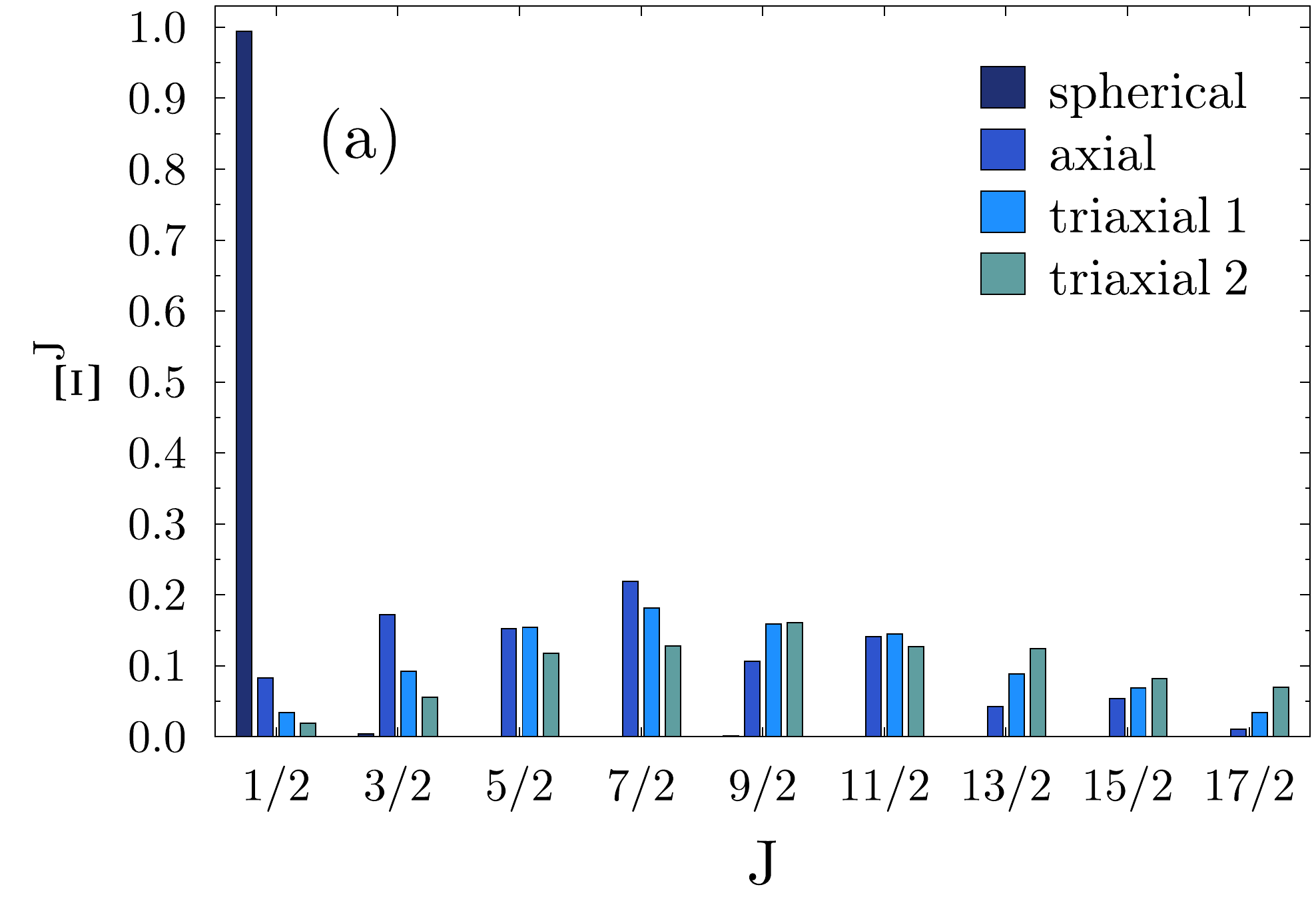} \\
  \includegraphics[width=8.5cm]{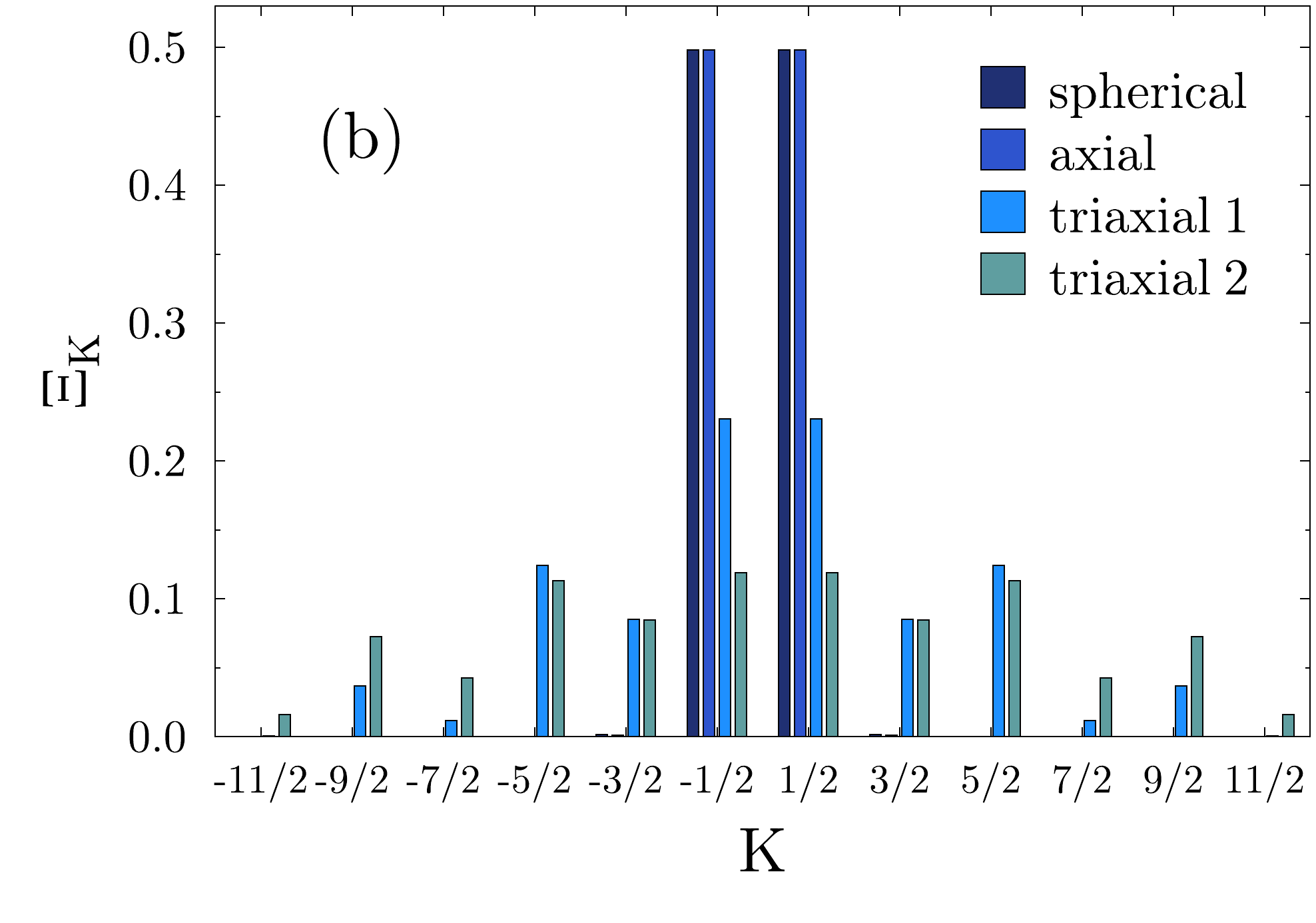} 
\caption{
\label{fig:phy1}
 (Color online) Top: quantity $\Xi^{J}$ for the selected quasiparticles states
 discussed in the text.
 Bottom: same for the quantity $\Xi_{J}$.
}
\end{figure}

In Fig.~\ref{fig:phy1}, we plot the two sums of components 
\begin{align} 
 \label{eq:sumofJK}
 \Xi^{J} & \equiv \sum_{K=-J}^{J} \elma{\Phi}{\hat{{P}}^{J}_{ K K} \hat{P}^Z \hat{P}^N}{\Phi} \, , \\
 \Xi_{K} & \equiv \sum_{J} \elma{\Phi}{\hat{{P}}^{J}_{ K K} \hat{P}^Z \hat{P}^N}{\Phi} \, , 
\end{align}
for the four quasiparticle vacua as specified in Tab.~\ref{tab:qpphy}. 
In addition to angular-momentum restoration, all states were also projected on
proton number $Z=12$ and neutron number $N=13$.

As can been observed, the decompositions of the four states are very different one from another,
but they all respect the symmetry $\Xi_{K}= \Xi_{-K}$, which is a consequence of the conserved 
$x$-signature of the reference state. This is to be contrasted with the more general 
asymmetric decomposition of state (b) discussed in Sec.~\ref{sec:numAMP} whose expectation value of
angular momentum is not constrained by symmetry to point into a specific direction.

For the ``spherical'' state, we can see that it has an almost pure $J=1/2$ and $|K| = 1/2$ character, 
the values being a fingerprint of the blocked $2s_{1/2}$ single-particle state in the canonical basis
of the quasiparticle vacuum \cite{RS80a}. Because of the constraint on vanishing quadrupole moments, polarization 
effects are tiny as components with different $J$ and $K$ are barely visible on the scale of Fig.~\ref{fig:phy1}.
It is also remarkable that, although the numerical representation on a three-dimensional cartesian mesh is not
closed under rotations, an almost pure eigenstate of angular momentum $J$ state can be created simply 
through a constraint on its average deformation.

The state labeled ``axial'' in Tab.~\ref{tab:qpphy} that was built by blocking 
a single-particle state that has a decoupling vector $d_z = 1/2$. We see from the decomposition 
that the many-body state is also of almost pure $|K|=1/2$ character. Because of its finite 
quadrupole moment, however, this state is a superposition of components with different 
values of $J$ that are spread to values larger than $17/2$.

Finally, looking at the two triaxial cases in Fig.~\ref{fig:phy1}, one observes that the states are now 
superpositions of components with different values of both $J$ and $K$. In general one finds that the spread
of the $K$ components increases with the degree of non-axiality as measured by the angle $\gamma$ of Bohr's
($\beta$, $\gamma$) parameterization of the quadrupole moment. Also, with increasing quadrupole deformation 
$\beta$, the distribution of $J$-components is more evenly spread and pushed towards larger values of the 
angular momentum \cite{Bender08a,Rodriguez10a}. While the state ``triaxial 1'' still conserves a certain fingerprint of the blocked 
single-particle state in its distribution with a dominant $|K|=1/2$ component, the state ``triaxial 2'' built at 
non-axiality angle $\gamma=30^\circ$ does not. 

We recall that the $K$ components projected out from a given state $\ket{\Phi}$ can in general not 
be interpreted as physical states. The two major reasons are that the $K$ components with given $J$ are 
in general not orthogonal, and that the spectrum of $K$ components depends on the orientation of the 
reference state, see the examples for even-even nuclei discussed in Refs.~\cite{Bender08a,Rodriguez10a}. 
The only exception is therefore the decomposition of a state that is axial with the $z$ axis as symmetry axis and that 
in addition is an eigenstate of $J_z$, such that projection yields at most only one $K$ component per 
irrep. With the choice of conserved $x$-signature as assumed throughout this Section, this special case 
is even limited to states that only have $K=0$ components.

%
%-----------------------------------------------------------------------
%
\subsubsection{Examples of $K$-mixing}

\begin{figure}[t!]
\centering  
  \includegraphics[width=7.0cm]{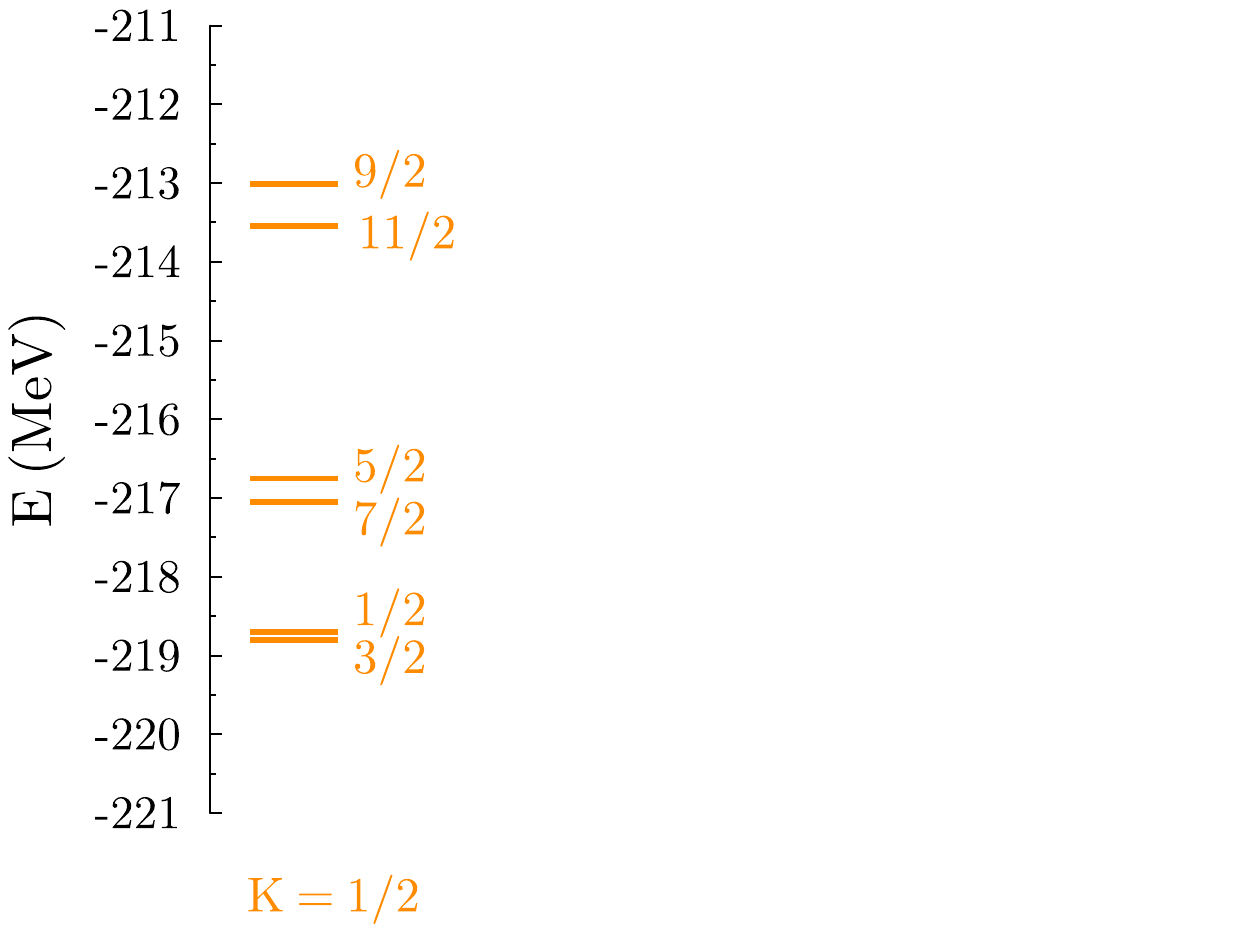}
\caption{\label{fig:phy2:ax}
(Color online) 
 Energy spectrum built from the expectation values $E^{JNZ}_{i K}$ for different values of $K$ obtained in the decomposition
 of the ``axial state'' as defined in Tab.~\ref{tab:qpphy}.
 }
\end{figure}

\begin{figure}[t!]
\centering  
  \includegraphics[width=7.0cm]{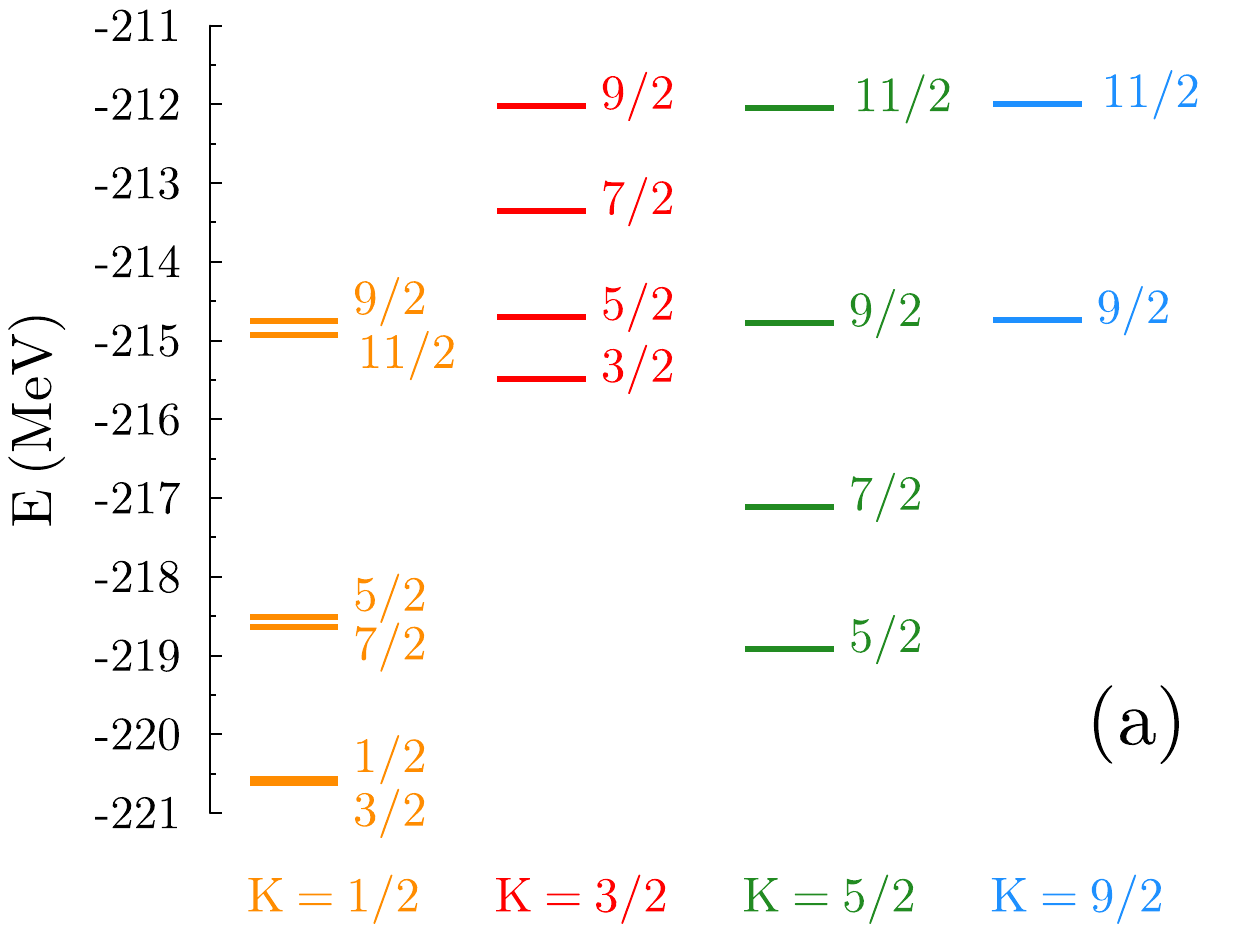} \\[0.2cm]
  \includegraphics[width=7.0cm]{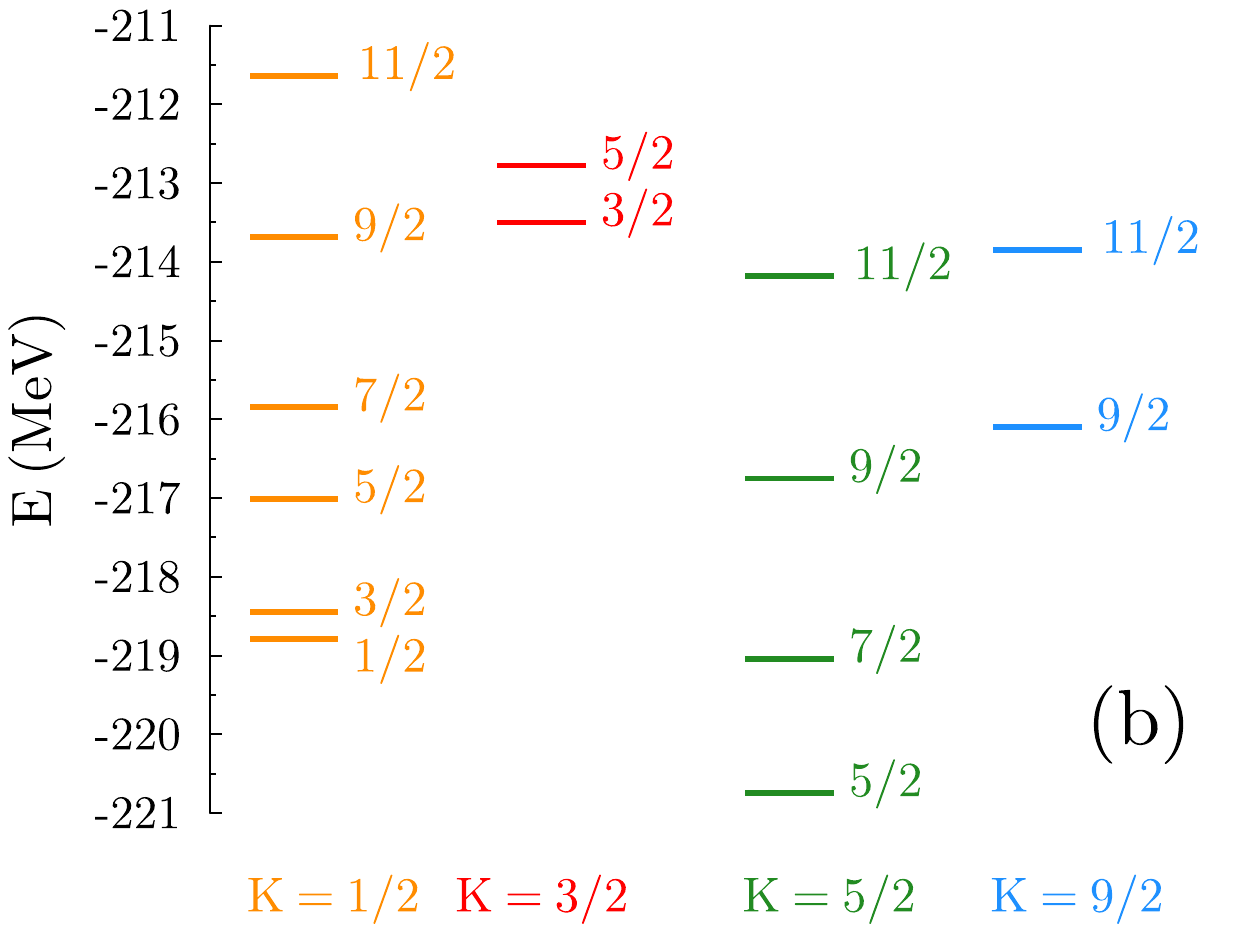}
\caption{\label{fig:phy2}
(Color online) 
Same as Fig.~\ref{fig:phy2:ax} for the decomposition states of the states $\ket{\Phi_a}$ (panel (a))
and $\ket{\Phi_b}$ (panel (b))
as defined in Tab.~\ref{tab:qpmix}. The color of the levels indicates the value of $K$, whereas the labels next to the
levels indicate the value of $J$.
}
\end{figure}

Whenever a state $\ket{\Phi}$ can be decomposed into several $K$ components with given $J$,
there is a necessary second step in the restoration of angular momentum that consists in
the mixing of the projected wave functions with same $M$ but different $K$, i.e.\ the states 
$\big\{ \hat{P}^{J}_{MK}\ket{\Phi}, K \in \llbracket -J, J \rrbracket \big\}$. It is only after this mixing 
that one obtains an orthogonal set of correlated states of given $J$ in the vector space $\text{span}(G\ket{\Phi})$ 
that diagonalize the Hamiltonian and that are independent on the orientation of the reference state $\ket{\Phi}$.
The effect of $K$ mixing sensitively depends on the properties of the state $\ket{\Phi}$ that is projected. It
may be very mild or change drastically the energy spectrum at hand. In Figs.~\ref{fig:phy2:ax} and~\ref{fig:phy2}
we display the spectrum built based on the projected energies 
\begin{equation}
 \label{eq:enkm}
 E^{JNZ}_{i K} 
 =  \frac{\elma{\Phi_i}{\hat{H} \, \hat{P}^{J}_{KK} \, \hat{P}^Z \, \hat{P}^N}{\Phi_i}}{\elma{\Phi_i}{\hat{P}^{J}_{KK} \, \hat{P}^Z \, \hat{P}^N}{\Phi_i}} \, ,
\end{equation}
for three different states,
considering only positive values of $K$ as the components with $\pm K$ are degenerate because of the $x$-signature imposed 
on all states $\ket{\Phi_i}$ considered throughout this Section. As explained in Sect.~\ref{sec:kmixing}, these pairs of states will 
contribute with weights of equal absolute value to the $K$-mixed states.

Figure~\ref{fig:phy2:ax} shows the energy spectrum for the ``axial state'' as defined in Tab.~\ref{tab:qpphy} that only 
has $K = \pm 1/2$ components. Applying the suitable transformation $ \boldsymbol{W}^J_{\eta}$ as defined in Eq.~\eqref{eq:WJeta:odd}
accomplishes the necessary $K$ mixing without the need to solve the GEP~\eqref{eq:gepsim}, and without changing the energy spectrum. 
In fact, this example illustrates a case that can also be described with the phenomenological Unified Model as defined in 
Refs.~\cite{Bohr76a,Rowe70a,BM98a,RS80a} that is widely used to interpret rotational bands of odd-mass nuclei in terms of simple 
intrinsic configurations relying on the schematic picture of
a single-particle state coupled to a deformed core. In the Unified Model, the 
treatment of the special case of rotational bands built on $k=1/2$ orbits requires the introduction of an additional 
$\Delta J = 1$ Coriolis interaction with a so-called ``decoupling parameter'', that in some cases pairwise switches the 
relative order of the energy levels compared to their usual scaling with $J(J+1) - K^2$, see also Ref.~\cite{Ring70a}. 
In an angular-momentum projected approach this effect is automatically described with a size that is entirely determined 
by the properties of the reference state, and the state used to prepare Fig.~\ref{fig:phy2:ax} is an example where the 
order of the states is indeed changed.

\begin{table}[b!]
\begin{ruledtabular}
\begin{tabular}{cccccccc}
label & $\langle \hat{Z} \rangle$ & $\langle \hat{N} \rangle$ & $(\beta_2, \gamma)$ & $\eta$ & $p$ & $\langle \hat{d}_z \rangle$ & $\langle \hat{d}_z \rangle_{\text{axial}}$ \\
\hline
a & 12 & 13 & $(0.60,14^\circ)$ & $-i$ & 1 &  0.65   & 0.50 \\
b & 12 & 13 & $(0.47,24^\circ)$   & $-i$ & 1 & 1.82  & 2.50
\end{tabular}
\caption{Characteristics of the quasiparticle vacua used to
prepare Figs.~\ref{fig:phy2} and~\ref{fig:phy3}.
}
\label{tab:qpmix}
\end{ruledtabular}
\end{table}

For two other triaxial quasiparticle vacua, whose characteristics are given in Tab.~\ref{tab:qpmix}, we plot 
in Fig.~\ref{fig:phy2} the spectra of $(J, K)$ components up to $K=9/2$. Their respective decomposition in 
energy is quite different from the one of Fig.~\ref{fig:phy2:ax} and also among each other. These two 
states have non-vanishing components for each possible value of $|K|$ for given $J$. Without $K$ mixing,
the levels of same $K$ but different $J$ do not necessarily form the usual pattern of a rotational band, which
is again most obvious for $K=1/2$ components for which the order of levels is again pairwise switched as
for the state discussed in Fig.~\ref{fig:phy2:ax}. At low $J$, the decomposition of state $\ket{\Phi_{\text{a}}}$ 
in general energetically favors the band based on $K=1/2$ components and gives an excited band based 
on $K=5/2$. For state $\ket{\Phi_{\text{b}}}$ one finds the opposite. This dominance is a fingerprint 
of the single-particle configuration that has been self-consistently blocked. Although the final many-body state 
is not pure, i.e.\ many bands appear in its decomposition, the dominant component is usually energetically 
favored. The other bands based on higher $K$ components are rather similar.

\begin{figure}[t!]
\centering  
  \includegraphics[width=7.7cm]{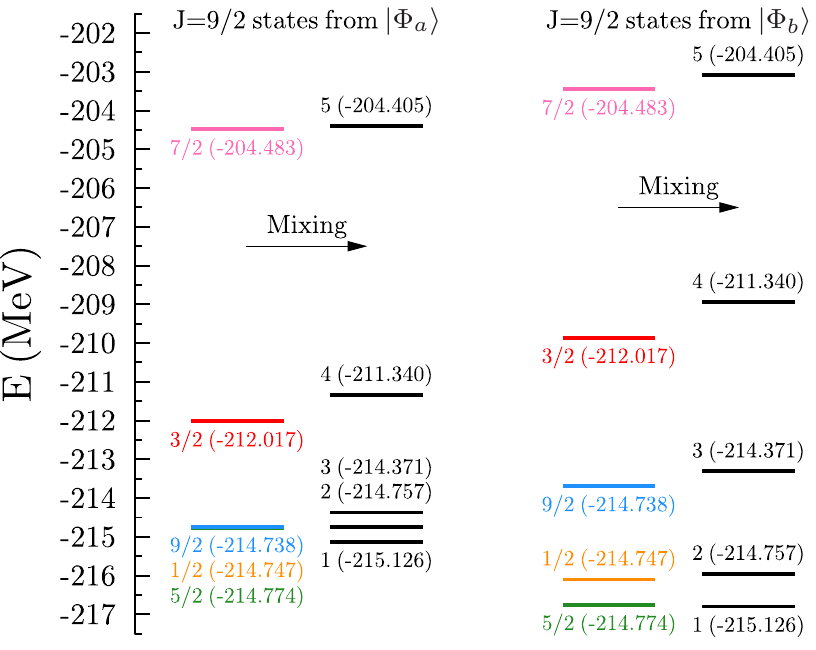} 
\caption{
\label{fig:phy3}
(Color online) Spectrum of $J=9/2$ states before and after $K$-mixing for the state $\ket{\Phi_{\text{a}}}$
and $\ket{\Phi_b}$ as defined in Tab.~\ref{tab:qpmix}. The energies $E^{JNZ}_{i K}$~\eqref{eq:enkm} of $K$ components 
are drawn in the same color code as used in Fig.~\ref{fig:phy2}, whereas energy levels after $K$ mixing 
are drawn in black. Note that the three lowest $K$ components of state $\ket{\Phi_{\text{a}}}$ are 
quasi-degenerate, such that their levels fall on top of each other on the scale of the plot.
}
\end{figure}

Both states have the particularity that one finds near-degenerate $J=9/2$ components with different $K$ in their decomposition. 
In Fig.~\ref{fig:phy3}, we compare the energies of these $K$ components with the eigenvalues of the projected states after $K$-mixing.
In the case of state $\ket{\Phi_{\text{a}}}$, we see that $K$-mixing has the effect of lifting the degeneracy of the three lowest states with 
$K=1/2$, $5/2$, and $9/2$, with one state visibly being lowered, another one being pushed up and the third staying in between
as can be expected from the general principles of configuration mixing. The $K=3/2$ and $K=7/2$ components are also clearly 
mixed with the others. The fourth state in the spectrum is actually more shifted energetically than the first three states despite 
the $K=3/2$ component being more isolated. For the state $\ket{\Phi_{\text{b}}}$, the values of the diagonal matrix elements
before mixing are more spread 
such that the diagonalisation does not change the global aspect of the spectrum. Nevertheless, the $K$-mixing still
has an effect on the energies and even the highest energy state, well above the others, is changed. All other observables are 
also obviously affected by the $K$ mixing, which we will not discuss. In both cases, there are no redundant components in the 
decomposition after the transformation $ \boldsymbol{W}^J_{\eta}$ as defined in Eq.~\eqref{eq:WJeta:odd} has been applied, 
such that one obtains five different $J=9/2$ states after $K$ mixing. 

%
%-----------------------------------------------------------------------
%
\subsubsection{Projection of cranked states}
\label{subsec:cranked}

The self-consistent cranking approach has been used for a long time to construct collective rotational 
bands in single-reference valence-space \cite{Ring70a} and EDF calculations~\cite{Bonche87a}. 
It can be motivated either as a self-consistent state in a rotating frame subject to Coriolis forces \cite{Thouless62a}, 
or as the first-order approximation to a VAP on angular momentum \cite{Kamlah68a,Beck70a,villars71a,Mang75a,islam79a}, 
see also Sect.~\ref{AMP:general}. The PAV on angular momentum of cranked Slater determinants and
Bogoliubov quasiparticle vacua has been used in exploratory studies using valence-space Hamiltonians \cite{Hara82a},
the BKN EDF \cite{Baye84a}, the Gogny force \cite{Borrajo15a,Shimada15a,Egido16b}, and the Skyrme EDF
\cite{Zdun07a,Zdunczuk07b}, in some cases combined with further configuration mixing of states with different
deformation and/or of states at different expectation value of angular momentum \cite{Egido16b}.

Practically speaking, assuming a good $x$-signature,\footnote{We recall that the minimization is actually made with a 
code in which the roles of the $x$ and $z$ axes are exchanged, see Sect.~\ref{subsubsect:applications:esperance:decomposition}.
For consistency with the rest of the discussion, Eq.~\eqref{eq:crconst} uses the orientation of the state as it is projected.} 
for states that are nearly axial around the $z$ axis, the self-consistent cranking approximation usually consists in adding 
an auxiliary condition on the expectation value of the form
\begin{equation} 
 \label{eq:crconst}
 \langle \hat{J}_x \rangle_{a(J_c)} \equiv \elma{\Phi_{a(J_c)}}{\hat{J}_x}{\Phi_{a(J_c)}} = \sqrt{J_{c}(J_{c}+1) - K^2_{c}} \, ,
\end{equation}
during the minimization procedure of the state $\ket{\Phi_{a(J_c)}}$ \cite{Ring70a,Beck70a,Ring74a,Ejiri89a},
where $J_c$ is the targeted angular momentum  and $K_c$ is the assumed length of the component of angular momentum 
in $z$ direction, which depends on the characteristics of the blocked quasiparticles of the state $\ket{\Phi_{a(J_c)}}$. 
We recall that when imposing $\langle \hat{P}, \hat{R}_x , \hat{S}^{\mathcal{T}}_y \rangle$ symmetry,
only the angular momentum operator in $x$ direction can have non-zero expectation value, and $K_c$ is related to the 
$z$ component of the decoupling vector as defined above.

Because angular momentum changes sign under time reversal, the cranking constraint necessarily
breaks the time-reversal invariance of the self-consistent procedure. 
The constraint~\eqref{eq:crconst} introduces a Coriolis interaction that in general acts differently 
on two nucleons in time-reversed single-particle states and thereby splits Kramers degeneracy. The reason 
is that one out of the two single-particle states has a component of angular momentum that is aligned with the cranking
constrained, whereas the angular momentum of its time-reversed partner state is anti-aligned.
If the cranking constraint  is the only source of time-reversal breaking, which implies $K=0$ in Eq.~\eqref{eq:crconst}, then cranking a 
state to $- \langle \hat{J}_x \rangle_{a(J_c)}$ generates the time-reversed of the state that is obtained when
cranking to $+ \langle \hat{J}_x \rangle_{a(J_c)}$, meaning that both states are equivalent as discussed in
Sect.~\ref{subsec:timerev}. 

When cranking a blocked quasiparticle vacuum, however, one can always construct two such pairs of 
non-equivalent solutions:
one for which one blocks a quasiparticle whose measurable $x$-component of angular-momentum is aligned 
with the direction of the cranking constraint, and one for which one blocks its time-reversed partner that is anti-aligned. The resulting 
many-body states cannot be transformed one into the other by time-reversal and will in general have slightly 
different total energy and and other observables~\cite{Ring74a}. For reasons elaborated for example in
Ref.~\cite{Ring70a,Ring74a,Mang75a} and recalled in what follows, it is customary to construct states 
cranked to $J_c= 1/2$, $5/2$, $9/2$, $13/2$, \ldots by blocking a single-particle level with $x$-signature 
$\eta_a = -i$, whereas states cranked to $J_c = 3/2$, $7/2$, $11/2$, $15/2$, \ldots are constructed by 
blocking a single-particle level with $\eta_a = +i$. These states are usually grouped into two
different signature-partner bands, for which one often finds experimentally that the levels in one band do
not fall exactly in between those in the other, but are slightly shifted \cite{Szymanski84a,Ejiri89a}, an effect 
that is in general also found in cranked HFB calculations.

\begin{figure}[t!]
\centering  
  \includegraphics[width=8.0cm]{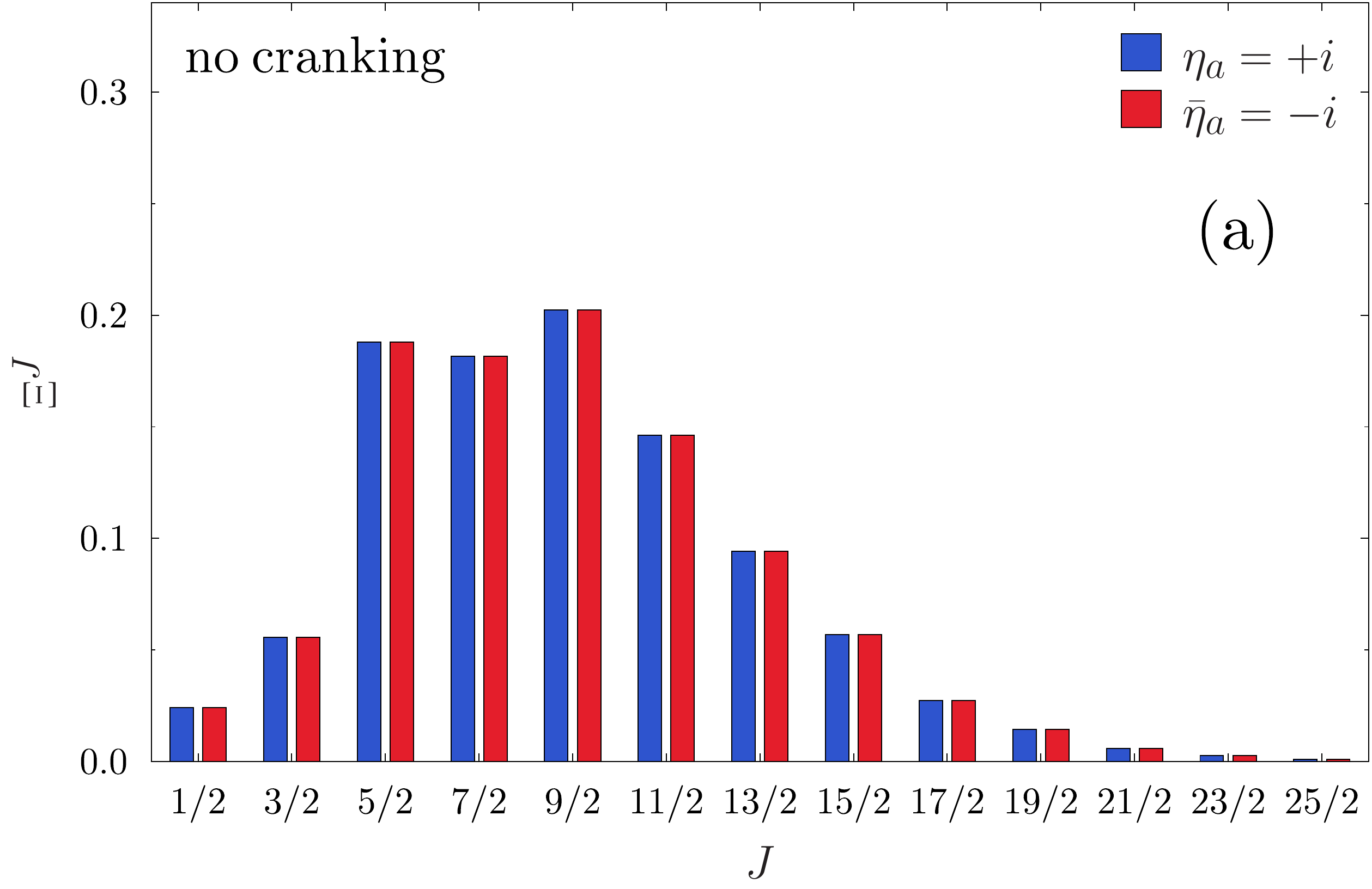}  \\ \vspace*{0.15cm} 
  \includegraphics[width=8.0cm]{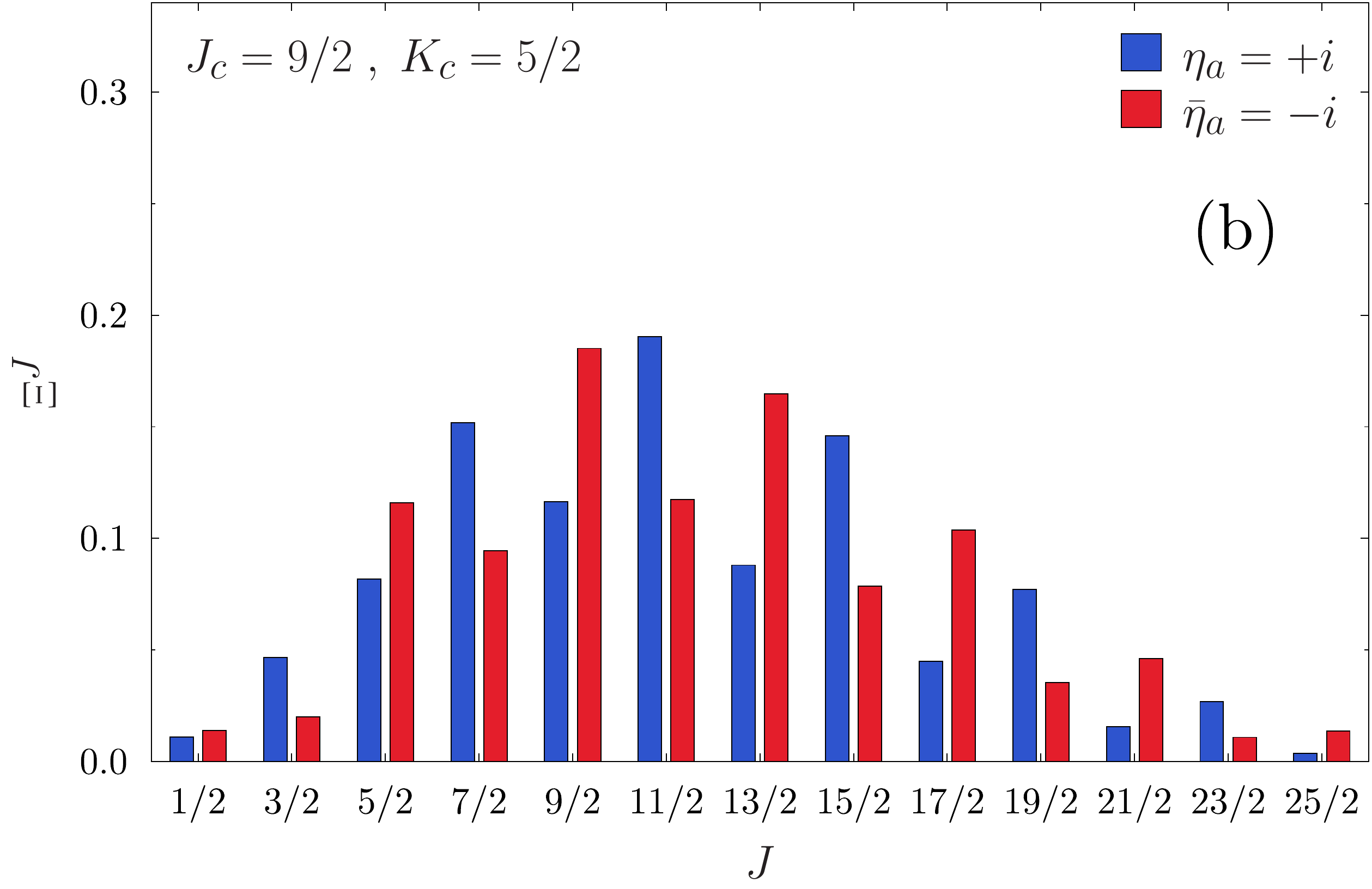} \\ \vspace*{0.15cm}
  \includegraphics[width=8.0cm]{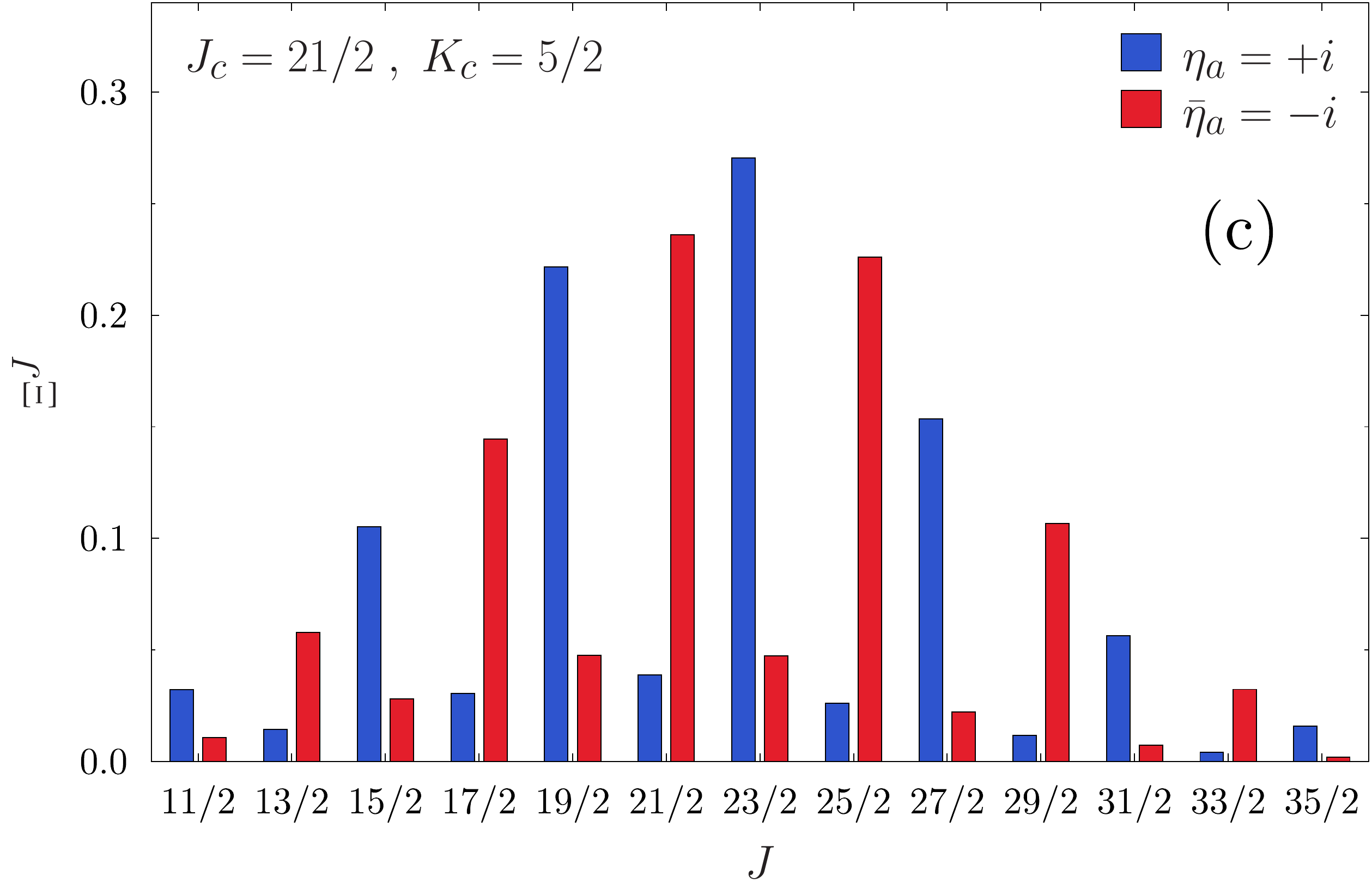}
\caption{
\label{fig:amrcr}
(Color online)
Summed weight $\Xi^{J}$ as defined in Eq.~\eqref{eq:weicr} of components with given $J$
projected out from one-quasiparticle states for \nuc{25}{Mg} obtained by blocking single particles in originally 
time-reversed partner states of opposite $x$-signature $\eta$, and cranked to different values 
of $J_c$.
}
\end{figure}

In order to illustrate this effect, we display in Fig.~\ref{fig:amrcr} the summed weight of the components
with given $J$ 
\begin{equation} 
 \label{eq:weicr}
 \Xi^{J} = \sum_{K=-J}^{J} N^{J}_{KK} \, , 
\end{equation}
obtained when projecting two one-quasiparticle states of \nuc{25}{Mg} obtained by blocking quasiparticles in 
(originally time-reversed) partner orbits of opposite  $x$-signature $\eta_a = +i$ and $\bar{\eta}_a = -i$, 
and which are cranked to different values of $J_c$ as defined in Eq.\ \eqref{eq:crconst}. These states are 
constructed in the same way as those discussed above.

As can be seen, in the absence of cranking constraint (panel (a)) the two distribution of the weights $\Xi^{J}$ 
are equal, up to numerical accuracy, which follows from the discussion in Sect.~\ref{subsec:timerev}.
With increasing value of the constrained angular momentum $J_c$ (panels (b) and (c)), the two distributions 
become different, with every second component becoming much smaller. In the limit of high collective angular 
momentum $J_c \gg K_c$ the two blocked states have a very different decomposition, and one then finds 
a distribution of components that mirrors the rules for the selection of the blocked single-particle levels 
mentioned above: cranked one-quasiparticle states $\ket{\Phi_{a(J_c)}}$ with $x$-signature $\eta_{a(J_c)} = +i$ 
are dominated by components with $J = 3/2$, $7/2$, $11/2$, $15/2$, etc, whereas one-quasiparticle states 
with $x$-signature $\eta_{a(J_c)} = -i$ are dominated by components with $J= 1/2$, $5/2$, $9/2$, $13/2$, etc.
We remark that this selection rule can also be understood using the Kamlah expansion \cite{Kamlah68a} at large spin
(see Ref.~\cite{BallyPHD} for more details).
 
Finally, we notice that as we increase the value of the constraint, the distribution of 
components move to larger values of $J$, and becomes wider. For the two cranked states, the 
distribution of the components is peaked near the cranked angular momentum, which is not
always the case \cite{islam79a,Baye84a,Zdun07a,Zdunczuk07b}.

\begin{figure}[t!]
\centering  
  \includegraphics[width=7.0cm]{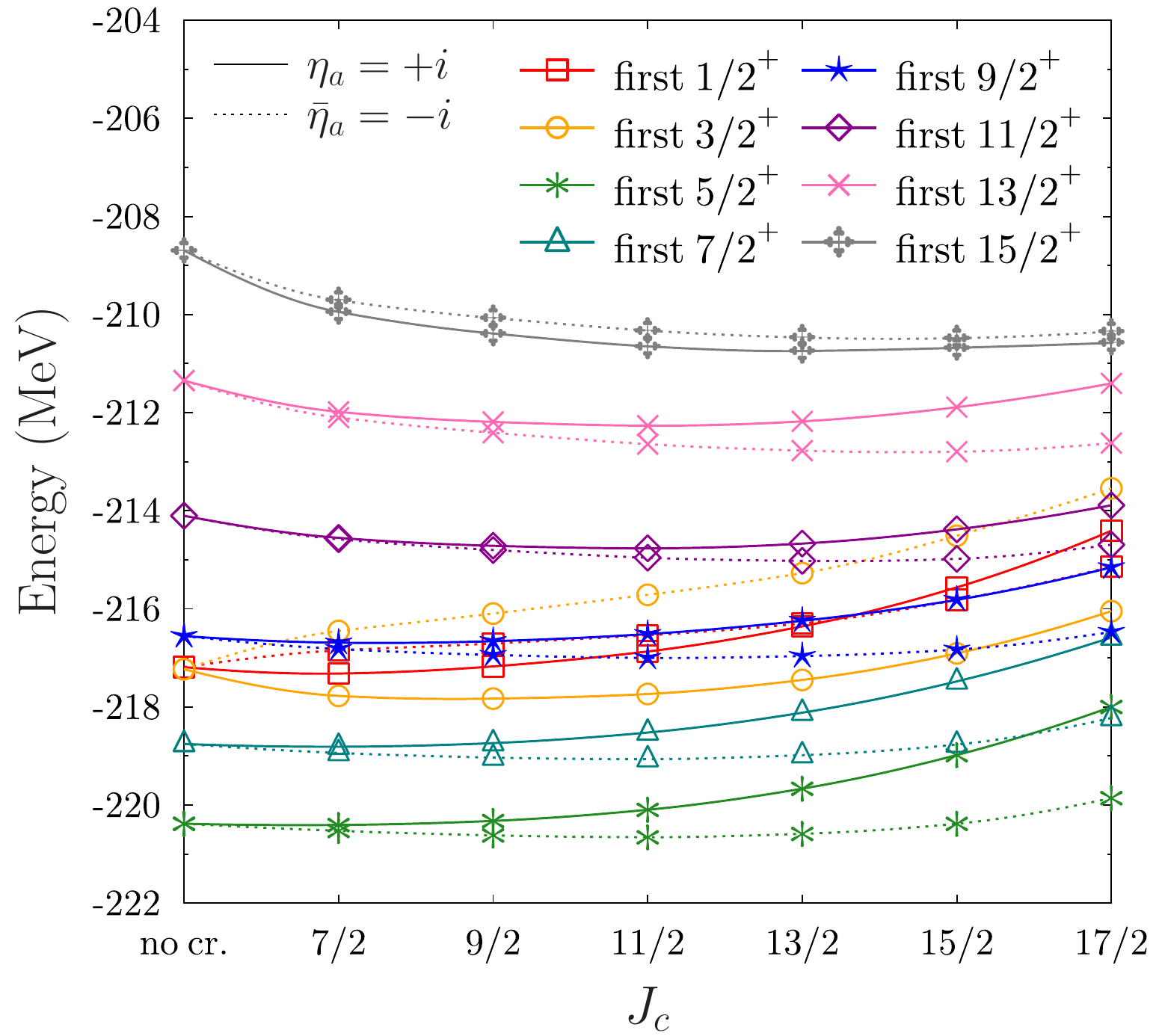}
\caption{
\label{fig:amrce}
(Color online)
Projected energies, as defined in Eq.~\eqref{eq:encr}, of yrast states projected from two
one-quasiparticle states for \nuc{25}{Mg} obtained from blocking two originally time-reversed partners 
of opposite $x$-signature $\eta_a$ cranked to different values of $J_c$ as indicated.
}
\end{figure}

The effect of cranking on blocked one-quasiparticle states can also be seen on the energies of the symmetry-restored states
\begin{equation}
 \label{eq:encr}
 E^{J N Z}_{a \epsilon(J_c)} = \elma{\Psi_{a \epsilon(J_c)}^{JMNZ}}{\hat{H}}{\Psi_{a \epsilon(J_c)}^{JMNZ}} \, ,
\end{equation}
displayed in Fig.~\ref{fig:amrce} for yrast states obtained when cranking the one-quasiparticle states to different 
values of $J_c$. The cranked states used to prepare this figure have the same slightly triaxial shape as the
states used to prepare Fig.~\ref{fig:amrcr}, but cover also intermediate values of the cranked angular momentum $J_c$.
First, we observe that in the absence of a cranking constraint the energies of the 
projected states built from the states of opposite signatures are strictly equal for a given $J$, which 
is expected as the two states are in that case connected by time-reversal. By contrast, when a cranking constraint 
is added during the minimization, the projected energies for given $J$ split. Their difference varies greatly
depending on $J$ and $J_c$. For all values of $J$ the cranked states always give 
lower projected energies than the non-cranked ones, with the gain in energy usually increasing with $J$ 
as already found earlier \cite{Borrajo15a,Shimada15a,Egido16b,Zdun07a,Zdunczuk07b}. This confirms that 
self-consistent cranking optimizes the states for subsequent projection on high angular momentum as should
be expected from an approximation to a VAP calculation. But for the values of cranked and projected angular momenta
covered by Fig.~\ref{fig:amrce}, we do not observe any clear pattern indicating 
which $x$-signature will be favored energetically.

%
%=======================================================================
%
\section{Summary and Conclusions}

The quantum-number projection technique is a powerful tool that permits to build symmetry-adapted
wave functions starting from any arbitrary many-body reference state.\footnote{However, this has to be 
complemented by an
efficient technique to evaluate the operator kernels, a question we have not addressed here as it 
depends on the nature of the states to be projected and their numerical representation. For an overview
over the various techniques to evaluate such kernels for multi-reference energy density functionals, see 
\cite{Bender19EDF3}. 
}
In addition, when defined as the diagonalization of the nuclear Hamiltonian in the subspace spanned by the set
of all rotated states (built by rotating the initial reference state in all possible ways) as elaborated 
in Section~\ref{sec:proj}, the method provides a procedure to grasp correlation energy that 
is absent in a symmetry-breaking calculation The eigenstates thus obtained represent variationally 
optimal wave functions (in this subspace) that respect the symmetries of the problem and therefore can be 
used to compute unambiguously the expectation values of observables of interest. 
In particular, the projection restores selection rules for the matrix elements of tensor operators.

In Sections~\ref{sec:PNP} and~\ref{sec:AMP}, we illustrated the method in the case of 
particle-number and angular-momentum projections. In each case, we presented the general
principles of the projection as well as the practical aspects of its numerical implementation.
While the group $U(1)$ associated with a good number particles is abelian,
the group $SU(2)$ associated with a good angular momentum is not, which
makes the latter projection much more involved. 
Nevertheless, it is possible in both cases to use efficient numerical discretizations of 
the integrals over group elements. When a sufficiently large number of discretization points
is used, the integration is performed exactly up to the numerical noise. 

We have limited our discussion to these two continuous symmetries. Another case of 
interest not addressed here is the breaking and restoration of parity \cite{Egido91a}, 
which from a formal and practical point of view is much simpler and straightforward: as it is a discrete 
abelian symmetry, parity projection of a reflection-symmetry-breaking state always corresponds 
to the mixing of exactly two states, the original one and its parity-inverted image. 
Their superposition span a vector space of dimension two that is the direct product of 
two one-dimensional subspaces corresponding to the irreps with parity $\pm 1$. As a 
consequence, there are no questions about the discretization of the numerical 
representation of the parity projection operator in order to achieve optimal convergence.

Finally, in section \ref{sec:simple}, we gave formal results that permit a great deal of 
simplifications when dealing with reference states that have the symmetry properties of the 
subgroup $\langle \hat{P}, \hat{R}_x , \hat{S}^{\mathcal{T}}_y \rangle$
of the group $D_{2h}^{TD}$, often employed in nuclear physics to describe collective phenomena
in axially and triaxially deformed nuclei.
For this class of states, we derived the most general expressions that can be applied 
to any reference state, whatever its number parity or the angular momentum considered.

Up to now, the projection method has been used extensively within multi-reference formalism 
based on Slater determinants and Bogoliubov quasiparticle vacua, either using valence-space
Hamiltonians or energy density functionals, to tackle a large variety of phenomena 
\cite{Bender03a,Egido16a,Robledo18a,Sheikh19a}. But there are also new ongoing developments
to design \textit{ab-initio} schemes that integrate symmetry restoration as one of their 
fundamental principles \cite{Duguet14a,Duguet17a,Hermes17a,Qiu17a,Qiu19a,Yao19a}.
For that reason, we precise that although we illustrated our discussion making use of
Bogoliubov quasiparticle reference states, most of the results presented here can be 
directly carried over to more general types of symmetry-breaking reference states.

%
%===================================================================
%
\section*{Acknowledgments}

This work has profited from many stimulating discussions that we had 
over the years with many practitioners of projection techniques, among 
whom we would like to single out the late Paul-Henri Heenen, who turned
our attention to many of the practical aspects of this study that can be used
to boost the numerical efficiency of projection methods.
B.\ B.\ would like to thank T.\ R.\ Rodr{\'i}guez for his work on the code \textsf{TAURUS}.
This project has received funding from the European Union's Horizon 2020 
research and innovation programme under the Marie Skłodowska-Curie grant 
agreement No.~839847.
%
%===================================================================
%
\appendix
%
%===================================================================
%
\section{Discretizations for particle-number projection}
\label{sec:discpnr}

In Sec.~\ref{sec:PNP}, we introduced a discretized projection operator for 
particle-number projection based on the original idea by Fomenko~\cite{Fomenko70a}. 
Actually, there exist other choices that behave similarly 
but differ by the position of the quadrature points in the interval $[0,\pi]$. 
Indeed, choosing $\alpha \in [0,1]$ one defines the general discretized projection operator 
\begin{equation}
\label{eq:disopappa}
 \hat{\mathbb{P}}^{N_0}_{\alpha, M_\varphi} 
 \equiv \frac{1}{M_\varphi}\,\sum_{m=1}^{M_\varphi} e^{-\iunit  \pi \frac{m-\alpha}{M_\varphi} (\hat{N} - N_0)} \, .
\end{equation}
All the operators corresponding to different values of $\alpha$ have in common that two consecutive angles are always separated by a step 
of $\pi/{M_\varphi}$, with the main difference being the position of the first angle.
In this section we will analyze their action on a given wave function, focusing in particular on four specific values of $\alpha$.
For the sake of keeping the discussion simple, we will consider only 
one generic particle species $N$ without any loss of generality.

The four possible choices for the discretized projection operator we will focus on are
\begin{align}
\label{eq:disopapp1}
\hat{\mathbb{P}}^{N_0}_{1, M_\varphi} &\equiv \frac{1}{M_\varphi}\,\sum_{n=1}^{M_\varphi} e^{-\iunit  \pi \frac{n-1}{M_\varphi} (\hat{N} - N_0)} \, , \\ 
\label{eq:disopapp2}
\hat{\mathbb{P}}^{N_0}_{1/2, M_\varphi} &\equiv \frac{1}{M_\varphi}\,\sum_{n=1}^{M_\varphi} e^{-\iunit  \pi \frac{n-\frac12}{M_\varphi} (\hat{N} - N_0)} \, ,\\ 
\label{eq:disopapp3}
\hat{\mathbb{P}}^{N_0}_{1/4, M_\varphi} &\equiv \frac{1}{M_\varphi}\,\sum_{n=1}^{M_\varphi} e^{-\iunit  \pi \frac{n-\frac14}{M_\varphi} (\hat{N} - N_0)} \, ,\\ 
\label{eq:disopapp4}
\hat{\mathbb{P}}^{N_0}_{3/4, M_\varphi} &\equiv \frac{1}{M_\varphi}\,\sum_{n=1}^{M_\varphi} e^{-\iunit  \pi \frac{n-\frac34}{M_\varphi} (\hat{N} - N_0)} \, , 
\end{align}
where $\hat{\mathbb{P}}^{N_0}_{1, M_\varphi}$ is the standard one used in the core of the text.
 
Although they are otherwise very similar, the first of these discretized operators \eqref{eq:disopapp1}
has the practical advantage that it reduces to the identity operator when taking 
$M_\varphi = 1$. By contrast, the third and fourth discretized operators never 
require the evaluation of the angle $\frac{\pi}{2}$ that might become
numerically problematic in pure particle-number projection \cite{Anguiano01a}, see also 
Sect.~\ref{subsect:PN:num:imp}. However, this problem can be avoided anyway by taking 
an odd number of points $M_\varphi$ for \eqref{eq:disopapp1}, respectively an even
value of $M_\varphi$ for \eqref{eq:disopapp2}.
In addition, when combined with other projections or a configuration 
mixing calculation the problem might appear at any gauge angle, not just $\frac{\pi}{2}$.

We will now analyze the action of operator $\hat{\mathbb{P}}^{N_0}_{\alpha, M_\varphi}$ 
acting on a reference state $\ket{\Phi}$ that is an eigenstate of number parity
with an eigenvalue compatible with $N_0$, i.e.\ $\pi_{n} = (-)^{N_0}$.
Using the decomposition~\eqref{eq:decompoN} of $\ket{\Phi}$, we obtain that
\begin{equation}
\label{eq:appA:decomp:1}
 \hat{\mathbb{P}}^{N_0}_{\alpha,M_\varphi} \ket{\Phi} 
 = \sum_{N_1 \ge 0} c^{N_1} \Bigg( \frac{1}{{M_\varphi}} \sum_{m=1}^{{M_\varphi}} 
   e^{-\iunit  \pi \frac{m-\alpha}{{M_\varphi}} (N_1 - N_0)} \Bigg) \ket{\Psi^{N_1}} \, . 
\end{equation}
The factor 
\begin{displaymath}
\frac{1}{{M_\varphi}} \sum_{m=1}^{{M_\varphi}} e^{-\iunit  \pi \frac{m-\alpha}{{M_\varphi}} (N_1 - N_0)}
\end{displaymath}
on the r.h.s.\ of Eq.~\eqref{eq:appA:decomp:1} is a geometric progression, i.e.\ a sum of 
the type $\sum_{k=i}^j \, a \, r^k $, with $r$ being the common ratio and $a$ the scale factor. 
Depending on the value of $r$, the result of a geometric progression can be expressed 
in a closed form as
\begin{displaymath}
 \sum_{k=i}^j \, a \, r^k =  
  \left\{  \begin{array}{cc}
        a \, (j - i + 1)   & \text{if } r = 1  \, , \\
             & \\
        a \, \frac{r^i - r^{j+1}}{1 - r} & \text{if } r \neq 1 \, .
  \end{array}  \right. 
\end{displaymath}
For the discretized projection operator $\hat{\mathbb{P}}^{N_0}_{\alpha,M_\varphi}$,
we have
\begin{displaymath}
\begin{split}
 i &= 1 \, , \\
 j &= M_\varphi \, , \\
 r &= e^{-\iunit  \pi \frac{1}{{M_\varphi}} (N_1 - N_0)} \, ,\\
 a &= \frac{e^{+\iunit  \pi \frac{\alpha}{M_\varphi}(N_1 - N_0)}}{{M_\varphi}} \, . 
\end{split}
\end{displaymath}
The case $r=1$ is realized if and only if $N_1 = N_0 + 2 \ell {M_\varphi}$ with $\ell \in \mathbb{Z}$. 
For those values of $N_1$, one thus obtains
\begin{equation}
\begin{split}
 \hat{\mathbb{P}}^{N_0}_{\alpha, M_\varphi} \ket{\Psi^{N_1}} 
 &= \frac{e^{+\iunit  2 \pi \alpha \ell}}{M_\varphi} \, \big( M_\varphi - 1 + 1 \big) \, \ket{\Psi^{N_1}} \\
 &= e^{+\iunit  2 \pi \alpha \ell} \, \ket{\Psi^{N_1}} \, .
\end{split}
\end{equation}
For any other value of $N_1$, one obtains instead
\begin{equation}
\begin{split}
\label{eq:flamvanish}
 \hat{\mathbb{P}}^{N_0}_{\alpha, M_\varphi} \ket{\Psi^{N_1}} 
 & = a \frac{1-e^{-\iunit  \pi (N_1 - N_0)}}{1 - e^{-\iunit  \pi \frac{1}{{M_\varphi}} (N_1 - N_0)}} \, \ket{\Psi^{N_1}} \\
 & = 0 \, ,
\end{split}
\end{equation}
where we have used that $\ket{\Phi}$ is an eigenstate of number parity,
such that we  always have the relation $N_1 = N_0 + 2n$, $n \in \mathbb{Z}$, 
for the components in~\eqref{eq:decompoN}, which implies that
$1-e^{-\iunit  \pi (N_1 - N_0)} = 0$. But note that this is true if and only
if the state $\ket{\Phi}$ has a number parity equal \mbox{to $(-)^{N_0}$,} 
hence the assumption made earlier.

To summarize, by applying $\hat{\mathbb{P}}^{N_0}_{\alpha, M_\varphi}$ on 
$\ket{\Phi}$, we remove all components with 
$N_1 \ne N_0 + 2 \ell {M_\varphi}, \, \ell \in \mathbb{Z}$,  
\begin{equation}
\begin{split}
 \hat{\mathbb{P}}^{N_0}_{\alpha, M_\varphi} \ket{\Phi} 
 &=  \sum_{\ell \in \mathbb{Z}} e^{+\iunit  2 \pi \alpha \ell} \, c^{N_0 + 2 \ell M_\varphi} \, \ket{\Psi^{N_0 + 2 \ell M_\varphi }} \\
 &\equiv \sum_{\ell \in\mathbb{Z}} \sum_{N_1 \ge 0} e^{+\iunit  2 \pi \alpha \ell} \, c^{N_1} \, \ket{\Psi^{N_1 }} \, 
          \delta_{N_1  N_0 + 2 \ell M_\varphi} \, .
\end{split}
\end{equation}
We can notice in particular that the discretized projection operator exhibits a periodicity of $2M_\varphi$, i.e.
\begin{equation}
 \forall \, \ell_1 \in \mathbb{Z} , \, \hat{\mathbb{P}}^{N_0 + 2 \ell M_\varphi}_{\alpha, M_\varphi} \ket{\Phi} 
 = e^{- \iunit 2 \pi \alpha \ell_1} \hat{\mathbb{P}}^{N_0}_{\alpha, M_\varphi} \ket{\Phi} \, . 
\end{equation}
As a consequence, the projection on any particle number $N_2$ that is such that 
$N_2 = N_0 + 2 \ell_1 M_\varphi$, $\ell_1 \in \mathbb{Z}$, will yield same 
result, up to a phase, as the projection on $N_0$, even if the component
with $N_2$ particles is not present in the wave function that is projected.
In particular, if $\alpha \ell_1 \in \mathbb{Z}$, the projected wave functions are strictly equal.  

Replacing the value of $\alpha$, we directly obtain the action of the four discretization operators of interest
\begin{alignat}{3}
 \hat{\mathbb{P}}_{1  ,M_\varphi}^{N_0} \ket{\Phi} &= \sum_{l\in\mathbb{Z}} &                   & c^{N_0 + 2 \ell M_\varphi} \, \ket{\Psi^{N_0 + 2lM_\varphi}} \, , \\
 \hat{\mathbb{P}}_{1/2,M_\varphi}^{N_0} \ket{\Phi} &= \sum_{l\in\mathbb{Z}} & (-)^\ell          & c^{N_0 + 2 \ell M_\varphi} \, \ket{\Psi^{N_0 + 2lM_\varphi}} \, , \\
 \hat{\mathbb{P}}_{1/4,M_\varphi}^{N_0} \ket{\Phi} &= \sum_{l\in\mathbb{Z}} & (\iunit)^\ell     & c^{N_0 + 2 \ell M_\varphi} \, \ket{\Psi^{N_0 + 2lM_\varphi}} \, , \\
 \hat{\mathbb{P}}_{3/4,M_\varphi}^{N_0} \ket{\Phi} &= \sum_{l\in\mathbb{Z}} & (-\iunit)^\ell \, & c^{N_0 + 2 \ell M_\varphi} \, \ket{\Psi^{N_0 + 2lM_\varphi}} \, .
\end{alignat}
As previously discussed, such discretized projection operators can only be used for states with a number parity equal to $(-)^{N_0}$.
It is straightforward to define more general discretized projection operators that can be used regardless of the number parity of
the state to be projected by simply replacing the factor $\pi$ in the exponential of any of these operators by a 
factor $2\pi$. Indeed, in that case we can lift the number parity condition for Eq.\ \eqref{eq:flamvanish} as for any values of 
$N_0$ and $N_1$, we always have in the numerator: $1-e^{-\iunit  2\pi (N_1 - N_0)} = 0$.
But using the same rationale as above, it is straightforward to derive that such a discretized operators
select out of the reference state all the components with $N_1 = N_0 + l {M_\varphi}, \, l\in \mathbb{Z}$. 
Consequently, from a convergence point of view, these operators are less effective than the ones defined with a factor $\pi$.
More precisely, the periodicity of the generalized operators being twice as small as the one of the operators 
defined in Eqs.\ \eqref{eq:disopapp1}-\eqref{eq:disopapp4}, this would require a number of points ${M_\varphi}$, 
i.e.\ a number of rotations over gauge angles to be performed, twice as large to remove the same number of undesired 
components in the superposition \eqref{eq:decompoN}.

%
%===================================================================
%

\bibliography{biblio,biblio_mb.bib}

%
%===================================================================
%
\end{document}